\documentclass[prd,aps,floats,floatfix,superscriptaddress,preprintnumbers,showpacs,eqsecnum,nofootinbib,twocolumn]{revtex4}
\usepackage{amsfonts,amsmath,epsfig,amssymb,latexsym,color,graphicx,array,subfigure,bm,float,nameref}\usepackage{hyperref} 

\definecolor{red}{rgb}{1,0,0}
\definecolor{blue }{rgb}{0,0,1}
\definecolor{green}{rgb}{0,1,0}


\usepackage{amsfonts}
\usepackage{dcolumn}
\usepackage {diagbox}
\allowdisplaybreaks[1]
\usepackage{stackengine}
\usepackage{appendix}

\newcommand{\bea}{\begin{eqnarray}}
\newcommand{\ena}{\end{eqnarray}}
\newcommand{\beann}{\begin{eqnarray*}}
\newcommand{\enann}{\end{eqnarray*}}

\newcommand{\dsl}{\pa \kern-0.5em /}

\newcommand{\pa}{\partial}

\newcommand{\nn}{\nonumber\\}

\newcommand{\vect}[1]{\!\!\mbox{ \boldmath $#1$}}
\newcommand{\gsim}{\, \mbox{\raisebox{-1.ex}
{$\stackrel{\textstyle>}{\textstyle\sim}$}}\,}

\allowdisplaybreaks

\begin{document}

\date{\today}

\title{Dynamical von Zeipel-Lidov-Kozai Oscillations of a Binary \\
on a Spherical Orbit around a Rotating Supermassive Black Hole}

\author{Kei-ichi Maeda}
\affiliation{
Department of Physics, Waseda University, 3-4-1 Okubo, Shinjuku, Tokyo 169-8555, Japan}
\affiliation{Center for Gravitational Physics and Quantum Information, Yukawa Institute for Theoretical Physics, Kyoto University, 606-8502, Kyoto, Japan}

\author{Hirotada Okawa}
\affiliation{Faculty of Software and Information Technology, Aomori University, Seishincho, Edogawa, Tokyo 134-0087, Japan
}

\begin{abstract}
 
 We study the dynamics of a binary system orbiting a rotating supermassive black hole (SMBH).
Using Fermi-Walker transport, we construct a local inertial reference frame in the Kerr spacetime and set up a Newtonian binary system.
Assuming the binary moves on a spherical orbit with constant radius around the black hole, we derive the equations of motion governing its dynamics.

We focus on von Zeipel-Lidov-Kozai (vZLK) oscillations, which arise when the binary is compact and its initial inclination exceeds a critical angle.
In our previous work on a circular orbit in the equatorial plane, we found that for hard binary systems, these oscillations in eccentricity and inclination follow a regular pattern, whereas in soft binaries, they exhibit chaotic behavior with irregular periods and amplitudes, yet remain stable.

In this study, we extend our analysis to a spherical orbit in the Kerr background.
The libration of the binary’s orbit in the latitudinal direction affects the vZLK oscillations: as the libration angle increases, the oscillation period shortens, and the maximum eccentricity grows, particularly when the oscillations become chaotic.
Notably, when the binary is sufficiently soft yet remains stable, the oscillation period is reduced to the dynamical timescale rather than the secular timescale.
This effect arises due to the interaction between the SMBH spin and the binary’s angular momentum.
As the Kerr rotation parameter increases or the radius of the spherical orbit decreases, both the enhancement of maximum eccentricity and the reduction in oscillation period become more pronounced.

\end{abstract}

\maketitle




\section{Introduction}
\label{Introduction}

The groundbreaking discovery of gravitational waves (GWs) by the LIGO-Virgo-KAGRA (LVK) Collaboration~\cite{GW150914,Abbott_2020,Abbott_2021} has ushered in a new era in astronomy and physics, revolutionizing our understanding of the Universe. One remarkable outcome is the identification of exceptionally massive stellar-mass black holes (BHs)\cite{GW150914}. With the expected increase in GW detections, we have a unique opportunity to test gravity in strong-field regimes and explore the redshift distribution of BHs and their environments~\cite{test1,test2,test3,test4,test5}. Realizing this potential, however, requires
 precise modeling of expected GW waveforms.

While current LVK detections primarily involve isolated binary systems, hierarchical triple systems may also play a significant role~\cite{Martinez_2020,Gerosa_2021}. In dense galactic nuclei surrounding supermassive black holes (SMBHs), binaries can naturally evolve into hierarchical triples~\cite{Heggie1975,Hut1993,Samsing2014,Riddle2015,Fabio2016,stephan2019}. Recent LIGO data suggest hierarchical mergers could be a key formation channel for binary BH coalescence~\cite{sym13091678,Gayathri_2020,Gerosa_2021}. Motivated by these findings, this paper explores the dynamics of such systems and their potential as GW sources.

Hierarchical triples consist of an ‘inner’ binary and a distant third body. The dynamics of these systems  were first analyzed by von Zeipel in 1910~\cite{vonZeipel10} and later rediscovered independently by Lidov and Kozai in 1962~\cite{Lidov62,Kozai62}, leading to what is now known as the von Zeipel-Lidov-Kozai (vZLK) resonance. This mechanism induces periodic exchanges between orbital eccentricity and inclination on secular timescales~\cite{Shevchenko17}, which can drive eccentricities  close to  unity and significantly enhance GW emission within 
the sensitivity range of space-based detectors.  While ground-based detectors typically observe binaries after their eccentricity has decayed, hierarchical interactions can still influence GW detection rates~\cite{Kimball_2021}.

The dynamics of hierarchical triples have been extensively
studied in Newtonian and post-Newtonian 
frameworks~\cite{naoz13b,Naoz12,Naoz16,Naoz2020,tey13,Li15,Will14a,Will14b}. 
Previous studies have analyzed the role of a massive tertiary using 
double-averaged equations of motion to explore relativistic effects such 
as de Sitter and Lense-Thirring precession~\cite{Liu:2019tqr,Liu:2021uam}. Additionally, three-body post-Newtonian (3BpN) secular effects have been shown to impact eccentricity and inclination evolution~\cite{Lim2020}. The influence of SMBH spin, particularly via Lense-Thirring precession and gravitomagnetic forces, has been examined in the context of binary BH merger timescales~\cite{Fang_2019a,Fang_2019b}.

Recent studies have further explored three-body systems
 and their GW emissions~\cite{Amaro-Seoane2010, Antonini2012,hoang18,Antonini2016,Meiron2017,Robson2018,Lisa2018,Lisa2019,Hoang2019,Loeb2019,Gupta_2020,kuntz2022transverse,Chandramouli_2022}. Notably, when a massive tertiary is present, the vZLK timescale can be reduced to a few years, making recurrent GW signals from vZLK oscillations detectable~\cite{Hoang2019, Gupta_2020}.
 Indirect GW observations from triple systems have also been explored via periastron shifts in binary pulsars undergoing vZLK oscillations~\cite{Haruka2019,Suzuki:2020zbg}. 

In this study, we consider a binary system orbiting an SMBH, treating it as a perturbation within the SMBH spacetime. While a single object behaves as a test particle under SMBH gravity, binary dynamics is more complex  due to strong mutual interactions. To describe the binary’s motion,
we construct  a local inertial frame and employ Newtonian gravitational dynamics, assuming moderate internal interactions.
Local inertial frames are established using Fermi normal coordinates or 
Fermi-Walker transport~\cite{1963JMP.....4..735M,Nesterov_1999,Delva:2011abw}.
This method has  applied to tidal effects near SMBHs~\cite{Banerjee_2019,PhysRevD.71.044017,Cheng_2013,Kuntz_2021} 
as well as binary dynamics~\cite{Gorbatsievich_Bobrik,Chen_Zhang,camilloni2023tidal,Maeda:2023tao,Maeda:2023uyx,Zhang_2024,Camilloni_2024}.

In our previous work~\cite{Maeda:2023tao,Maeda:2023uyx}, 
we studied vZLK oscillations in a binary system confined to the equatorial plane of Kerr spacetime. 
We found that vZLK oscillations can persist even near the innermost stable circular orbit (ISCO) for sufficiently inclined compact binaries. While highly compact binaries exhibit regular oscillations, softer binaries display chaotic vZLK oscillations with irregular periods and amplitudes. For binaries with high initial inclination, we even observed orbital flips of the relative inclination between the inner and outer orbits. Notably, we found that these results were largely independent of the black hole spin, as the Riemann curvature on the equatorial plane in Kerr spacetime is identical to that in Schwarzschild spacetime.

However, binaries around SMBHs are not necessarily confined to the equatorial plane, particularly if they form via dynamical capture. To address this, we extend our analysis in this paper to binary systems orbiting on a spherical orbit with constant radius in Kerr spacetime. This generalization allows us to explore how the SMBH’s spin and the binary’s latitude affect vZLK oscillations and their impact on GW emission (see also \cite{Zhang_2024}, in which they discuss the case of a parabolic orbit with arbitrary inclination).

This paper is organized as follows: 
Section II summarizes the properties of spherical orbits for test particles in Kerr spacetime, including their innermost stable orbits.
Section III derives the equations of motion for a binary system 
around a Kerr SMBH, assuming its center of mass (CM) moving along a spherical orbit. We construct a local inertial frame and derive the Lagrangian describing the binary’s motion.
Section IV reformulates the equations in dimensionless form, introduces key orbital parameters, and provides stability conditions for hierarchical triples.
Section V presents numerical analyses of vZLK oscillations. After briefly summarizing previous results for binaries confined to the equatorial plane, we explore cases where the binary is on a spherical orbit. We analyze the effects of latitudinal libration, eccentricity enhancement, and vZLK period reduction. We also examine the dependence of these effects on binary radius and SMBH spin.
Section VI provides a summary and discussion.

Appendices A–C include additional details: Appendix A presents a general framework for binary dynamics in arbitrary background spacetimes, Appendix B summarizes the Riemann curvature in the Carter tetrad system, and Appendix C derives Lagrange planetary equations and orbital parameter evolution equations using averaging techniques over binary cycles. Additionally, we present analytic solutions for the eccentricity and vZLK oscillation period in the equatorial case, aiding in the interpretation of our numerical results.

\textit{Notation}: Greek indices range from 0 to 3, Roman indices from 1 to 3. Tilde indices denote tetrad components in an observer’s proper reference frame, while barred symbols correspond to the Carter's tetrad-frame quantities.  We set $G = c = 1$ unless otherwise specified.

\begin{widetext}

\section{Spherical Orbits in Kerr Spacetime}
\label{EOM_Kerr}
\subsection{Test particle in Kerr spacetime}
First we consider a test particle with a unit mass in the Kerr spacetime, which metric is written  in Boyer-Lindquist (BL) coordinates as
\bea
d\bar s^2
&=&
-\frac{\Delta}{\Sigma}\left[dt-a(1-\zeta^2)d\phi\right]^2
+\frac{1-\zeta^2}{\Sigma}\left[
(\mathfrak{r}^2+a^2)d\phi-adt\right]^2+\frac{\Sigma}{\Delta}d\mathfrak{r}^2+\frac{\Sigma}{1-\zeta^2} d\zeta^2
\,,
\ena
\end{widetext}
where
\bea
\zeta&=&\cos\theta
\\
\Sigma&=& \mathfrak{r}^2+a^2 \zeta^2
\\
\Delta&=& \mathfrak{r}^2-2M\mathfrak{r}+a^2
\,.
\ena
 $M$ is a gravitational mass of an SMBH
 and $a$ is its proper angular momentum.

There are two Killing vectors, which guarantee two conserved quantities
in a test particle motion; the proper energy $E$ and the $z$-component of the proper 
angular momentum $L$.
In addition, there exists one more conserved quantity, 
so-called Carter constant ${\cal C}$ or $K$.
These constants are  defined by the Killing tensor $K_{\mu\nu}$ as
\bea
K&=&K_{\mu\nu} p^\mu p^\nu
\\
{\cal C}&=&K-\ell^2
\ena
where 
$p_\mu\equiv u_\mu$ is the proper 4-momentum of a test particle and $\ell\equiv L-aE$.

The Carter constant ${\cal C}$ in BL coordinates is given by
\bea
{\cal C}\equiv p_\theta^2+\cos^2\theta\left[a^2(1-E^2)+{L^2\over \sin^2\theta}\right]
\,,
\ena
where  $p_\theta$ is the $\theta$-component of the proper 4-momentum.
Note that $K$ is always non-negative, whereas ${\cal C}$ can be either positive or negative. It vanishes when the orbit is confined to the equatorial plane ($\theta=\pi/2$).

In order to decouple the equations of motion (EOM),
 we shall introduce the Mino time $\mathsf{t}$,
which is defined by
\bea
d\mathsf{t}\equiv {1 \over \Sigma} d\tau
\,,
\ena
where $\tau$ is a proper time of a test particle.

\begin{widetext}

We then find two decoupled equations
 for $\mathfrak{r}$ and $\zeta$ as
\bea
\dot{\mathfrak{r}}^2 &=&\left[E\mathfrak{r}^2-a\ell \right]^2-\Delta \left[ \mathfrak{r}^2+\ell^2+{\cal C}\right]
\label{eq_r}
\\
\dot \zeta^2 &=&-\zeta^2\left[a^2\left(1-E^2\right)\left(1-\zeta^2\right)+L^2\right]
+{\cal C}\left(1-\zeta^2\right)
\label{eq_th}
\,,
\ena
where a dot denotes the derivative with respect to the Mino time $\mathsf{t}$.
The other two EOM for $t$ and $\phi$ are given by 
\bea
\dot{t} &=&{\mathfrak{r}^2+a^2\over \Delta}\left[E\mathfrak{r}^2-a\ell\right]-a^2 E\left(1-\zeta^2\right)+aL
\label{eq_t}
\\
\dot \phi&=&{a\over \Delta}\left[E\mathfrak{r}^2-a\ell\right]+{L\over 1-\zeta^2}-aE
\,,
\label{eq_phi}
\ena
which can be integrated once we find the solutions of $\mathfrak{r}(\mathsf{t})$ and $\zeta(\mathsf{t})$.

In this paper we consider only a bound orbit. 
The above differential equations are analytically  integrated by use of the elliptic functions as follows:\cite{vandeMeent:2019cam}

Eqs. (\ref{eq_r}) and (\ref{eq_th}) are then  rewritten as
\bea
\dot{\mathfrak{r}}^2 &=&(1-E^2)(\mathfrak{r}_a-\mathfrak{r})(\mathfrak{r}-\mathfrak{r}_p)(\mathfrak{r}-\mathfrak{r}_3)(\mathfrak{r}-\mathfrak{r}_4)
\label{dot_r}
\\
\dot \zeta^2 &=&(\zeta^2-\zeta_{\textsf L}^2)[a^2(1-E^2)\zeta^2-\zeta_2^2]
\,,
\label{dot_zeta}
\ena
where $\mathfrak{r}_a$ and $\mathfrak{r}_p$ are two turning points; apoapsis and periapsis ($\mathfrak{r}_a\geq \mathfrak{r}_p$)  in the radial direction, while 
$\zeta_{\textsf L}$ is the turning point in the latitudinal direction, which gives the latitudinal libration angle 
$\chi_{\textsf L}$ by $\zeta_{\textsf L}=\sin \chi_{\textsf L}$. 

Comparing the coefficients, we find
\bea
\mathfrak{r}_3&=&{M\over 1-E^2}-{\mathfrak{r}_a+\mathfrak{r}_p\over 2}+\sqrt{\left({\mathfrak{r}_a+\mathfrak{r}_p\over 2}-{M\over 1-E^2}
\right)^2-{a^2{\cal C}\over \mathfrak{r}_a\mathfrak{r}_p(1-E^2)}}
\label{r3}
\\
\mathfrak{r}_4&=&{a^2{\cal C}\over \mathfrak{r}_a\mathfrak{r}_p\mathfrak{r}_3(1-E^2)}
\label{r4}
\\
\zeta_2
&=&\sqrt{a^2(1-E^2)+{L^2\over 1-\zeta_{\textsf L}^2}}
\label{zeta2}
\ena
and
\bea
&&
(1-E^2)\left[\mathfrak{r}_a\mathfrak{r}_p-(\mathfrak{r}_a+\mathfrak{r}_p)^2\right]+2M(\mathfrak{r}_a+\mathfrak{r}_p)+{a^2{\cal C}\over \mathfrak{r}_a\mathfrak{r}_p}=L^2+a^2(1-E^2)+{\cal C}
\label{eq1}
\\
&&-(1-E^2)\mathfrak{r}_a\mathfrak{r}_p(\mathfrak{r}_a+\mathfrak{r}_p)+2M\mathfrak{r}_a\mathfrak{r}_p+a^2{\cal C}{\mathfrak{r}_a+\mathfrak{r}_p\over \mathfrak{r}_a\mathfrak{r}_p}=2M(\ell^2+{\cal C})
\label{eq2}
\\
&&
\zeta_{\textsf L}^2\zeta_2^2={\cal C}
\label{Q}
\,.
\ena
For given values of $\mathfrak{r}_a, \mathfrak{r}_p$ and $\zeta_{\textsf L}$, the last three equations (\ref{eq1})-(\ref{Q}) determine $E, \ell$, and ${\cal C}$,
and then $\mathfrak{r}_3, \mathfrak{r}_4$ and $\zeta_2$
by Eqs. (\ref{r3})-(\ref{zeta2}), 
 too. 

\end{widetext}

Assuming $\mathfrak{r}=\mathfrak{r}_p$ and $\zeta=0 (\theta=\pi/2)$ at $\mathsf{t}=0$, 
the solutions for $\mathfrak{r}$ and $\zeta$ are described as
\beann
\mathfrak{r}&=&{\mathfrak{r}_3(\mathfrak{r}_a-\mathfrak{r}_p){\rm sn}^2\left(\mathfrak{w}_r\mathsf{t}/2 ; k_r^2\right)-\mathfrak{r}_p(\mathfrak{r}_a-\mathfrak{r}_3)\over 
(\mathfrak{r}_a-\mathfrak{r}_p){\rm sn}^2\left(\mathfrak{w}_r\mathsf{t}/2 ; k_r^2\right)-(\mathfrak{r}_a-\mathfrak{r}_3) }
\\
\zeta&=&\zeta_{\textsf L} \, {\rm sn}\left(\zeta_2\mathsf{t}; k_\zeta^2\right)
\,,
\enann
where
${\rm sn}(x;k)$ is the Jacobi elliptic sine function with the modulus $k$, and 
\beann
\mathfrak{w}_r&\equiv &\sqrt{(1-E^2)(\mathfrak{r}_a-\mathfrak{r}_3)(\mathfrak{r}_p-\mathfrak{r}_4)}
\\
k_r^2&\equiv&{(\mathfrak{r}_a-\mathfrak{r}_p)(\mathfrak{r}_3-\mathfrak{r}_4)\over (\mathfrak{r}_a-\mathfrak{r}_3)(\mathfrak{r}_p-\mathfrak{r}_4)}
\\
 k_\zeta^2&\equiv&
 a^2(1-E^2){\zeta_{\textsf L}^2\over \zeta_2^2}
 \,.
\enann
 
The frequencies of oscillations in radial and polar directions 
are given by
 \beann
 \Upsilon_r&\equiv&{\pi\mathfrak{w}_r\over 2\mathsf{K}(k_r)}={\pi\over 2\mathsf{K}(k_r)}\sqrt{(1-E^2)(\mathfrak{r}_a-\mathfrak{r}_3)(\mathfrak{r}_p-\mathfrak{r}_4)}
 \\
 \Upsilon_\zeta &\equiv&{\pi\zeta_2 \over 2\mathsf{K}(k_\zeta)}
  \,,
 \enann
 where
 $\mathsf{K}(k_r)$ and $\mathsf{K}(k_\zeta)$ are the complete elliptic integrals of the first kind 
 with the moduli $k_r$ and $k_\zeta$, respectively.
 
 The proper time $\tau$ is given by the integration
 \beann
 \tau=\int_0^\mathsf{t} d\mathsf{t}\, \Sigma=\int_0^\mathsf{t} d\mathsf{t} \left[ r^2(\mathsf{t})+a^2\zeta^2(\mathsf{t})\right]
 \,.
 \enann

\subsection{Spherical orbit ($\mathfrak{r}=\mathfrak{r}_0$) in Kerr spacetime }
Here we consider only a spherical orbit with the radius $r=\mathfrak{r}_0$ (constant), i.e., $r_a=r_p=\mathfrak{r}_0$.
From Eqs. (\ref{eq1}), and (\ref{eq2}) with Eq. (\ref{Q}),
which gives 
\beann
{\cal C}&=&\zeta_{\textsf L}^2\left[a^2(1-E^2)+{L^2\over 1-\zeta_{\textsf L}^2}\right]
\,,
\enann
we find
 the equations for $E$ and $L$ as
\bea
&&
A_1 E^2+B_1 L^2+C_1E L=D_1
\label{eq:EL1}
\\
&&
A_2 E^2+B_2 L^2+C_2E L=D_2
\,,
\label{eq:EL2}
\ena
where
\beann
A_1&=&-\left[\mathfrak{r}_0^4 + a^2 \mathfrak{r}_0 (\mathfrak{r}_0+2 M )+a^2\Delta_0\zeta_{\textsf L}^2\right]
\\
B_1&=&{\mathfrak{r}_0^2-2 M \mathfrak{r}_0+  a^2 \zeta_{\textsf L}^2\over 1-\zeta_{\textsf L}^2}
\\
C_1&=&4 a M \mathfrak{r}_0
\\
D_1&=&-\Delta_0 (\mathfrak{r}_0^2 + a^2 \zeta_{\textsf L}^2)
\\
A_2&=&-\left[2 \mathfrak{r}_0^3 + a^2 (M (1 -\zeta_{\textsf L}^2) +\mathfrak{r}_0 (1 + \zeta_{\textsf L}^2))\right]
\\
B_2&=&{\mathfrak{r}_0-M\over 1-\zeta_{\textsf L}^2}
\\
C_2&=&2 a M 
\\
D_2&=&-\mathfrak{r}_0^2 (2 \mathfrak{r}_0-3 M )- a^2\left[\mathfrak{r}_0+(\mathfrak{r}_0- M)\zeta_{\textsf L}^2\right]
\,,
\enann
with $\Delta_0\equiv \Delta(\mathfrak{r}_0)$.

Eqs. (\ref{eq_t}) and (\ref{eq_phi}) for $t$ and $\phi$ are integrated as
\bea
t&=&{1\over \Delta(\mathfrak{r}_0)}\left[E\mathfrak{r}_0^2(\mathfrak{r}_0^2+a^2)-2Ma\ell \mathfrak{r}_0\right]\mathsf{t}
\nn
&&
+{a^2E\zeta_{\textsf L}^2\over \zeta_2 k_\zeta^2}\left[\zeta_2
\mathsf{t}-{\cal E}\left(\zeta_2\mathsf{t},k_\zeta\right)\right]
\\
\phi&=&{a\over \Delta(\mathfrak{r}_0)}\left(2ME\mathfrak{r}_0-aL\right)\mathsf{t}
\nn
&&
+{L\over \zeta_2}\mathsf{\Pi}(\zeta_{\textsf L}^2,{\rm am}(\zeta_2\mathsf{t}, k_\zeta),k_\zeta^2)
\ena
where ${\cal E}(x,k)$ is the Jacobi epsilon function with the modulus $k$, which is defined by
\beann
{\cal E}(x,k)\equiv \int_0^x {\rm dn}^2(u,k) du
\enann
and 
$ {\rm am}(x, k)$ and $\mathsf{\Pi}(n,x,m)$ are  the Jacobi  amplitude  with the modulus $k$ and 
the elliptic integral of the third kind with the characteristic $n$ and the parameter $m=k^2$ ($k$: modulus), which are defined by
\beann
{\rm am}(x, k) &\equiv&\sin^{-1}({\rm sn}(x,k))
\\
\mathsf{\Pi}(n,x,m)&\equiv&\int_0^x {du\over (1-nu^2)\sqrt{(1-u^2)(1-m u^2)}}
\,.
\enann

Giving $a/M$ and $\mathfrak{r}_0/M$, we can solve the equations (\ref{eq:EL1}) and (\ref{eq:EL2}) for $E$ and $L/M$ in terms of $\zeta_{\textsf L}$.
We then find all parameters including the Carter constant ${\cal C}/M^2$  for a spherical orbit.

\begin{figure}[h]
\includegraphics[width=6cm]{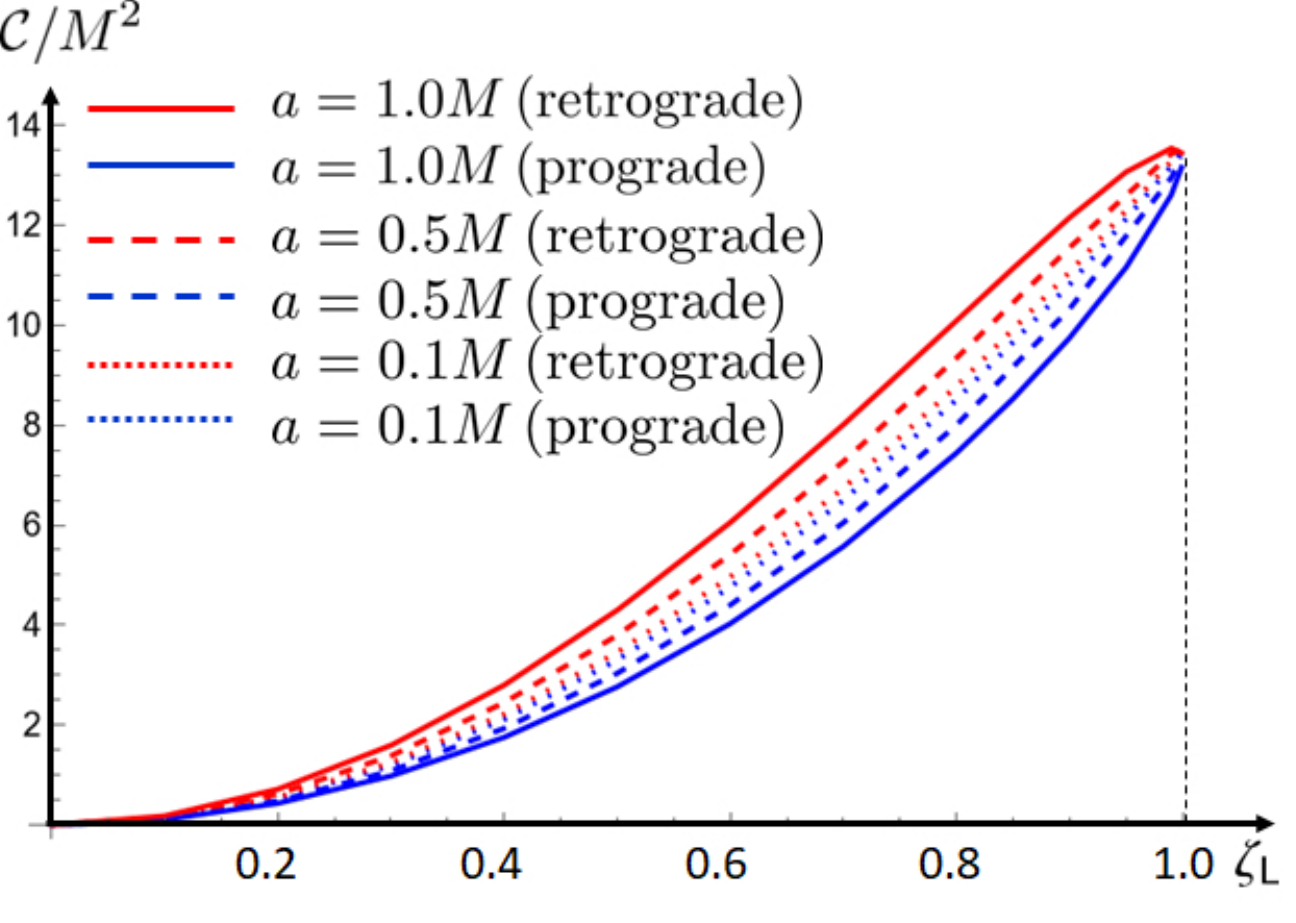}~
\caption{Carter constant in terms of $\zeta_{\textsf L}$ for $a=0.1M$ (the dotted curves), $0.5M$ (the dashed curves) and $1.0M$ (the solid curves). The red and blue curves correspond to the retrograde and prograde orbits, respectively. 
We choose $\mathfrak{r}_0=9M$. In the limit of $a\rightarrow 0$, the Cater constants for the retrograde and prograde orbits coincide.}
\label{Carter_constant}
\end{figure}

The Carter constant ${\cal C}/M^2$ increases as $\zeta_{\textsf L}$ becomes large as shown in Fig. \ref{Carter_constant}.
The behavior of the Carter constant looks very similar
 although it depends on  the sign of $L$ as well as $a$.
 In the limit of $a\rightarrow 0$, the difference between the prograde and retrograde orbits vanishes as expected.

\begin{figure}[h]
\includegraphics[width=5cm]{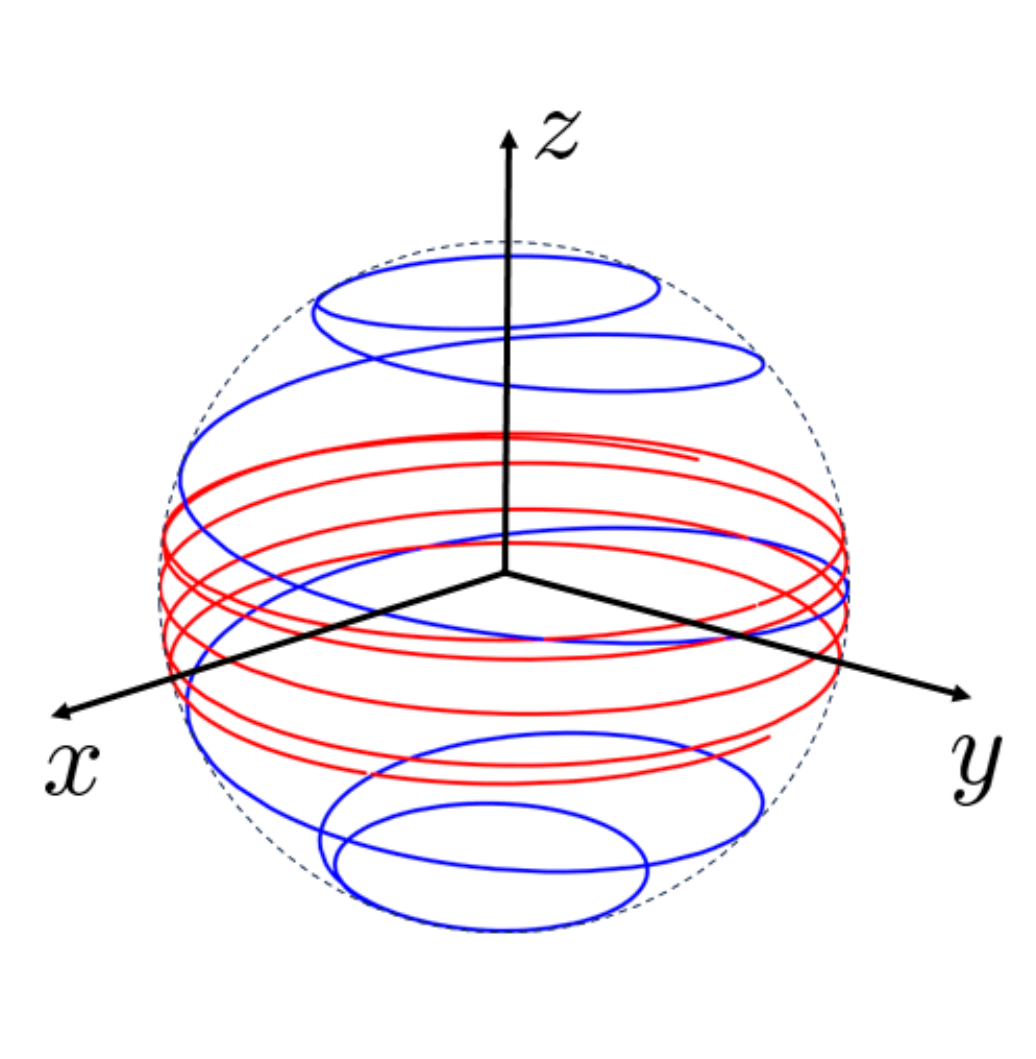}~
\caption{The spherical orbits for $\zeta_{\textsf L}=0.2$(red) and $\zeta_{\textsf L}=0.9$(blue). The latitudinal libration 
angles $\chi_{\textsf L}$ are $11.5^\circ$ and $64.2^\circ$, respectively.
 We choose $a=1.0M$ and $\mathfrak{r}_0=9M$.}
\label{CM_motion}
\end{figure}

We also show typical spherical orbits in Fig. \ref{CM_motion}.
We assume that $a=1.0M$ and $\mathfrak{r}_0=9 M$.
We also choose $\zeta_{\textsf L}=0.2$ (red) and 0.9 (blue), which correspond to $\chi_{\textsf L}\approx 11.5^\circ$ and $64.2^\circ$, respectively, where we use $\chi_{\textsf L}$, which denotes the latitudinal libration angle, i.e., the maximum oscillation angle in the latitudinal direction.

 
The proper time $\tau$ is given by
\beann
\tau&\equiv& \int_0^\mathsf{t} d\mathsf{t}\left[\mathfrak{r}_0^2+a^2\zeta^2(\mathsf{t})\right]
\nn
&=&\left(\mathfrak{r}_0^2+{a^2\zeta_{\textsf L}^2\over k_\zeta^2}\right)\mathsf{t}-{a^2\zeta_{\textsf L}^2\over \zeta_2k_\zeta^2}{\cal E}(\zeta_2\mathsf{t},k_\zeta)
\,.
\enann

\begin{widetext}

\subsection{Innermost Stable Spherical Orbit (ISSO)}
Just as the same as the innermost stable circular orbit (ISCO) for circular motions in a Kerr spacetime, there exists an innermost stable spherical orbit (ISSO), below which there is no stable spherical orbits.
The ISSO radius $\mathfrak{r}_{\rm ISSO}$ is obtained  by the following three equations 
\bea
\dot{\mathfrak{r}}&=&
-(1-E^2)\mathfrak{r}_{\rm ISSO}^4+2M\mathfrak{r}_{\rm ISSO}^3-\left[L^2+{\cal C}+a^2(1-E^2)\right]\mathfrak{r}_{\rm ISSO}^2+2M(\ell^2+{\cal C})\mathfrak{r}_{\rm ISSO}-a^2{\cal C}=0\,,~~~~~
\label{ISSO1}
\\
\ddot{\mathfrak{r}}
&=&-2(1-E^2)\mathfrak{r}_{\rm ISSO}^3+3M\mathfrak{r}_{\rm ISSO}^2-\left[L^2+{\cal C}+a^2(1-E^2)\right]\mathfrak{r}_{\rm ISSO}+M(\ell^2+{\cal C})=0
\,,
\label{ISSO2}
\\
\dddot{\mathfrak{r}}
&=&-6(1-E^2)\mathfrak{r}_{\rm ISSO}^2+6M\mathfrak{r}_{\rm ISSO}-\left[L^2+{\cal C}+a^2(1-E^2)\right]=0
\,.
\label{ISSO3}
\ena
From the first two equations, we find
\beann
3(1-E^2)\mathfrak{r}_{\rm ISSO}^4-4M\mathfrak{r}_{\rm ISSO}^3+\left[L^2+{\cal C}+a^2(1-E^2)\right]\mathfrak{r}_{\rm ISSO}^2-a^2{\cal C}=0
\,.
\enann 

\end{widetext}
Eliminating ${\cal C}$, 
we also find the equations for the ISSO energy and ISSO angular momentum  in terms of $\zeta_{\textsf L}$ as
\beann
&&
\tilde A_1 (1-E^2)+\tilde B_1 L^2= \tilde D_1
\\
&&
\tilde A_2 (1-E^2)+\tilde B_2 L^2= \tilde D_2
\enann
where 
\beann
\tilde A_1&=&6\mathfrak{r}_{\rm ISSO}^2+a^2(1+\zeta_{\textsf L}^2)
\\
\tilde B_1&=&-{1\over 1-\zeta_{\textsf L}^2}
\\
\tilde D_1&=&6\mathfrak{r}_{\rm ISSO}(\mathfrak{r}_{\rm ISSO}-M)+a^2(1+\zeta_{\textsf L}^2)
\\
\tilde A_2&=&3\mathfrak{r}_{\rm ISSO}^2+a^2(1+\zeta_{\textsf L}^2)-{a^4\zeta_{\textsf L}^2\over \mathfrak{r}_{\rm ISSO}^2}
\\
\tilde B_2&=&-{1\over 1-\zeta_{\textsf L}^2}\left(1-{a^2\zeta_{\textsf L}^2\over  \mathfrak{r}_{\rm ISSO}^2}\right)
\\
\tilde D_2&=&\mathfrak{r}_{\rm ISSO}(3\mathfrak{r}_{\rm ISSO}-4M)+a^2(1+\zeta_{\textsf L}^2)-{a^4\zeta_{\textsf L}^2\over \mathfrak{r}_{\rm ISSO}^2}
\,,
\enann
\begin{widetext}
which solve 
\bea
E_{\rm ISSO}&=&\sqrt{1-{2M\mathfrak{r}_{\rm ISSO}(\mathfrak{r}_{\rm ISSO}^2-3a^2\zeta_{\textsf L}^2)\over 3\mathfrak{r}_{\rm ISSO}^4-6a^2\zeta_{\textsf L}^2\mathfrak{r}_{\rm ISSO}^2-a^4\zeta_{\textsf L}^4}
}
\\
L_{\rm ISSO}&=&\pm \sqrt{2M\mathfrak{r}_{\rm ISSO}\left(1-\zeta_{\textsf L}^2\right){3\mathfrak{r}_{\rm ISSO}^4-a^2(1+\zeta_{\textsf L}^2)\mathfrak{r}_{\rm ISSO}^2+3a^4\zeta_{\textsf L}^2 \over 3\mathfrak{r}_{\rm ISSO}^4-6a^2\zeta_{\textsf L}^2\mathfrak{r}_{\rm ISSO}^2-a^4\zeta_{\textsf L}^4}}
\,,
\ena
where $\pm$ correspond to the prograde and retrograde orbits, respectively.

Inserting $E_{\rm ISSO}$ and $L_{\rm ISSO}$ into Eq.(\ref{ISSO2}), 
we find the equation for $\mathfrak{r}_{\rm ISSO}$
as
\beann
&&
\mathfrak{r}_{\rm ISSO}^6-3a^2(1-2\zeta_{\textsf L}^2)\mathfrak{r}_{\rm ISSO}^4+3a^4\zeta_{\textsf L}^2(2-\zeta_{\textsf L}^2)\mathfrak{r}_{\rm ISSO}^2+a^6\zeta_{\textsf L}^4
-2M\mathfrak{r}_{\rm ISSO}\left[3\mathfrak{r}_{\rm ISSO}^4-2a^2\mathfrak{r}_{\rm ISSO}^2
+3a^4\zeta_{\textsf L}^2(2-\zeta_{\textsf L}^2)\right]
 \\
&&
\pm 2a\sqrt{3\mathfrak{r}_{\rm ISSO}^4-6a^2\zeta_{\textsf L}^2\mathfrak{r}_{\rm ISSO}^2-a^4\zeta_{\textsf L}^4-2M\mathfrak{r}_{\rm ISSO}(\mathfrak{r}_{\rm ISSO}^2-3a^2\zeta_{\textsf L}^2)
}\sqrt{2M\mathfrak{r}_{\rm ISSO}\left(1-\zeta_{\textsf L}^2\right)\left[3\mathfrak{r}_{\rm ISSO}^4-a^2(1+\zeta_{\textsf L}^2)\mathfrak{r}_{\rm ISSO}^2+3a^4\zeta_{\textsf L}^2\right]}
\\
&&=0
\label{ISSO_radius2}
\enann

\end{widetext}

The Carter constant at the ISSO radius is 
\bea
{\cal  C}_{\rm ISSO}={2M\mathfrak{r}^3_{\rm ISSO}
\zeta_{\textsf L}^2(3\mathfrak{r}_{\rm ISSO}^2-a^2\zeta_{\textsf L}^2)\over 3\mathfrak{r}_{\rm ISSO}^4-6a^2\zeta_{\textsf L}^2\mathfrak{r}_{\rm ISSO}^2-a^4\zeta_{\textsf L}^4}
\label{Carter_ISSO}
\ena

\begin{figure}[ht]
\includegraphics[width=6.5cm]{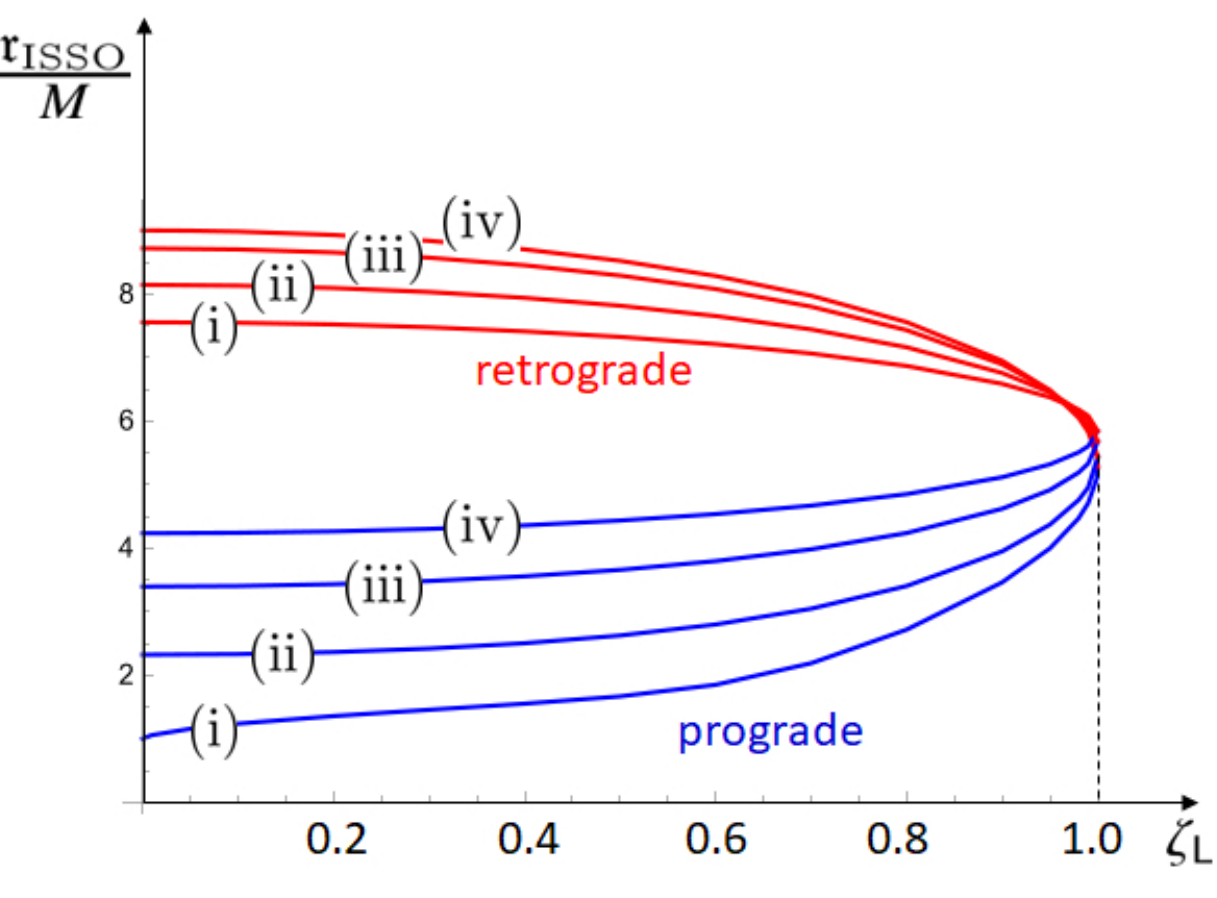}
\caption{The ISSO radius $\mathfrak{r}_{\rm ISSO}$ with respect to $\zeta_{\textsf L}$. We choose $a=1.0M {\rm (i)} \,, 0.9M{\rm (ii)}\,, 0.7M{\rm (iii)}\,, 0.5M{\rm (iv)}$. The blue and red curves correspond to the prograde and retrograde orbits, respectively.}
\label{rISSO}
\end{figure}

We show some values of $\mathfrak{r}_{\rm ISSO}$ in terms of $\zeta_{\textsf L}$ for given $a$ in Fig. \ref{rISSO}.
As $\zeta_{\textsf L}$ increases, the ISSO radius $\mathfrak{r}_{\rm ISSO}$ increases  for the prograde orbit, while it decreases for the retrograde orbits, and 
in the limit of $\zeta_{\textsf L}\rightarrow 1(\chi_{\textsf L}\rightarrow 90^\circ)$, 
they become the same value, which is close to the Schwarzschild's one (6$M$) although it slightly  depends on the Kerr parameter $a$.

\begin{figure}[htb]
\includegraphics[width=5.cm]{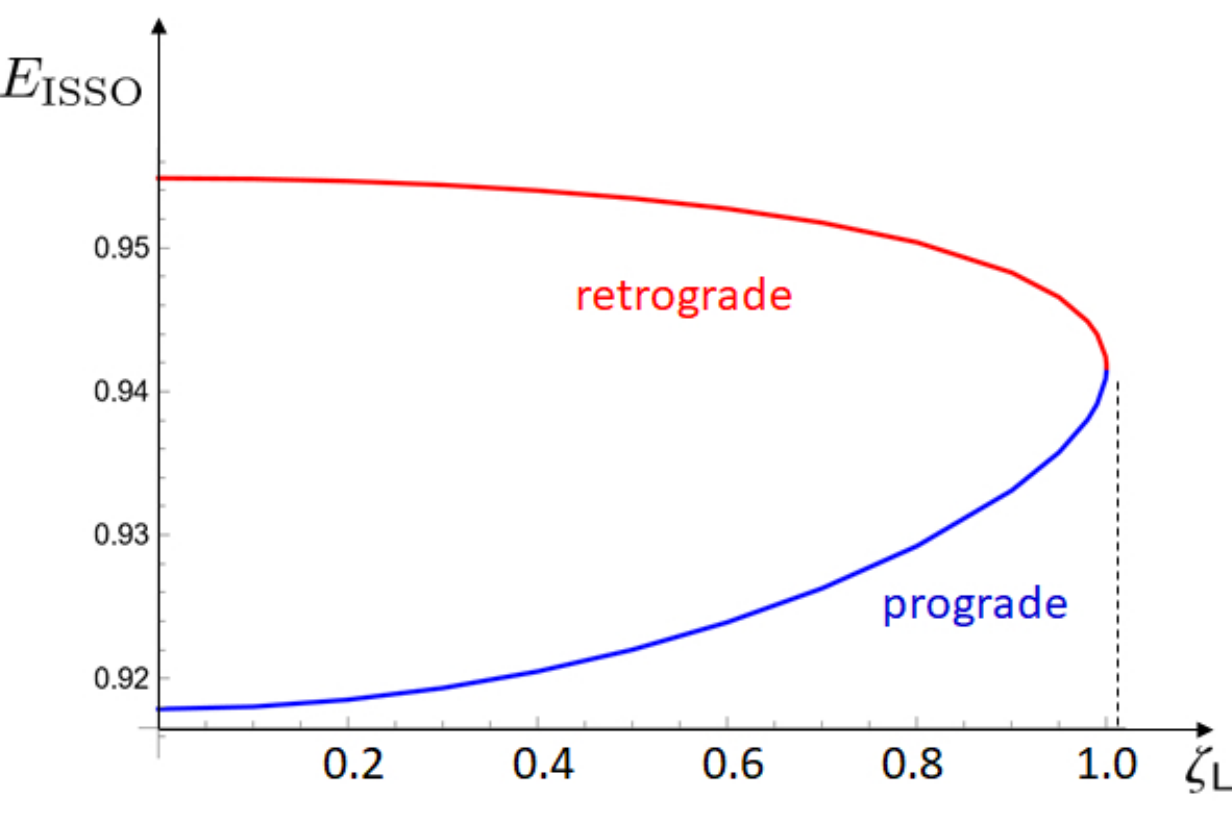}\\
(a)
\\
\includegraphics[width=5.cm]{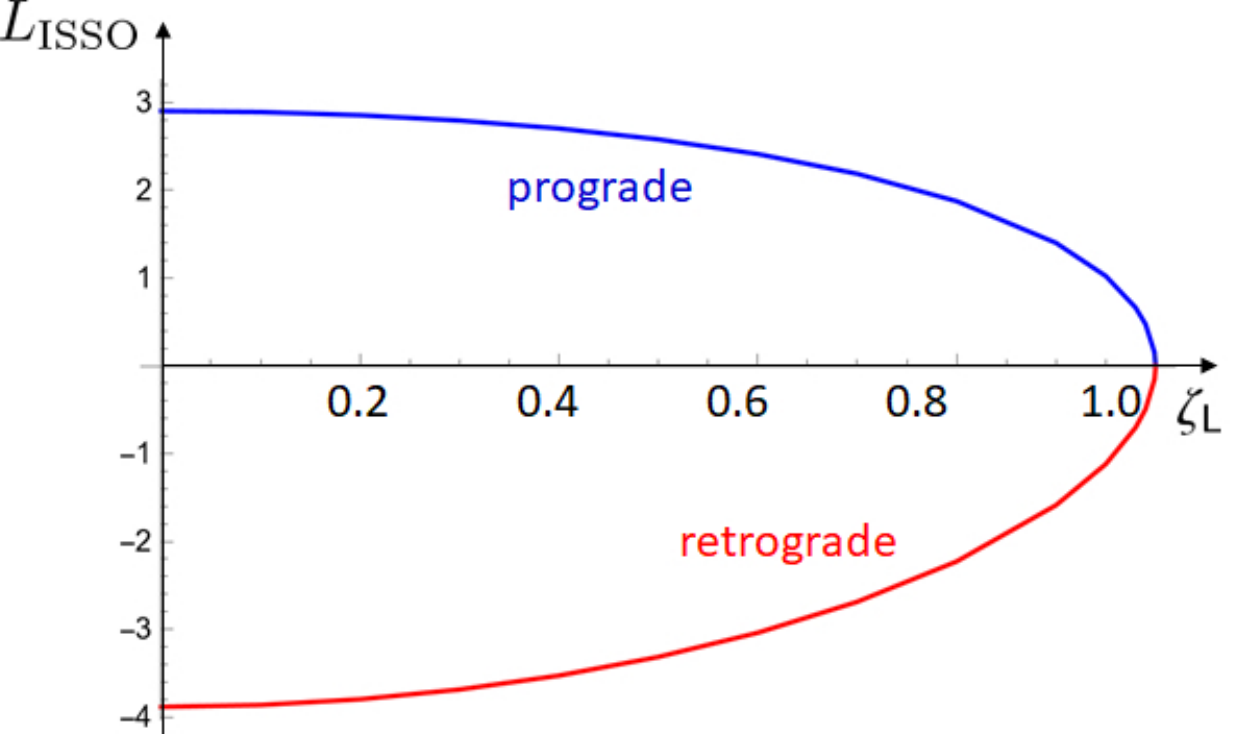}\\
(b)
\\
\hskip -0.5cm \includegraphics[width=5.5cm]{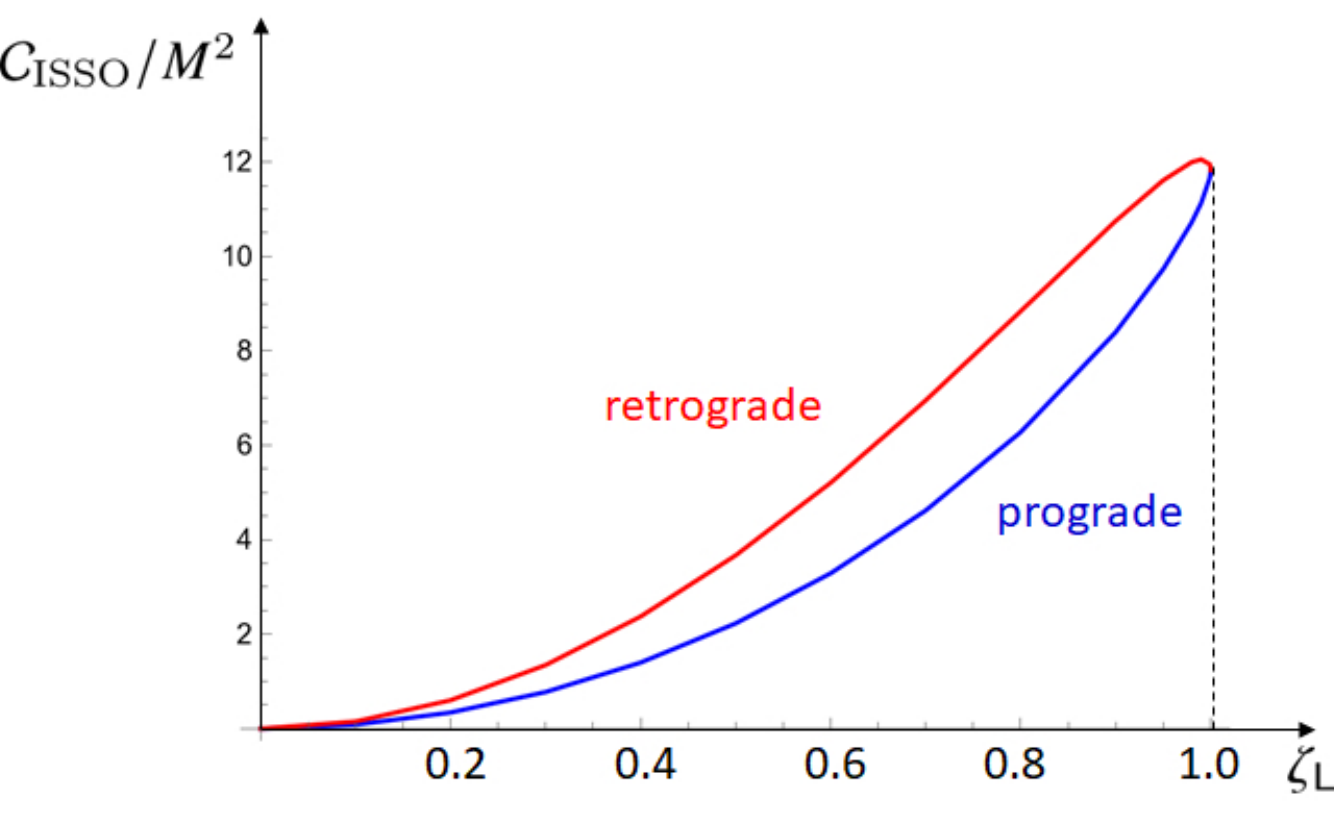}
\\
(c)
\caption{The energy $E_{\rm ISSO}$, angular momentum $L_{\rm ISSO}$ and Carter constant ${\cal C}_{\rm ISSO}$ of the ISSO
 with respect to $\zeta_{\textsf L}$ are shown at the top (a), middle (b) and bottom (c) figures, respectively. We choose $a=0.5M$}
\label{ELCISSO}
\end{figure}

We also present the energy, $z$-component of the angular momentum, and Carter constant of the particle on ISSO in terms of $\zeta_{\textsf L}$ in Fig. \ref{ELCISSO}.

As $\zeta_{\textsf L}$ increases, the energy rises while the $z$-component of the angular momentum decreases for prograde orbits. In contrast, for retrograde orbits, the energy decreases and the $z$-component of the angular momentum increases. In the limit $\zeta_{\textsf L} \rightarrow 1$ (i.e., $\chi_{\textsf L} \rightarrow 90^\circ$), both quantities converge to the same values for prograde and retrograde orbits.
The Carter constant also increases with $\zeta_{\textsf L}$; however, for retrograde orbits, it reaches a maximum near $\chi_{\textsf L} \sim 90^\circ$, then decreases, matching the value for prograde orbits at $\zeta_{\textsf L} = 1$.


\section{A Binary on Spherical Orbit}

As summarized in Appendix \ref{binary_in_curved_ST}, in order to discuss a binary motion in a Kerr background, we first have to construct a local inertial frame, and then put a self-gravitating binary system.
The tidal force by the Kerr black hole is evaluated by the Riemann curvature in this frame.

\subsection{Local Inertial System}
We first consider on a local inertial reference frame 
along a general geodesic motion in Kerr spacetime.
To construct a local inertial frame, we have to find three spatial tetrad vectors $e_{\tilde i}^{~\mu}~(i=1, 2, 3)$
parallel transported along 4-velocity 
$u^\mu\equiv e_{\tilde 0}^{~\mu}$
\cite{vandeMeent:2019cam}.
For a given proper energy $E$ and  $z$ component of 
angular momentum $L$, the 4-velocity is given by
\bea
e_{\tilde 0\mu}=u_\mu=\left(-E\,,{\dot{\mathfrak{r}}\over \Delta}\,,{\dot \zeta\over 1-\zeta^2}\,, L\right)
\,.
\ena

In Kerr spacetime, there exists the Killing-Yano tensor $f_{\mu\nu}$, which is defined by 
$\nabla_{(\rho}f_{\mu)\nu}=0$ and related to the Killing tensor 
as $K_{\mu\nu}=f_{\mu\rho}f^{\rho}_{~\nu}$.
Using this tensor, we find 
one natural vector $f_{\mu\rho}u^\rho$, which is parallelly transported along the geodesic. In fact,
\beann
{D(f_{\mu\rho}u^\rho)\over d\tau}
&=&
u^\nu u^\rho \nabla_\nu \left(f_{\mu\rho}\right)+\left(u^\nu \nabla_\nu u^\rho\right) f_{\mu\rho}=0
\,.
\enann

We then obtain one spatial unit vector parallelly transported along the geodesic as
\beann
e_{\tilde 3\, \mu}&\equiv& {f_{\mu\rho}u^\rho\over \sqrt{ K}}
\nn
&=&
{1\over \sqrt{K}}
\left(
\begin{array}{c}
-{a\over \Sigma}\left(\mathfrak{r}
\dot\zeta+\zeta \dot{\mathfrak{r}}\right)\\
{a\zeta\over \Delta}\left(E\mathfrak{r}^2-a\ell\right)\\
-{\mathfrak{r}\over 1-\zeta^2}\left(\ell+Ea\zeta^2\right)\\
{1\over \Sigma}\left(a^2\zeta(1-\zeta^2)\dot{\mathfrak{r}}+\mathfrak{r}(\mathfrak{r}^2+a^2)\dot \zeta\right)\\
\end{array}
\right)\,,
\enann
where $K$ is the Carter constant and 
the prefactor $1/\sqrt{K}$ is just for normalization.

The other two spatial unit tetrad vectors ($e_{\tilde 1\, \mu}\,, e_{\tilde 2\, \mu}$) 
are given by 
the tetrad vectors of rotating inertial frame ($e_{\hat 1\, \mu}\,, e_{\hat 2\, \mu}$) 
such that 
\beann
e_{\tilde 1\, \mu}&=&e_{\hat 1\, \mu}\cos \Psi-e_{\hat 2\, \mu}\sin\Psi
\\
e_{\tilde 2\, \mu}&=&e_{\hat1\, \mu}\sin\Psi+e_{\hat  2\, \mu}\cos\Psi
\,,
\enann
where
\beann
e_{\hat 1\, \mu}&\equiv & 
{1\over \sqrt{K}}
\left(
\begin{array}{c}
{1\over \Xi \Sigma}\left(-\Xi^2 \mathfrak{r}\dot{\mathfrak{r}}+a^2\zeta \dot \zeta\right)\\
{\Xi \mathfrak{r}\over \Delta}\left(E\mathfrak{r}^2-a\ell\right)\\
{a\zeta\over \Xi (1-\zeta^2)}\left(\ell+Ea \zeta^2\right)\\
{a\over \Xi \Sigma}\left(\Xi^2(1-\zeta^2)\mathfrak{r} \dot{\mathfrak{r}}- (\mathfrak{r}^2+a^2)\zeta \dot \zeta\right)\\
\end{array}
\right)
\\
e_{\hat 2\, \mu}&\equiv & 
\left(
\begin{array}{c}
-{E\over \Xi}+{(1-\Xi^2)\over \Xi\Sigma}\left(E\mathfrak{r}^2 -a\ell \right)\\
{\Xi\over \Delta} \dot{\mathfrak{r}} \\
{1\over \Xi(1-\zeta^2)}\dot\zeta\\
\Xi L+{1-\Xi^2\over \Xi \Sigma}(\mathfrak{r}^2+a^2)\left(\ell+Ea\zeta^2\right)\\
\end{array}
\right)
\,,
\enann
with
\beann
\Xi\equiv \sqrt{K-a^2\zeta^2\over K+\mathfrak{r}^2}.
\enann
The rotation angle $\Psi$ satisfies the following evolution equation:
\beann
\dot \Psi=\sqrt{K}\left[{E\mathfrak{r}^2-a\ell\over K+\mathfrak{r}^2}+{a\left(\ell+Ea\zeta^2\right)\over K-a^2\zeta^2}\right]\,.
\enann

We then obtain the transformation matrix 
$\Lambda_{\hat\lambda}^{~\bar \alpha}$ 
from the Carter's tetrad ($e_{\bar \alpha}^{~\mu}$) to 
the above rotating inertial frame tetrad 
($e_{\hat\alpha}^{~ \mu}$)
as
\bea
\Lambda_{\hat\lambda}^{~\bar \alpha}&\equiv& e_{\hat\lambda \mu}
\omega^{\bar \alpha\mu}
\,,
\ena
which explicit form is
written as
\beann
\Lambda_{\hat 0}^{~\bar \lambda}&=&\left({P\over \Xi},{Q\over \Xi}, \Xi  R,  \Xi S\right)
\\
\Lambda_{\hat 1}^{~\bar \lambda}
&=&{1\over \sqrt{K}}\left(\mathfrak{r}Q, 
\mathfrak{r}P, a\zeta S, 
-a\zeta R
\right)
\\
\Lambda_{\hat 2}^{~\bar \lambda}&=&
\left(P, Q, R, S\right)
\\
\Lambda_{\hat 3}^{~\bar \lambda}&=&{1\over \sqrt{K}}\left(a\zeta{Q\over \Xi}, a\zeta{P\over \Xi}, -\mathfrak{r}\Xi  S,  r \Xi R\right)
\enann
where
\beann
&&
P\equiv {\Xi(E\mathfrak{r}^2-a\ell)\over \sqrt{\Sigma\Delta}}
\,,~
Q\equiv {\Xi\dot{\mathfrak{r}}\over \sqrt{\Sigma\Delta}}
\,,~
\\
&&
R\equiv {\dot\zeta\over \Xi\sqrt{\Sigma(1-\zeta^2)}}
\,,~
S\equiv {\ell+Ea\zeta^2 \over \Xi\sqrt{\Sigma(1-\zeta^2)}}
\,.
\enann

The transformation matrix 
$\Lambda_{\tilde\lambda}^{~\bar \alpha}= e_{\tilde \lambda \mu}
\omega^{\bar \alpha\mu}$ 
from the Carter's tetrad
to the  non-rotating inertial frame tetrad is 
given by
\beann
\Lambda_{\tilde 1}^{~\bar \lambda}&=&\Lambda_{\hat1}^{~\bar \lambda}\cos\Psi-\Lambda_{\hat 2}^{~\bar \lambda}\sin\Psi
\\
\Lambda_{\tilde 2}^{~\bar \lambda}&=&\Lambda_{\hat 1}^{~\bar\lambda}\sin\Psi+\Lambda_{\hat 2}^{~\bar \lambda}\cos\Psi
\enann
and
\beann
\Lambda_{\tilde 3}^{~\bar \lambda}&=&\Lambda_{\hat 3}^{~\bar \lambda}
\enann

This transformation gives the curvature components in the local inertial frame, and then we find the basic Lagrangian for a binary system, which CM follows general geodesic in Kerr spacetime.



\subsection{Curvature components for the spherical orbits  ($r=\mathfrak{r}_0$)}
\label{Curvature_spherical orbits}
For a spherical orbit with $\mathfrak{r}=\mathfrak{r}_0$,
the transformation matrix to the rotating and 
non-rotating inertial frames are
 rewritten as
\beann
\Lambda_{\hat 0}^{~\bar \lambda}&=&\left({P\over \Xi},0, \Xi  R,  \Xi S\right)
\\
\Lambda_{\hat 1}^{~\bar \lambda}
&=&{1\over \sqrt{K}}\left(0, 
\mathfrak{r}_0P, a\zeta S, 
-a\zeta R
\right)
\\
\Lambda_{\hat 2}^{~\bar \lambda}&=&
\left(P, 0, R, S\right)
\\
\Lambda_{\hat 3}^{~\bar \lambda}&=&{1\over \sqrt{K}}\left(0, a\zeta{P\over \Xi}, -\mathfrak{r}_0\Xi  S,  \mathfrak{r}_0\Xi R\right)
\enann
where
\beann
P&\equiv& {\Xi(E\mathfrak{r}_0^2-a\ell)\over \sqrt{\Delta_0(\mathfrak{r}_0^2+a^2\zeta^2)}}
\,,~
\\
R&\equiv& {\dot\zeta\over \Xi\sqrt{(1-\zeta^2)(\mathfrak{r}_0^2+a^2\zeta^2)}}
\,,
\\
S&\equiv& {\ell+Ea\zeta^2 \over \Xi\sqrt{(1-\zeta^2)(\mathfrak{r}_0^2+a^2\zeta^2)}}
\enann
with
\beann
\Xi&=&\sqrt{K-a^2\zeta^2\over K+\mathfrak{r}_0^2}
\enann
and 
\beann
\Lambda_{\tilde 0}^{~\bar \lambda}&=&\Lambda_{\hat 0}^{~\bar \lambda}
\\
\Lambda_{\tilde 1}^{~\bar \lambda}
&=&\Lambda_{\hat 1}^{~\bar \lambda}\cos\Psi
-\Lambda_{\hat 2}^{~\bar \lambda}\sin\Psi
\\
\Lambda_{\tilde 2}^{~\bar \lambda}&=&
\Lambda_{\hat 1}^{~\bar \lambda}\sin\Psi
+\Lambda_{\hat 2}^{~\bar \lambda}\cos\Psi
\\
\Lambda_{\tilde 3}^{~\bar \lambda}&=&\Lambda_{\hat 3}^{~\bar \lambda}
\,,
\enann
respectively.
The rotation angle $\Psi$ satisfies the evolution equation
\beann
\dot \Psi=\sqrt{K}\left[{E\mathfrak{r}_0^2-a\ell\over K+\mathfrak{r}_0^2}+{a(\ell+Ea\zeta^2)\over K-a^2\zeta^2}\right]
\,.
\enann

We then find the curvature components in the non-rotating inertial frame as
 \beann
 \bar{\cal R}_{\tilde 0\tilde 1\tilde 0\tilde 1} &=&\bar{\cal R}_{\hat 0\hat 1\hat 0\hat 1}\cos^2\Psi +\bar{\cal R}_{\hat 0\hat 2\hat 0\hat 2}\sin^2\Psi
 \\
\bar{\cal R}_{\tilde 0\tilde 2\tilde 0\tilde 2} &=&\bar{\cal R}_{\hat 0\hat 1\hat 0\hat 1}\sin^2\Psi+\bar{\cal R}_{\hat 0\hat 2\hat 0\hat 2}\cos^2\Psi 
 \\
\bar{\cal R}_{\tilde 0\tilde 3\tilde 0\tilde 3} &=&\bar{\cal R}_{\hat 0\hat 3\hat 0\hat 3}
 \\
\bar{\cal R}_{\tilde 0\tilde 1\tilde 0\tilde 2} &=&\left(
\bar{\cal R}_{\hat 0\hat 1\hat 0\hat 1}-\bar{\cal R}_{\hat 0\hat 2\hat 0\hat 2}\right)\cos\Psi\sin\Psi
 \\
\bar{\cal R}_{\tilde 0\tilde 2\tilde 0\tilde 3} &=&\bar{\cal R}_{\hat 0\hat 1\hat 0\hat 3}
\sin\Psi
 \\
\bar{\cal R}_{\tilde 0\tilde 3\tilde 0\tilde 1} &=&\bar{\cal R}_{\hat 0\hat 1\hat 0\hat 3}
\cos\Psi
\,,
 \enann
 where the curvature components in the rotating inertial frame are given by
\begin{widetext}
\beann
\bar{\cal R}_{\hat 0\hat 1\hat 0\hat 1}&=&{1\over K\Xi^2}\Big{[}\Big{(}-2\mathfrak{r}_0^2P^4-P^2(R^2+S^2)(\mathfrak{r}_0^2\Xi^4-a^2\zeta^2)
+2a^2(R^2+S^2)^2\Xi^4\zeta^2\Big{)}{\cal Q}_1-6a\mathfrak{r}_0\zeta \Xi^2P^2(R^2+S^2) {\cal Q}_2\Big{]}
\\
\bar{\cal R}_{\hat 0\hat 1\hat 0\hat 3}&=&\bar{\cal R}_{\hat 0\hat 3\hat 0\hat 1}
\\
&=&{1\over K\Xi^3}\Big{[}
-a\mathfrak{r}_0\zeta \Big{(}2P^4+P^2(R^2+S^2)\Xi^2(1+\Xi^2)
+2(R^2+S^2)^2\Xi^6\Big{)}{\cal Q}_1+3P^2(R^2+S^2)\Xi^2\left(\mathfrak{r}_0^2\Xi^2-a^2\zeta^2\right){\cal Q}_2\Big{]}
\\
\bar{\cal R}_{\hat 0\hat 2\hat 0\hat 2}&=&{1\over \Xi^2}P^2(R^2+S^2)\left(\Xi^2-1\right)^2 {\cal Q}_1
\\
\bar{\cal R}_{\hat 0\hat 3\hat 0\hat 3}&=&{1\over K\Xi^4}\left[\left(-2a^2P^4\zeta^2+P^2\Xi^4(R^2+S^2)(\mathfrak{r}_0^2-a^2\zeta^2)+2\mathfrak{r}_0^2(R^2+S^2)^2\Xi^8
\right){\cal Q}_1+6a\mathfrak{r}_0\zeta P^2(R^2+S^2)\Xi^4 {\cal Q}_2\right]
\,,
\enann
with 
\beann
{\cal Q}_1\equiv 
{M\mathfrak{r}(\mathfrak{r}_0^2-3a^2\zeta^2)\over \Sigma^3(\mathfrak{r}_0)}
\, ~{\rm and}~~
{\cal Q}_2\equiv 
{Ma\zeta(3\mathfrak{r}_0^2-a^2\zeta^2)\over \Sigma^3(\mathfrak{r}_0)}
\,.
\enann

\subsection{Lagrangian of A Binary System on a Spherical Orbit and EOM}

The components of non-rotating inertial frame are a little complicated.
Instead we first write down the Lagrangian ${\cal L}_{\rm rel\mathchar`-\bar R}$ in rotating frame, and then 
 transform the coordinates $(x,y)$ to $(\mathsf{x},\mathsf{y})$ by the transformation 
 \beann
 x&=&\mathsf{x}\cos\Psi+\mathsf{y}\sin\Psi
 \\
 y&=&-\mathsf{x}\sin\Psi+\mathsf{y}\cos\Psi
 \enann
 
Using the reduced mass of a binary, $\mu=m_1m_2/(m_1+m_2)$, the Lagrangian  ${\cal L}_{\rm rel\mathchar`-\bar R}$ in the rotating frame  is given by
 \beann
 {\cal L}_{\rm rel\mathchar`-\bar R}&=&-{\mu\over 2}\bar{\cal R}_{\hat 0\hat k\hat 0\hat \ell}r^k r^\ell
 \\
&= &
 -{\mu\over 2}\left[\bar{\cal R}_{\hat 0\hat 1\hat 0\hat 1}x^2+\bar{\cal R}_{\hat 0\hat 2\hat 0\hat 2}y^2+\bar{\cal R}_{\hat 0\hat 3\hat 0\hat 3}z^2
 +2\bar{\cal R}_{\hat 0\hat 1\hat 0\hat 3}xz\right]
 \\
&= &
  -{\mu\over 2}\Big{[}\bar{\cal R}_{\hat 0\hat 1\hat 0\hat 1}\left(\mathsf{x}\cos\Psi+\mathsf{y}\sin\Psi\right)^2
  +\bar{\cal R}_{\hat 0\hat 2\hat 0\hat 2}\left(-\mathsf{x}\sin\Psi+\mathsf{y}\cos\Psi\right)^2+\bar{\cal R}_{\hat 0\hat 3\hat 0\hat 3}\mathsf{z}^2
   +2\bar{\cal R}_{\hat 0\hat 1\hat 0\hat 3}\left(\mathsf{x}\cos\Psi+\mathsf{y}\sin\Psi\right)\mathsf{z}\Big{]}
\enann
which gives  the Lagrangian in terms of the non-rotating inertial frame coordinates $(\mathsf{x}, \mathsf{y}, \mathsf{z})$
 as
 \beann
 {\cal L}_{\rm rel\mathchar`-\bar R}
 &=&
 -{\mu\over 2}\left[\bar{\cal R}_{\tilde 0\tilde 1\tilde 0\tilde 1} \mathsf{x}^2
 +\bar{\cal R}_{\tilde 0\tilde 2\tilde 0\tilde 2} \mathsf{y}^2
 +\bar{\cal R}_{\tilde 0\tilde 3\tilde 0\tilde 3}\mathsf{z}^2
  +2\bar{\cal R}_{\tilde 0\tilde 1\tilde 0\tilde 2}\mathsf{x}\mathsf{y} 
  +2\bar{\cal R}_{\tilde 0\tilde 2\tilde 0\tilde 3} \mathsf{y}\mathsf{z}
  +2\bar{\cal R}_{\tilde 0\tilde 3\tilde 0\tilde 1} \mathsf{z}\mathsf{x}\right]
  \,.
 \enann

\end{widetext}

Since the CM spherical  orbit is described by the analytic functions of Mino time  $\mathsf{t}$,
it may be better to discuss a binary motion with respect to the Mino time.
The action $S$ is given by
\beann
S=\int d\tau {\cal L}=\int d\mathsf{t} \tilde  {\cal L}
\enann
where
\beann
\tilde  {\cal L}&=& \tilde  {\cal L}_{\rm N}+\tilde  {\cal L}_{1/2}
\enann
with
\beann
\tilde  {\cal L}_{\rm N}&\equiv&\Sigma {\cal L}_{\rm N}={\mu\over 2\Sigma} \dot{\vect{\mathsf{r}}}^2+  {G\Sigma m_1m_2\over \mathsf{r}}
-{\mu\Sigma\over 2} 
\bar{\cal R}_{\tilde 0\tilde k \tilde 0 \tilde \ell}\mathsf{r}^{\tilde k}\mathsf{r}^{\tilde\ell}
\\
\tilde  {\cal L}_{1/2}&\equiv&\Sigma {\cal L}_{1/2}=- {2\over 3} \mu{(m_1-m_2)\over (m_1+m_2)}\bar{\cal R}_{\tilde 0\tilde k \tilde j \tilde \ell}\mathsf{r}^{\tilde k}\mathsf{r}^{\tilde \ell}\dot{\mathsf{r}}^{\tilde j}
\enann

We then introduce the momentum $\vect{\pi}$ as
\beann
\vect{\pi}\equiv {\pa \tilde {\cal L}\over \pa \dot{\vect{\mathsf{r}}}}=\vect{\pi}_{\rm N}+\vect{\pi}_{1/2}
\enann
where
\beann
\vect{\pi}_{\rm N}&=&{\mu\over \Sigma} \dot{\vect{\mathsf{r}}}
\\
{\pi_{1/2}}_{\tilde j}&=&- {2\over 3} \mu{(m_1-m_2)\over (m_1+m_2)}
\bar{\cal R}_{\tilde 0\tilde k \tilde j \tilde \ell}\mathsf{r}^{\tilde k}\mathsf{r}^{\tilde \ell}
\enann]

The Hamiltonian $\tilde {\cal H}$ is defined by
\beann
\tilde {\cal H}&=&\vect{\pi}\cdot \dot{\vect{\mathsf{r}}}-\tilde {\cal L}
\\
&=&\Sigma\left[
{1\over 2\mu}\vect{\pi}_{\rm N}^2-  {Gm_1m_2\over \mathsf{r}}
+{\mu\over 2} 
\bar{\cal R}_{\tilde 0\tilde k \tilde 0 \tilde \ell}\mathsf{r}^{\tilde k}\mathsf{r}^{\tilde\ell}
\right]
\\
&=&\Sigma\left[
{1\over 2\mu}\left(\vect{\pi}-\vect{\pi}_{1/2}\right)^2-  {Gm_1m_2\over \mathsf{r}}
+{\mu\over 2} 
\bar{\cal R}_{\tilde 0\tilde k \tilde 0 \tilde \ell}\mathsf{r}^{\tilde k}\mathsf{r}^{\tilde\ell}
\right]
\,.
\enann
The Hamilton equations are
\beann
\dot{\vect{\mathsf{r}}}&=&{\pa \tilde{\cal H}\over \pa \vect{\pi}}
={\Sigma\over \mu}
\left(\vect{\pi}-\vect{\pi}_{1/2}\right)={\Sigma\over \mu}
\vect{\pi}_{\rm N}
\\
\dot{\vect{\pi}}&=&-{\pa \tilde{\cal H}\over \pa \vect{\mathsf{r}}}
\enann
which are explicitly written as
\beann
\dot{\mathsf{r}}_{\tilde j}&=&{\Sigma\over \mu}
\left(\pi_{\tilde j}+ {2\over 3} \mu{(m_1-m_2)\over (m_1+m_2)}\bar{\cal R}_{\tilde 0\tilde k \tilde j \tilde \ell}\mathsf{r}^{\tilde k}\mathsf{r}^{\tilde \ell}\right)
\\
\dot{\pi}_{\tilde j}
&=&-\mu\Sigma\left[{G(m_1+m_2)\mathsf{r}_{\tilde j}\over \mathsf{r}^3}+\bar{\cal R}_{\tilde 0\tilde k \tilde 0 \tilde j}\mathsf{r}^{\tilde k}\right]
\\
&&
- {2\over 3} \mu{(m_1-m_2)\over (m_1+m_2)}\left(\bar{\cal R}_{\tilde 0\tilde k \tilde \ell \tilde j}+\bar{\cal R}_{\tilde 0\tilde j \tilde \ell \tilde k}\right)
\mathsf{r}^{\tilde k}\dot{\mathsf{r}}^{\tilde \ell}
\enann
For an equal mass binary $(m_1=m_2)$, we obtain
the simple equations of motion as
\bea
\dot{\mathsf{r}}_{\tilde j}&=&{\Sigma\over \mu}
\pi_{\tilde j}
\label{EOM1}
\\
\dot{\pi}_{\tilde j}&=&-\mu\Sigma\left[{G(m_1+m_2)\mathsf{r}_{\tilde j}\over \mathsf{r}^3}+\bar{\cal R}_{\tilde 0\tilde k \tilde 0 \tilde j}\mathsf{r}^{\tilde k}\right]
\label{EOM2}
\ena

\begin{widetext}

Before numerical analysis of a binary evolution, we have to evaluate  
the Riemann curvature components and the rotation variable $\Psi$.
In the  Riemann curvature components, we have $P^2$, $R^2+S^2$ and $\Xi$, which are given by
\beann
\Xi&=&\sqrt{K-a^2\zeta^2\over K+\mathfrak{r}_0^2}=\sqrt{K-a^2\zeta_{\textsf L}^2{\rm sn}^2(\zeta_2\mathsf{t}, k_\zeta)\over K+\mathfrak{r}_0^2}
\\
P^2&=&{(E\mathfrak{r}_0^2-a\ell)^2\over \Delta(\mathfrak{r}_0)(K+\mathfrak{r}_0^2)}{(K-a^2\zeta^2)\over (\mathfrak{r}_0^2+a^2\zeta^2)}
=
{(E\mathfrak{r}_0^2-a\ell)^2\over \Delta(\mathfrak{r}_0)(K+\mathfrak{r}_0^2)}{(K-a^2\zeta_{\textsf L}^2{\rm sn}^2(\zeta_2\mathsf{t}, k_\zeta))\over (\mathfrak{r}_0^2+a^2\zeta_{\textsf L}^2{\rm sn}^2(\zeta_2\mathsf{t}, k_\zeta))}
\\
R^2+S^2&=&{(K+\mathfrak{r}_0^2)(\dot \zeta^2+(\ell+Ea\zeta^2)^2\over  (1-\zeta^2)(K-a^2\zeta^2)(\mathfrak{r}_0^2+a^2\zeta^2)}
={(K+\mathfrak{r}_0^2)\left[\zeta_{\textsf L}^2\zeta_2^2{\rm cn}^2(\zeta_2\mathsf{t}, k_\zeta){\rm dn}^2(\zeta_2\mathsf{t}, k_\zeta)+(\ell+Ea\zeta_{\textsf L}^2{\rm sn}^2(\zeta_2\mathsf{t}, k_\zeta)
)^2\right]\over  (1-\zeta_{\textsf L}^2{\rm sn}^2(\zeta_2\mathsf{t}, k_\zeta)
)(K-a^2\zeta_{\textsf L}^2{\rm sn}^2(\zeta_2\mathsf{t}, k_\zeta)
)(\mathfrak{r}_0^2+a^2\zeta_{\textsf L}^2{\rm sn}^2(\zeta_2\mathsf{t}, k_\zeta))
}
\enann
We also find
\beann
\Psi&=&\sqrt{K}(EK+a\ell)\left[-{\mathsf{t}\over K+\mathfrak{r}_0^2}+{1\over K\zeta_2}
\mathsf{\Pi}\left({a^2\zeta_{\textsf L}^2\over K},{\rm am}(\zeta_2\mathsf{t}, k_\zeta),k_\zeta^2\right)
\right]
\enann

The curvature components in the Carter tetrad frame, ${\cal Q}_1$ and ${\cal Q}_2$, are evaluated as
\beann
{\cal Q}_1&=&{M\mathfrak{r}_0(\mathfrak{r}_0^2-3a^2\zeta^2)\over (\mathfrak{r}_0^2+a^2\zeta^2)^3}={M\mathfrak{r}_0[\mathfrak{r}_0^2-3a^2\zeta_{\textsf L}^2{\rm sn}^2(\zeta_2\mathsf{t}, k_\zeta)]\over [\mathfrak{r}_0^2+a^2\zeta_{\textsf L}^2{\rm sn}^2(\zeta_2\mathsf{t}, k_\zeta)]^3}
\\
{\cal Q}_2&=&{Ma\zeta(3\mathfrak{r}_0^2-a^2\zeta^2)\over (\mathfrak{r}_0^2+a^2\zeta^2)^3}={Ma\zeta_{\textsf L}{\rm sn}(\zeta_2\mathsf{t}, k_\zeta)[3\mathfrak{r}_0^2-a^2\zeta_{\textsf L}^2{\rm sn}^2(\zeta_2\mathsf{t}, k_\zeta)]\over [\mathfrak{r}_0^2+a^2\zeta_{\textsf L}^2{\rm sn}^2(\zeta_2\mathsf{t}, k_\zeta)]^3}
\,.
\enann

\end{widetext}
\section{preliminary considerations}
\label{preliminary_considerations}
For numerical analysis of the binary motion, we shall first rewrite
 the basic equations using dimensionless variables, 
 introduce the orbital parameters, and show 
 how to set up initial data. 
In what follows, we assume that $m_1=m_2$ for simplicity.

\subsection{Normalized EOM of A Binary System }

We introduce dimensionless variables as 
\beann
\tau_*&\equiv&n_0\tau
\\
\mathsf{t}_*&\equiv&n_0\mathfrak{r}_0^2\mathsf{t}
\\
\mathsf{r}_*^{\tilde j}&\equiv&{\mathsf{r}^{\tilde j}\over \mathfrak{a}_0}
\\
\pi^*_{\tilde{j}}&\equiv&{\pi_{\tilde{j}}\over \mathfrak{a}_0 n_0\mu}
\,,
\enann
where $\mathfrak{a}_0$ is the initial ``semi-major axis'',
which will be defined in the next subsection. 
The initial ``mean motion'' of a binary $n_0$ is defined by
\beann
n_0\equiv \sqrt{G(m_1+m_2)\over \mathfrak{a}_0^3}
\enann
We also use the dimensionless variables as
\beann
\Sigma_*&\equiv&{\Sigma(\mathfrak{r}_0)\over \mathfrak{r}_0^2}=1+{a^2\over \mathfrak{r}_0^2}\zeta^2
\\
{\cal R}^*_{\tilde 0 \tilde{k}\tilde 0\tilde{\ell}}&\equiv&\mathfrak{r}_0^2
\bar{\cal R}_{\tilde 0 \tilde{k}\tilde 0\tilde{\ell}}
\,.
\enann

The normalized EOM by use of the dimensionless variables 
are now
\beann
{d\mathsf{x}_*\over d\mathsf{t}_*}&=&\Sigma_*\pi^*_{\mathsf{x}}
\\
{d\mathsf{y}_*\over d\mathsf{t}_*}&=&\Sigma_*\pi^*_{\mathsf{y}}
\\
{d\mathsf{z}_*\over d\mathsf{t}_*}&=&\Sigma_*\pi^*_{\mathsf{z}}
\enann
and 
\beann
{d\pi^*_{\mathsf{x}}\over d\mathsf{t}_*}
&=&-\Sigma_*\left[{\mathsf{x}_*\over \mathsf{r}_*^3}+{1\over n_0^2\mathfrak{r}_0^2}
\left({\cal R}^*_{\tilde 0 \tilde 1\tilde 0\tilde 1} \mathsf{x}_*+{\cal R}^*_{\tilde 0 \tilde 2\tilde 0\tilde 1} \mathsf{y}_*+{\cal R}^*_{\tilde 0 \tilde 3\tilde 0\tilde 1} \mathsf{z}_*\right)\right]
\\
{d\pi^*_{\mathsf{y}}\over d\mathsf{t}_*}
&=&-\Sigma_*\left[{\mathsf{y}_*\over \mathsf{r}_*^3}+{1\over n_0^2\mathfrak{r}_0^2}
\left({\cal R}^*_{\tilde 0 \tilde 1\tilde 0\tilde 2} \mathsf{x}_*+{\cal R}^*_{\tilde 0 \tilde 2\tilde 0\tilde 2} \mathsf{y}_*+{\cal R}^*_{\tilde 0 \tilde 3\tilde 0\tilde 2}\mathsf{z}_*\right)\right]
\\
{d\pi^*_{\mathsf{z}}\over d \mathsf{t}_*}
&=&-\Sigma_*\left[{\mathsf{z}_*\over \mathsf{r}_*^3}+{1\over n_0^2\mathfrak{r}_0^2}
\left({\cal R}^*_{\tilde 0 \tilde 1\tilde 0\tilde 3} \mathsf{x}_*+{\cal R}^*_{\tilde 0 \tilde 2\tilde 0\tilde 3}\mathsf{y}_*+{\cal R}^*_{\tilde 0 \tilde 3\tilde 0\tilde 3}\mathsf{z}_*\right)\right]
\enann

\subsection{Orbital Parameters and initial data}
 In order to describe the properties of 
a binary orbit, it is more convenient to use the orbital parameters,
since we expect that
 the binary motion is close to an elliptic orbit, which 
 is described by
\bea
\mathsf{r}={\mathfrak{a}(1-e^2)\over 1+e\cos f}
\,,
\label{elliptic_orbit}
\ena
where $\mathfrak{a}, e$ and $f$ are the semi-major axis
of an elliptic orbit, the eccentricity, and the true anomaly, respectively.
We also introduce 
three angular variables; the argument of periapsis $\omega$, the ascending node $\Omega$ and the inclination angle $I$.

\begin{widetext}
For the elliptic orbit, the relation between the  Cartesian  coordinates $\bf{\mathsf{r}}=(\mathsf{x},\mathsf{y},\mathsf{z})$ of a binary and the orbital parameters $(\omega\,,\Omega\,,  \mathfrak{a}\,, e\,, I\,, f)$ is given by 
\bea
\begin{pmatrix}
\mathsf{x} \\
\mathsf{y} \\
\mathsf{z} \\
\end{pmatrix}
&=&
\mathsf{r}\begin{pmatrix}
\cos \Omega\cos(\omega+f)-\sin\Omega\sin(\omega+f)\cos I \\
\sin \Omega\cos(\omega+f)+\cos\Omega\sin(\omega+f)\cos I\\
\sin(\omega+f)\sin I \\
\end{pmatrix}
\label{orbital_parameters}
\ena
\end{widetext}

In order to extract the orbital parameters 
from the orbit given by the Cartesian 
coordinates, one can use the osculating orbit when the orbit is close to an ellipse. 
The 
normalized Laplace-Runge-Lenz vector is defined by
\bea
\vect{e}\equiv \vect{\mathsf{p}}^*\times (\vect{\mathsf{r}}_*
\times \vect{\mathsf{p}}^*)-{\vect{\mathsf{r}}_*\over \mathsf{r}}_*\,,
\label{LRL_vector}
\ena
and its magnitude $e=|\vect{e}|$
is commonly used for a measure of orbital eccentricity.
The inclination angle $I$ is defined as mutual inclination between angular momenta of the inner and outer binary.
\bea
I=\cos^{-1}\left({\mathfrak{l}^*_{\mathsf{z}} \over |\vect{\mathfrak{l}}^*|}\right)
\,,
\label{def_incrination}
\ena
where $\vect{\mathfrak{l}}^*\equiv \vect{\mathsf{r}}_*\times \vect{\mathsf{p}}^*$ is the normalized angular momentum of a binary, which is defined in the non-rotating local inertial frame.

The other two essential angles $\Omega$ and $\omega$ governing the orientation of the orbital plane. 
The longitude of ascending node $\Omega$ is the angle between the reference axis (say $\mathsf{x}$-axis) and node line vector $\vect{N}$, which is defined by $\vect{N} = \hat{\vect{\mathsf{z}}}\times \vect{\mathfrak{l}}^* $, where $\hat{\vect{\mathsf{z}}}$ is normal to the reference plane (the unit vector in the $\mathsf{z}$ direction). 
So $\Omega$ is computed as,
\bea
\Omega = \cos^{-1} (N_{\mathsf{x}}/N) \,.
\label{def_Omega}
\ena
The argument of periapsis $\omega$ is the angle between node line and periapsis measured in the direction of motion. Therefore,
\bea
\omega = \cos^{-1} \bigg(\frac{\vect{N} \cdot \vect{e}}{N\,e}\bigg) \,.
\label{def_omega}
\ena



In order to provide the initial data of a binary, i.e., ${\mathsf{x}_*}_0\,,{\mathsf{y}_*}_0\,,{\mathsf{z}_*}_0$ and $\mathsf{p}^*_{\mathsf{x}0}\,,\mathsf{p}^*_{\mathsf{y}0}\,,\mathsf{p}^*_{\mathsf{z}0}$, 
we shall give the initial orbital parameters  $(\omega_0\,,\Omega_0\,,  \mathfrak{a}_0\,, e_0)$.
From (\ref{orbital_parameters}), assuming $f=0$ at $\tau=0$, 
we find
\beann
{\mathsf{x}_*}_0
&=&(1-e_0)\left[\cos \Omega_0\cos\omega_0-\sin\Omega_0\sin\omega_0\cos I_0\right]\,,
\\
{\mathsf{y}_*}_0
&=&(1-e_0)\left[\sin \Omega_0\cos\omega_0+\cos\Omega_0\sin\omega_0\cos I_0\right]\,,
\\
{\mathsf{z}_*}_0
&=&(1-e_0)\sin\omega_0\sin I_0
\,.
\enann
As for the momentum $\mathsf{p}^*_{\mathsf{x}0}\,,\mathsf{p}^*_{\mathsf{y}0}\,,
\mathsf{p}^*_{\mathsf{z}0}$, 
we use the definitions of the orbital parameters 
of the osculating orbit, i.e., Eqs. (\ref{LRL_vector}), (\ref{def_incrination}), 
(\ref{def_Omega}) and (\ref{def_omega}).
The details are found in \cite{Maeda:2023uyx}.

\subsection{Validity and Stability}
\label{validity}
Before showing our numerical results, 
we discuss validity of the present approach and 
the stability conditions. The minimum curvature radius at the radius $\mathfrak{r}_0$ is evaluated as 
\beann
 \ell_{\bar{\cal R}} &\equiv& {\rm min} \left[|\bar{\cal R}_{\hat \mu\hat \nu\hat \rho\hat \sigma}|^{-{1\over 2}}, |\bar{\cal R}_{\hat \mu\hat \nu\hat \rho\hat \sigma ; \hat \alpha}|^{-{1\over 3}}, |\bar{\cal R}_{\hat \mu\hat \nu\hat \rho\hat \sigma ; \hat \alpha;\hat \beta}|^{-{1\over 4}}
\right]
\\
&\sim&
 {\rm min} \left[\left({M\over \mathfrak{r}_0^3}\right)^{-{1\over 2}}\,,\left({M\over \mathfrak{r}_0^4}\right)^{-{1\over 3}}\,,
\left( {M\over \mathfrak{r}_0^5}\right)^{-{1\over 4}}
\right]
\\
&\sim &\mathfrak{r}_0 \left( {\mathfrak{r}_0 \over M}\right)^{1/4}
\,.
\enann
When we put a binary at $\mathfrak{r}=\mathfrak{r}_0$,
 the  binary size $\ell_{\rm binary}$ should satisfy
\beann
\ell_{\rm binary}\ll  \ell_{\bar{\cal R}} 
\enann

The relativistic effect in a binary is not  important when
\beann
\ell_{\rm binary} &\gg& {G(m_1+m_2)\over c^2}
\,.
\enann

As for stability of a binary, the mutual gravitational interaction between a binary 
should be much larger than the tidal force by a third body.
The condition is given by
\beann
{Gm_1m_2\over \mathsf{r}^2}\gg {\mu M  \over \mathfrak{r}_0^3}\,\mathsf{r}
\,.
\enann
It gives the constraint on a binary size $\ell_{\rm binary}$ as
\bea
\ell_{\rm binary} 
&\ll &
\left({m_1+m_2\over M}\right)^{1\over 3}\,\mathfrak{r}_0.
\label{stability_tidal}
\ena
Hence, 
for a binary with the size of
\beann
{G(m_1+m_2)\over c^2 }\ll \ell_{\rm binary} 
 \ll 
 \left({m_1+m_2\over M}\right)^{1\over 3}\,\mathfrak{r}_0\,,
 \enann
 we may apply the present Newtonian approach.
 
We also have another criterion for stability
\cite{mardling2001tidal,myllari2018stability}.
In order to avoid a chaotic energy exchange 
instability, we may have to impose 
the condition for the ratio of the circular radius $\mathfrak{r}_0$ to 
the binary size $\ell_{\rm binary} $ 
such that
\bea
{\mathfrak{r}_0\over \ell_{\rm binary} }
\gsim C_{\rm chaotic}\left({M\over m_1+m_2}\right)^p\,,
\label{chaotic_stability}
\ena
when $M\gg m_1, m_2$. 
In the previous paper, we analysed the details in the case of the circular orbit and 
 found that 
$p=1/3$ and $ C_{\rm chaotic}\sim 2-4$.

\section{Numerical Analysis}
\label{numerical_analysis}
Now we show our numerical results.
In this paper, we focus on the von Zeipel-Lidov-Kozai (vZLK) mechanism, in which the oscillations between the inclination angle $I$ and the eccentricity $e$ of a binary are caused by
 the tidal force of the tertial body\cite{vonZeipel10, Lidov62, Kozai62}.

\begin{figure}[htbp]
\begin{center}
\includegraphics[width=7.5cm]{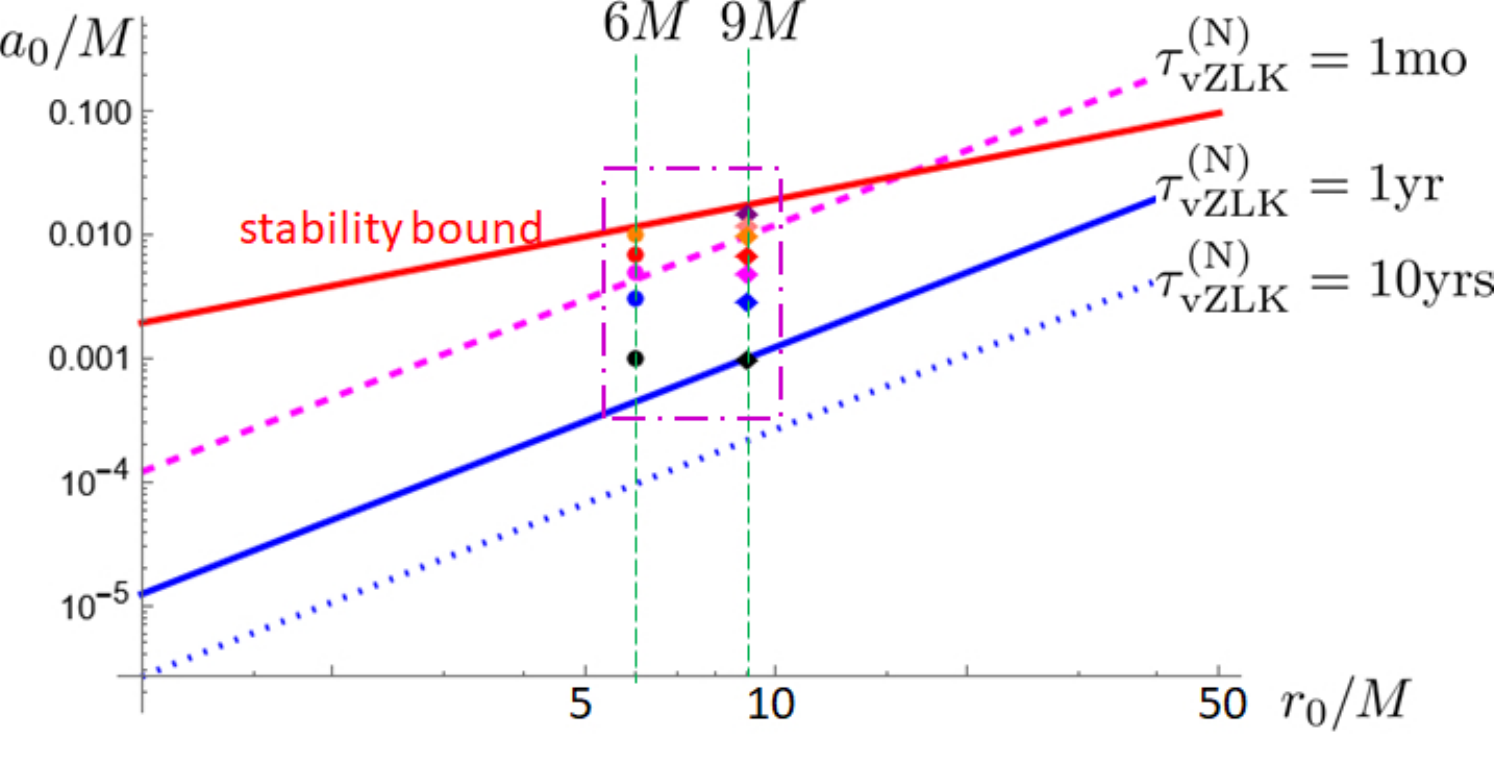}
\\(a)
\\[1em]
\includegraphics[width=7.5cm]{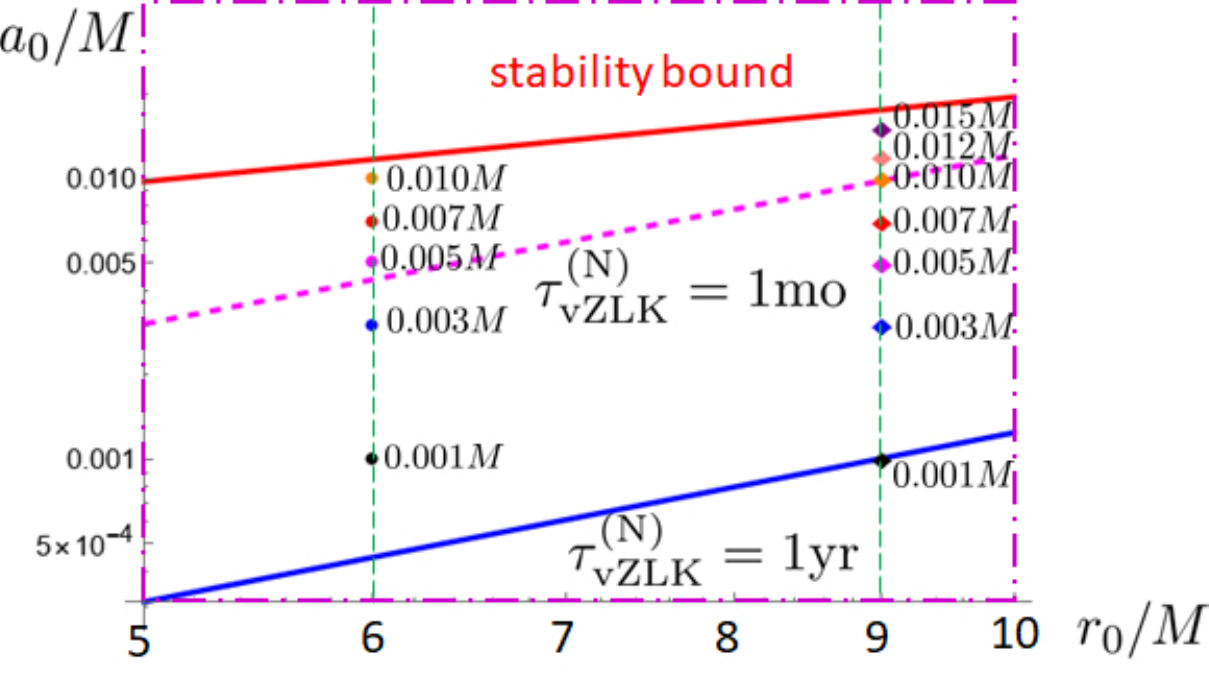}
\\
(b)
\caption{(a) The chaotic stability bound (the red curve) and The Newtonian vZLK oscillation time scales 
$\tau_{\rm vZLK}^{\rm (N)}$ of 
1 month (the Dashed Magenta), 1 year (the solid blue) and 10 years (the dotted blue). (b) The enlarged figure of the dotted-dash magenta box in  (a),  and the parameters of our models; 
  $\mathfrak{r}_0=6M,  9M$, and
 $\mathfrak{a}_0=0.001M$ (black), $0.003M$ (blue), $0.005M$ (magenta), $0.007M$ (red), $0.01M$ (orange), $0.012M$ (pink), and $0.015M$ (purple).
}
\label{fig:parameter_range}
\end{center}
\end{figure}

First we  show  the parameter range, where the above chaotic instability will be
avoided in Fig. \ref{fig:parameter_range}.
We also plot the curves of $\tau_{\rm vZLK}^{\rm (N)}=1$ mo, 1 yr  and 10 yrs
in order to get some idea of observability of vZLK oscillations, where 
$\tau_{\rm vZLK}^{\rm (N)}\equiv 16n_0\mathfrak{r}_0^3/M$ is 
the Newtonian vZLK oscillation period.

Since we will  focus on the dependence of $\zeta_{\textsf L}$ (or the latitudinal libration) in
our analysis, we consider compact and stable binary on a spherical orbit around a Kerr black hole.
We analyze the case such that the mass of SMBH is $10^8 M_\odot$ and $m_1=m_2=10M_\odot$.
We  then perform numerical analysis with wide range of parameters, i.e., 
$a=0.1 M, 0.5 M$ and $1.0 M$ for the Kerr parameter, $\mathfrak{r}_0=9M$, and $6M$ 
for the position of a binary. As for the initial orbital parameters of a binary, we choose 
$\mathfrak{a}_0=0.001M, 0.003M, 0.005M, 0.007M, 0.010M, 0.012M$, and $0.015M$ for the initial semi-major axis, $I_0 =30^\circ\,, 40^\circ\,, 45^\circ\,, 50^\circ\,, 55^\circ\,, 60^\circ\,, 70^\circ\,, 80^\circ$\,,  and $85^\circ$ for the initial inclination angle, and $\zeta_{\textsf L} = 0\,,  0.2\,,  0.4\,, 0.6\,, 0.8\,, 0.9$\,,  and $0.99$
for the latitudinal libration angle. The initial eccentricity,  the argument of periapsis, and the ascending node  are fixed at $e_0=0.01$, $\omega_0=60^\circ$, and $\Omega_0=30^\circ$, respectively. 


\subsection{vZLK oscillations of a binary on a circular orbit}
As reference, we first summarize the results in the case of a binary in the equatorial plane, which was studied in the previous work\cite{Maeda:2023tao,Maeda:2023uyx}.\\
(1) We find regular and stable vZLK oscillations for hard binary.\\
We show the case with $a=0.5 M, \mathfrak{r}_0=9M$ and $\mathfrak{a}_0=0.003M$ in Fig.  \ref{fig:vZLK_593}.
The regular vZLK oscillations are clearly found for 
$I_0=80^\circ, 60^\circ$ and $45^\circ$, which are larger than the critical 
angle $I_{\rm cr}\approx 40^\circ$. 
The maximum values of the eccentricity decreases and the vZLK oscillation period $T_{\rm vZLK}$ increases 
as the initial inclination angle $I_0$ decreases.
For $I_0\leq I_{\rm cr}$, the enhancement of the eccentricity does not occur.

\noindent
(2) For soft binary, the vZLK oscillations become irregular and chaotic.\\
We show the case with $a=0.5 M, \mathfrak{r}_0=9M$ and $I_0=60^\circ$ in Fig.  \ref{fig:vZLK_5960}.
The initial semi-major axis are chosen as  (a) $\mathfrak{a}_0=0.005M$, (b) $\mathfrak{a}_0=0.012M$, (c) $\mathfrak{a}_0=0.015M$. The vZLK oscillations of a hard binary (a)  are regular, while both amplitude and oscillation period 
for a soft binary (c) become irregular. The intermediate case (b) shows a small irregularity in the eccentricity.

\begin{widetext}

\begin{figure}[htbp]
\begin{center}
\includegraphics[width=12cm]{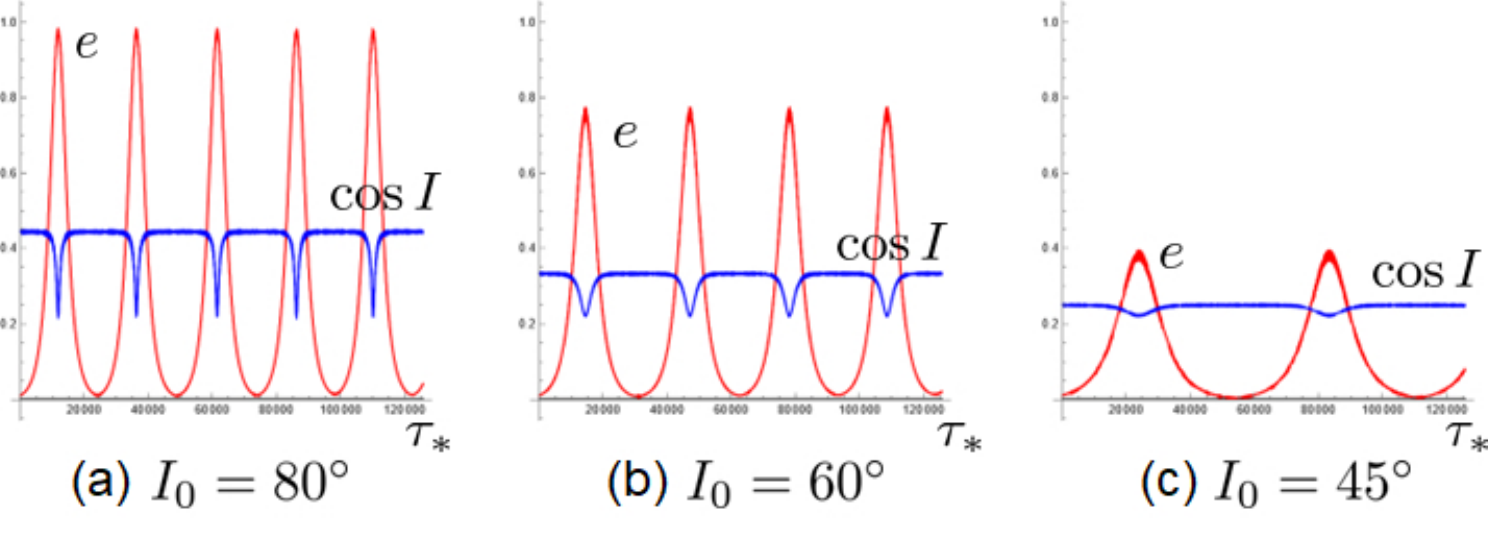}
\caption{vZLK oscillations for $a = 0.5M$, $\mathfrak{r}_0 = 9M$, and $\mathfrak{a}_0 = 0.003M$.
The red and blue curves show the time evolution of the eccentricity $e$ and inclination $I$, respectively.
Panels (a), (b), and (c) correspond to initial inclinations $I_0 = 80^\circ$, $60^\circ$, and $45^\circ$.
}
\label{fig:vZLK_593}
\end{center}
\end{figure}

\begin{figure}[htbp]
\begin{center}
\includegraphics[width=15cm]{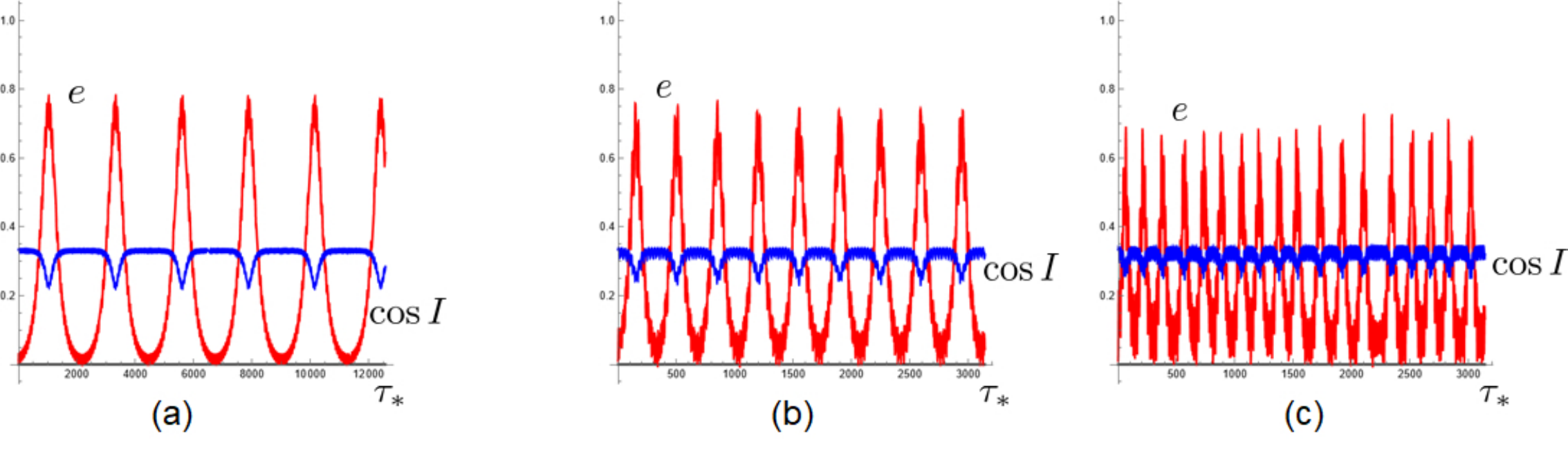}
\caption{Time evolution of the eccentricity (red curve) and the inclination (blue curve).
We choose $a=0.5 M, \mathfrak{r}_0=9M$ and $I_0=60^\circ$, and 
the initial semi-major axis as  (a) $\mathfrak{a}_0=0.005M$, (b) $\mathfrak{a}_0=0.012M$, (c) $\mathfrak{a}_0=0.015M$. 
We find a regular vZLK oscillation for a hard binary (a), while irregular oscillations both in the eccentricity and the inclination angle for a soft binary (c).
}
\label{fig:vZLK_5960}
\end{center}
\end{figure}

\subsection{Latitudinal libration  dependence on vZLK oscillations}

Next we consider a binary on a spherical orbit with the latitudinal libration angle $\chi_{\textsf L}$, where
 $\zeta_{\textsf L}=\sin \chi_{\textsf L}$. We analyze seven values of $\zeta_{\textsf L}=0, 0.2, 0.4, 
 0.6, 0.8, 0.9$ and $0.99$, which correspond to 
the latitudinal libration angles as
$\chi_{\textsf L}=0^\circ, 11.5^\circ, 23.6^\circ, 36.9^\circ, 53.1^\circ, 64.1^\circ$, and $81.9^\circ$, respectively.

In order to see the latitudinal libration dependence, 
in Fig. \ref{fig:vZLK_5960LL}, we present the evolution of the eccentricity $e$ and inclination $I$ for the case with 
$a=0.5 M, \mathfrak{r}_0=9M$ and $\mathfrak{a}_0=0.007M$. The initial inclination angle is fixed as $I_0=60^\circ$.  
We find the qualitative difference between zero libration and nonzero libration.
The  vZLK oscillations for $\zeta_{\textsf L}=0$ are regular [(a)] , whereas 
 the vZLK oscillations  become irregular for nonzero libration angles [(b)-(d)]. 
Although the oscillations for  $\zeta_{\textsf L}=0.2$ are slightly irregular,  those for  $\zeta_{\textsf L}=0.6$ and $0.9$ become highly chaotic. The deviation from regular oscillations are two ways: One is the oscillation period,
and the other is the oscillation amplitude of the eccentricity. 
The vZLK oscillation period becomes shorter
 and the maximum eccentricity gets slightly larger  for nonzero libration angle.
 These features are always found for nonzero libration oscillations.
 In particular, as we will see later, for highly chaotic case, the maximum eccentricity is enhanced to nearly unity, 
 and the vZLK period becomes close to the dynamical time scale of a binary orbit.
It is very important for observation. The high eccentricity yields the large emission of the gravitational waves from a binary, while the short vZLK oscillation period affects on the evolutionary history of a binary system.
    
\begin{figure}[htbp]
\begin{center}
\includegraphics[width=15cm]{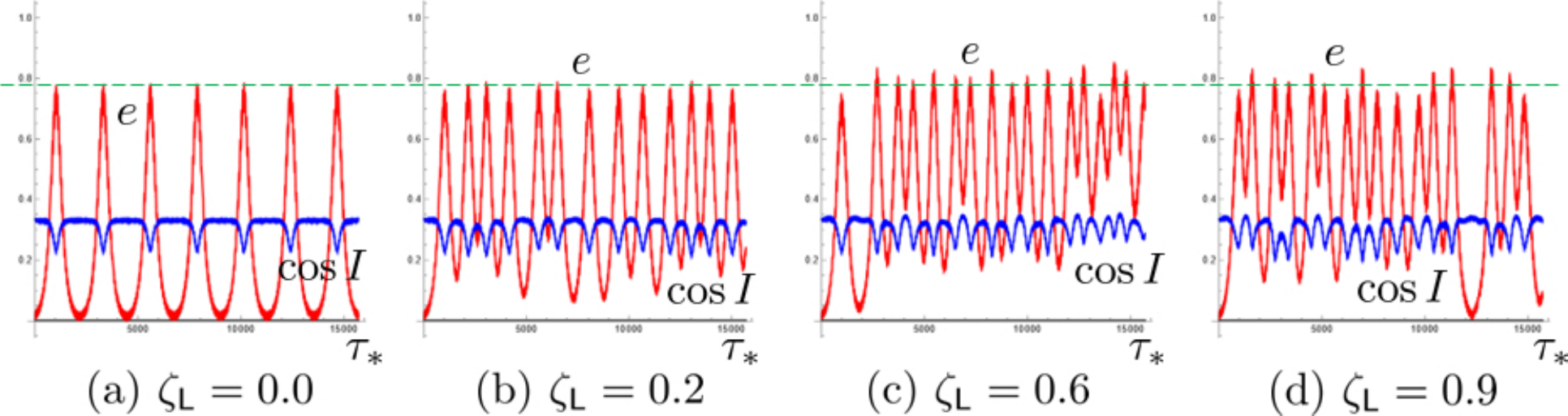}
\caption{Time evolution of the eccentricity (red curve) and the inclination (blue curve).
We choose $a=0.5 M, \mathfrak{r}_0=9M, I_0=60^\circ$, and 
the initial semi-major axis  $\mathfrak{a}_0=0.007M$.
The latitudinal libration angles are set 
as $\zeta_{\textsf L}=0, 0.2,  
 0.6, 0.9$, corresponding to 
the latitudinal libration angles 
$\chi_{\textsf L}=0^\circ, 11.5^\circ, 36.9^\circ, 64.1^\circ$, respectively.
The green dotted line denotes the maximum eccentricity  for $\zeta_{\textsf L}=0$.
We find a regular vZLK oscillation for $\zeta_{\textsf L}=0$, whereas 
 the vZLK oscillations  become irregular for nonzero libration angles. 
Although the oscillations for  $\zeta_{\textsf L}=0.2$ are slightly irregular,  those for  $\zeta_{\textsf L}=0.6$ and $0.9$ become highly chaotic.
}
\label{fig:vZLK_5960LL}
\end{center}
\end{figure}

\end{widetext}

Since we perform numerical analyses over a wide range of parameters, the main results are summarized below according to the following key aspects.

\subsection{Enhancement of the eccentricity}
\label{Enhancement_of_eccentricity}

Here we focus on the enhancement of the eccentricity.
 When a binary is hard enough, we can apply the double averaging approach as summarized in Appendix
 \ref{DA_approach} and \ref{DA_solution}.
 For a binary in the equatorial plane, we find the analytic solutions, in which the maximum eccentricity 
 $e_{\rm max}$ is
 given by two integration constants, $\vartheta$ and $C_{\rm vZLK}$, which are defined by
 \beann
 \vartheta&\equiv &\sqrt{1-e^2}\cos I
 \\
C_{\rm vZLK}&\equiv &e^2\left(1-{5\over 2}\sin^2 I\sin^2\omega\right)
\,.
 \enann
As a result,  as shown in Fig. \ref{fig:emaxminTvZLK} (a), 
$e_{\rm max}$ depends highly on the initial eccentricity $e_0$ and inclination $I_0$.
In our analysis, we analyze only the case of $e_0=0.01$ because 
we would like to see how the orbital eccentricity grows
by vZLK mechanism.
Using the analytic solution in DA approach as reference, we show our results in below.

\begin{figure}[htbp]
\begin{center}
\includegraphics[width=7cm]{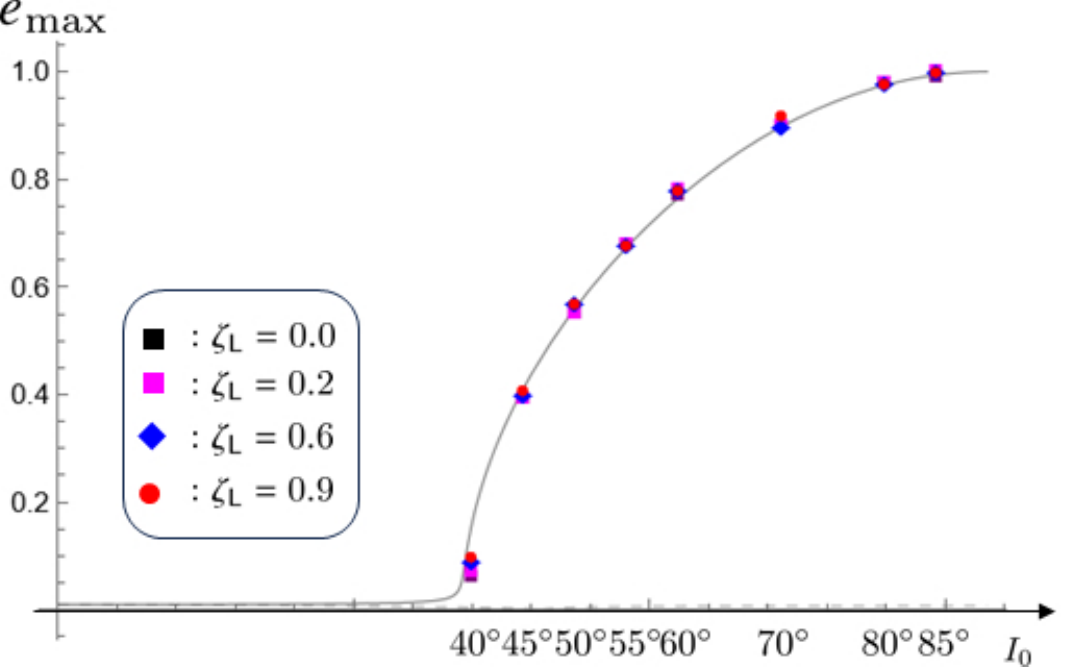}\\
(a)\\[1em]
\includegraphics[width=7cm]{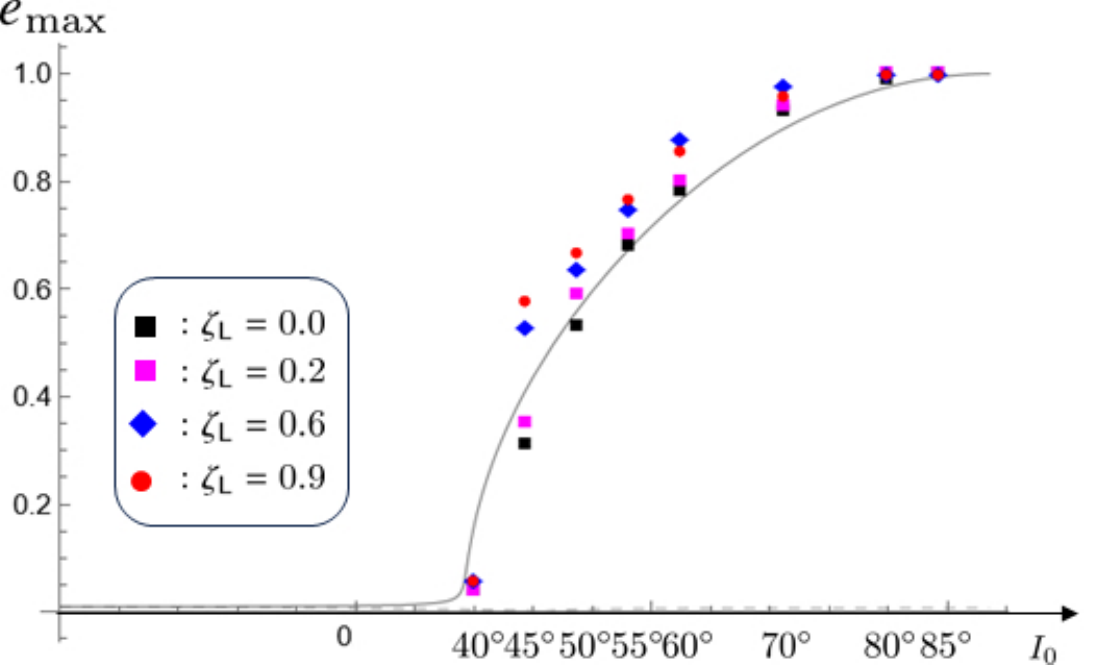}\\
(b)
\caption{The maximum values of the eccentricity are plotted in terms of the initial inclination $I_0$.
The latitudinal libration angles are set 
as $\zeta_{\textsf L}=0, 0.2,  0.6, 0.9$, which are marked by the filled black square, the filled magenta square, 
the filled blue diamond, and the filled red circle, respectively.
We choose $a=0.5 M, \mathfrak{r}_0=9M$, and  (a) $\mathfrak{a}_0=0.003M$ and (b) $\mathfrak{a}_0=0.007M$.
The black solid curve denotes the maximum eccentricity of a binary in the equatorial plane in the DA approach.
}
\label{fig:emax5937}
\end{center}
\end{figure}

In Fig. \ref{fig:emax5937}, setting $a=0.5 M$ and $\mathfrak{r}_0=9M$, 
we plot the maximum values of the eccentricity in terms of the initial inclination $I_0$
for the case of  (a) $\mathfrak{a}_0=0.003M$ and of (b) $\mathfrak{a}_0=0.007M$.

The latitudinal libration angles are set 
as $\zeta_{\textsf L}=0, 0.2,  0.6, 0.9$, which are marked by the filled black square, the filled magenta square, 
the filled blue diamond, and the filled red circle, respectively.
The black solid curve, which denotes the maximum eccentricity of a binary in the equatorial plane in the DA approach, is also plotted as reference.

For $\mathfrak{a}_0=0.003M$, 
the maximum values are degenerated to 
the black solid curve (the DA approach).
It confirms that the DA approximation is valid for a hard binary.
For $\mathfrak{a}_0=0.007M$, on the other hand, 
the maximum eccentricity becomes larger than the black solid curve for non-zero libration angle.
Note that for the zero libration angle, 
 the  maximum eccentricity is almost the same as the value in the DA approach.
 We can conclude that the latitudinal libration enhances the maximum eccentricity of 
 a binary.

\begin{figure}[htbp]
\begin{center}
\includegraphics[width=7cm]{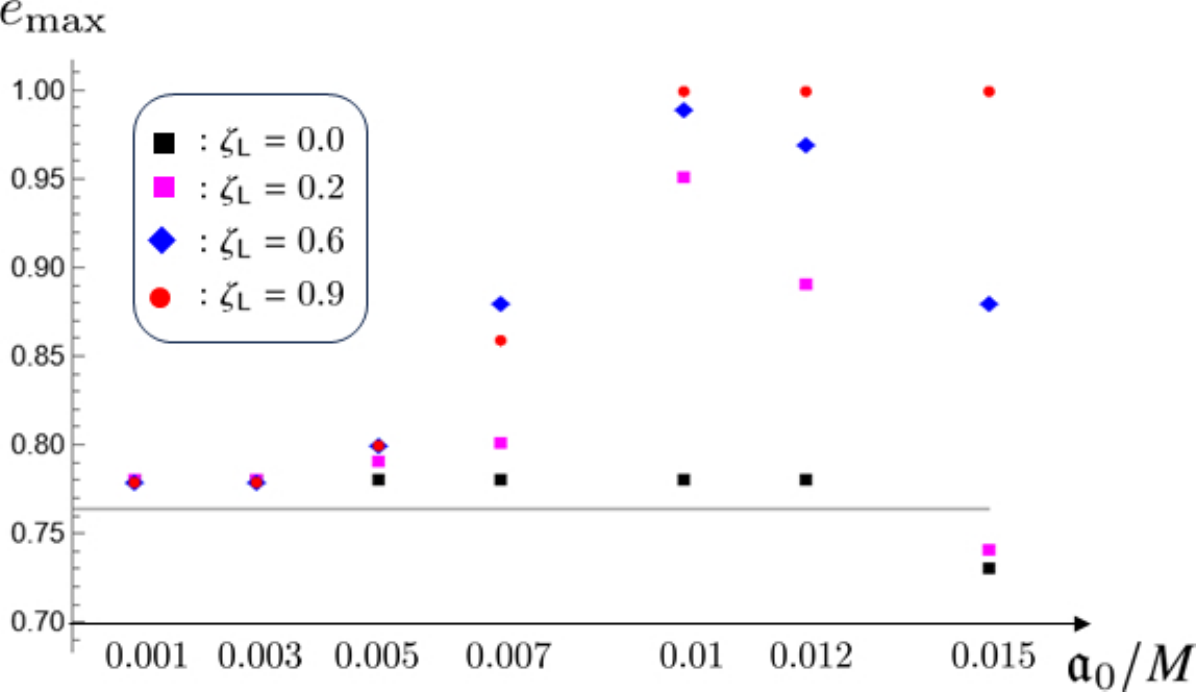}
\caption{The maximum values of the eccentricity are plotted in terms of the initial semi-major axis $\mathfrak{a}_0$.
The latitudinal libration angles are set 
as $\zeta_{\textsf L}=0, 0.2,  0.6, 0.9$, which are marked by the filled black square, the filled magenta square, 
the filled blue diamond, and the filled red circle, respectively.
We choose $a=0.5 M, \mathfrak{r}_0=9M$, and  $I_0=60^\circ$.
The black solid curve denotes the maximum eccentricity of a binary on the equatorial plane in the DA approach.
}
\label{fig:emax5960}
\end{center}
\end{figure}

In Fig. \ref{fig:emax5960}, setting $a=0.5 M$, $\mathfrak{r}_0=9M$ and $I_0=60^\circ$, 
we show the maximum eccentricity in terms of the initial semi-major axis $\mathfrak{a}_0 $.

For hard binaries ($\mathfrak{a}_0 =0.001, 0.003$), the enhancement is not so large,
while for soft binaries
 ($\mathfrak{a}_0 =0.007-0.015$), the enhancement gets large.
 In particular, for very soft binaries, the maximum value can reach to almost unity, and some binaries are  
 broken ($\mathfrak{a}_0 =0.015$ and $\zeta_{\textsf L}= 0.4, 0.8, 0.9, 0.99$).
 
\subsection{Dynamical vZLK oscillations}

Next we discuss the vZLK oscillation period.
As \S. \ref{Enhancement_of_eccentricity}, 
using the analytic solution in DA approach as reference, we show our numerical results.

\begin{figure}[htbp]
\begin{center}
\includegraphics[width=7cm]{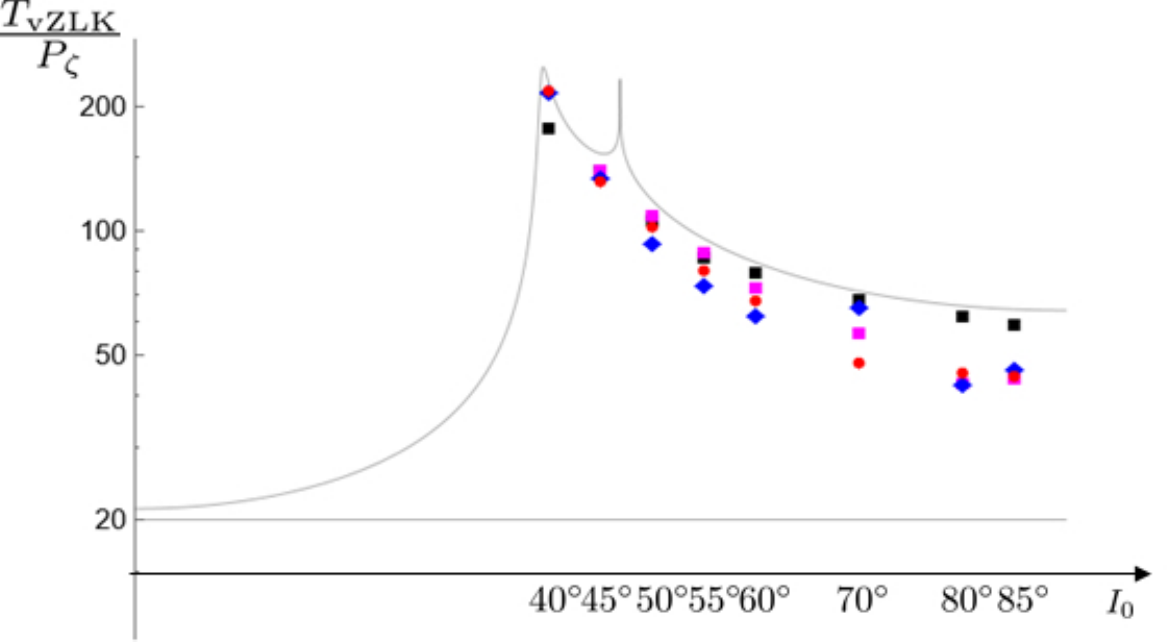}\\
(a)\\[1em]
\includegraphics[width=7cm]{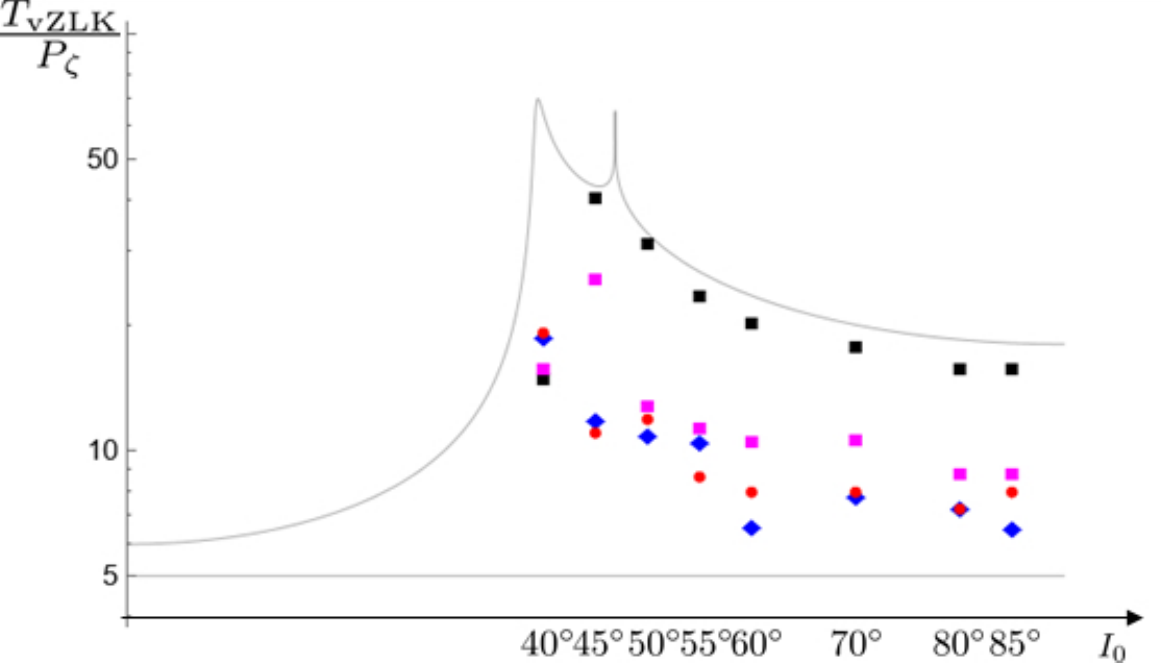}\\
(b)
\caption{The vZLK oscillation period $T_{\rm vZLK}$ normalized by the libration period 
$P_\zeta$ is
 plotted in terms of the initial inclination $I_0$.
The latitudinal libration angles are set 
as $\zeta_{\textsf L}=0, 0.2, 0.6, 0.9$, which are marked by the filled black square, the filled magenta square, 
the filled blue diamond, and the filled red circle, respectively.
We choose $a=0.5 M, \mathfrak{r}_0=9M$, and  (a) $\mathfrak{a}_0=0.003M$ and (b) $\mathfrak{a}_0=0.007M$.
The black solid curve denotes the vZLK oscillation period of a binary in the equatorial plane in the DA approach.
}
\label{fig:TvZLK5937}
\end{center}
\end{figure}

As shown in Fig. \ref{fig:TvZLK5937}, 
the vZLK oscillation period $T_{\rm vZLK}$ for a binary in the equatorial plane 
is almost the same as that in DA approximation, while 
the period for a binary with libration always
decreases.
This reduction rate gets larger for a softer binary ($\mathfrak{a}_0 =0.007$),
 although there is not so much difference between the libration angles.
  
 \begin{figure}[htbp]
\begin{center}
\includegraphics[width=7cm]{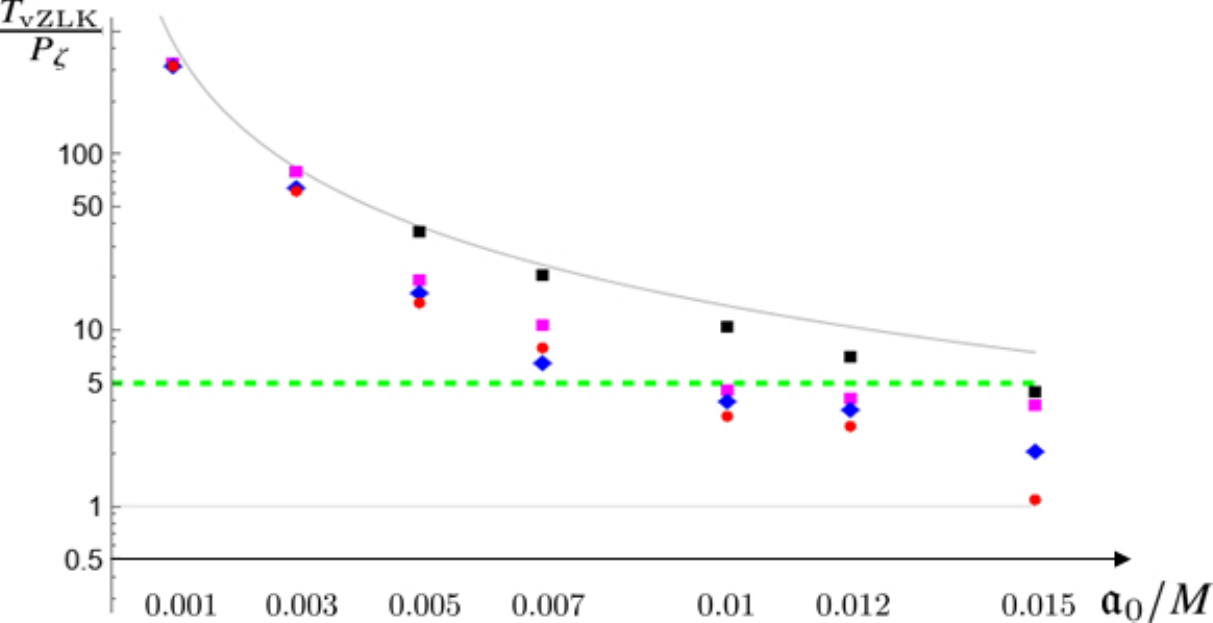}
\caption{The vZLK oscillation period $T_{\rm vZLK}$ normalized by the libration period 
$P_\zeta$ is
 plotted in terms of the initial semi-major axis $\mathfrak{a}_0$.
The latitudinal libration angles are set 
as $\zeta_{\textsf L}=0$ (the filled black square), 0.2 (the filled magenta square) , 0.6 (the filled blue diamond),  and 0.9 (the filled red circle).
We choose $a=0.5 M, \mathfrak{r}_0=9M$, and $I_0=60^\circ$.
The black solid curve denotes the vZLK oscillation period of a binary on the equatorial plane in the DA approach.
The green dotted line ($T_{\rm vZLK} = 5P_\zeta$) is given as a reference.
}
\label{fig:TvZLK5960Pz}
\end{center}
\end{figure}

We also show the vZLK oscillation period $T_{\rm vZLK}$ normalized by the libration period 
$P_\zeta (\equiv 2\pi/\Upsilon_\zeta)$  in terms of the initial semi-major axis $\mathfrak{a}_0$.
The latitudinal libration angles are set 
as $\zeta_{\textsf L}=0, 0.2, 0.6, 0.9$, which are marked by the filled black square, the filled magenta square, 
the filled blue diamond, and the filled red circle, respectively.
We choose $a=0.5 M, \mathfrak{r}_0=9M$, and $I_0=60^\circ$.
The black solid curve denotes the vZLK oscillation period of a binary in the equatorial plane in the DA approach.
The green dotted line ($T_{\rm vZLK} = 5P_\zeta$) is given as reference.
We find that the ``secular'' vZLK oscillations occurs on an almost dynamical timescale because 
$T_{\rm vZLK}$ is just several times the libration period 
$P_\zeta$. 

\subsection{Dependence of spherical orbit radius $\mathfrak{r}_0$}

Since the smallest radius of a stable spherical orbit in Kerr geometry 
is the ISSO radius, 
we can discuss  more relativistic orbit with smaller radius than $\mathfrak{r}_0=9M$.
In order to compare the dependence of the libration angles, 
we shall choose $\mathfrak{r}_0=6M$, 
which is always larger than $\mathfrak{r}_{\rm ISSO}$ for the prograde orbits.

\begin{figure}[htbp]
\begin{center}
\includegraphics[width=7cm]{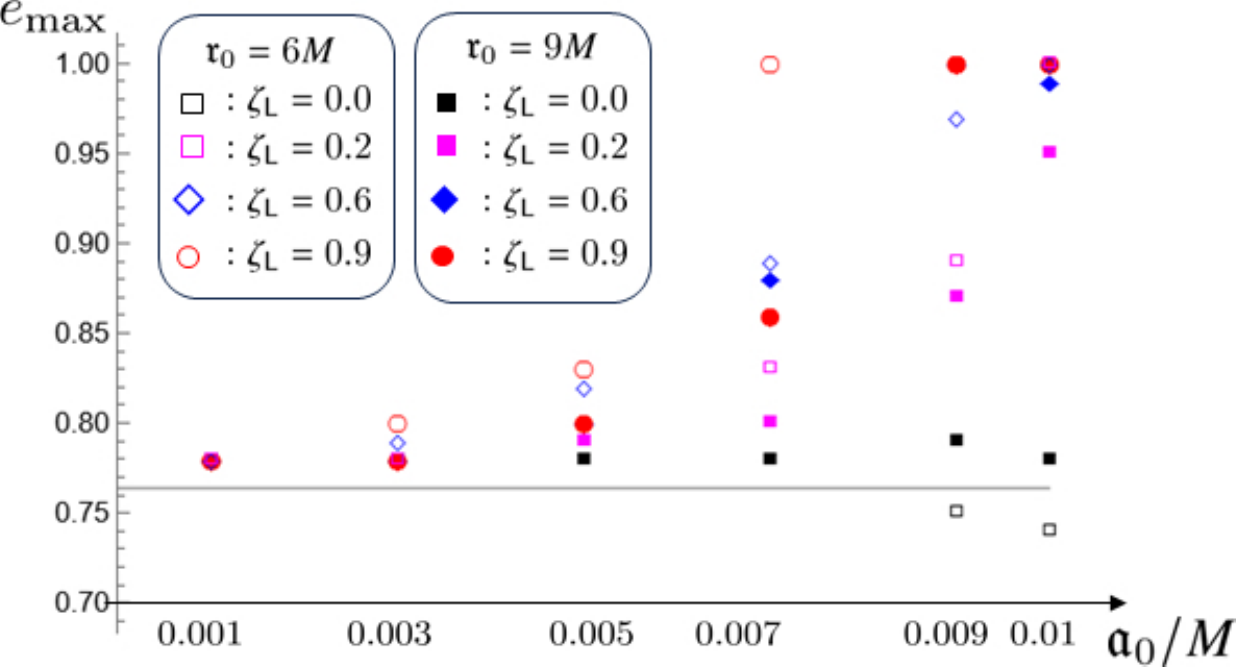}
\\
(a)
\\
\includegraphics[width=7cm]{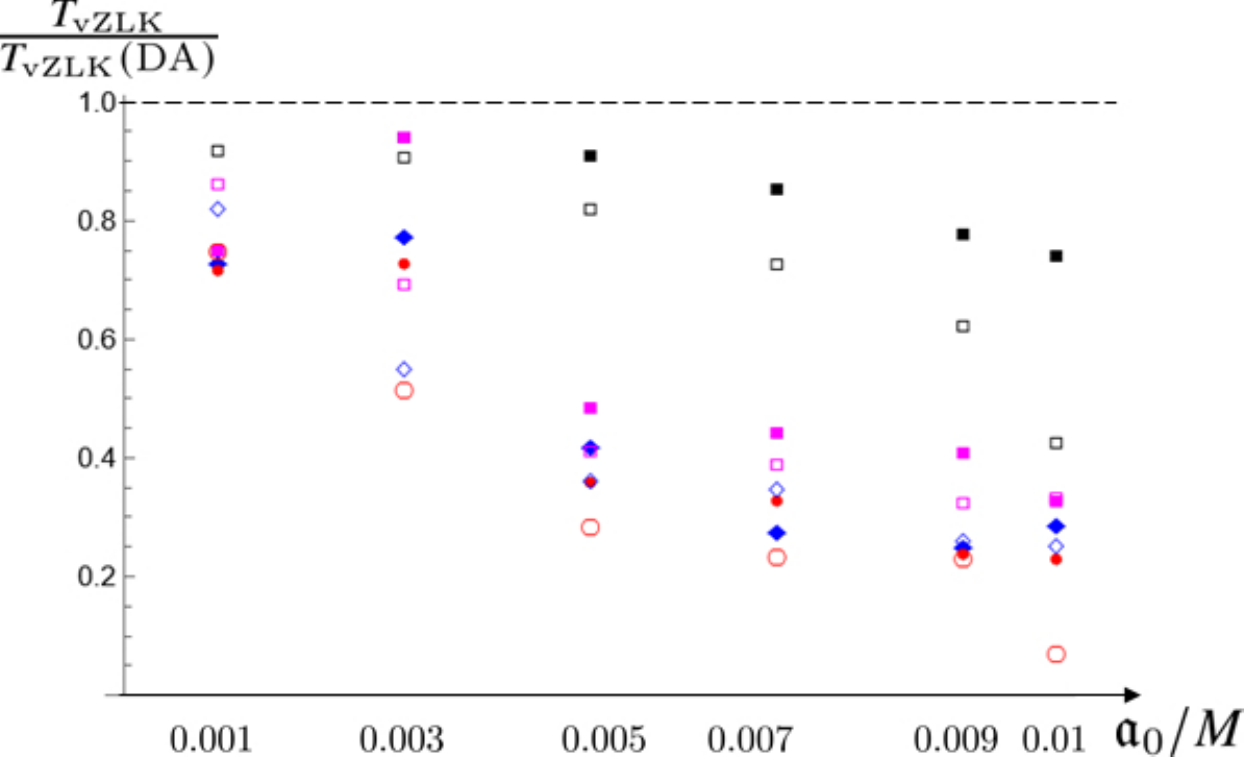}
\\
(b)
\caption{We depict the  maximum eccentricity (a) and the vZLK oscillation period
normalized by the value in the DA approximation for  $ \mathfrak{r}_0=6M $ and $9M$, 
by the empty marks and filled marks, respectively.
}
\label{fig:56&960}
\end{center}
\end{figure}

In Fig. \ref{fig:56&960}, we present the results for $ \mathfrak{r}_0=6M $ and $9M$, 
by the empty marks and filled marks, respectively.
In Fig. \ref{fig:56&960} (a), we show the maximum eccentricity. For the hard binary 
($\mathfrak{a}_0=0.001M$),
the enhancement of the eccentricity is not observed, but for the softer binary, it is clearly found.
The rate of enhancement increases as  $\mathfrak{a}_0$ gets larger.
For a smaller binary radius ($\mathfrak{r}_0=6M$), we find the larger enhancement.
For example, for $\mathfrak{a}_0=0.007M$ and $\zeta_{\textsf L}=0.9$, 
$e_{\rm max}=0.86$ for $ \mathfrak{r}_0=9M $, while $e_{\rm max}\approx 1.0$ for $ \mathfrak{r}_0=6M $,

In Fig. \ref{fig:56&960} (b), we show the vZLK oscillation period, which is normalized by that obtained in the DA approximation. 
Although we find a reduction of the vZLK timescale for nonzero libration angles, we do not find a big difference between $ \mathfrak{r}_0=6M $ and $9M$.

\subsection{Dependence of BH spin $a$}
As shown in the previous work\cite{Maeda:2023uyx}, 
in the case of a circular orbit, the results are largely insensitive to the BH spin parameter $a$. This is because the Riemann curvature components on the equatorial plane remain unchanged for any value of $a$. 
However, in the present case involving latitudinal libration, the Riemann curvature components vary with time along the binary orbit.
In a Schwarzschild BH, a spherical orbit remains circular even when $\zeta_{\textsf L} \neq 0$, and thus the vZLK oscillations are identical to those in an equatorial orbit.
By contrast, in a Kerr BH spacetime, the vZLK oscillations in spherical orbits exhibit dependence on the BH spin.

\begin{figure}[htbp]
\begin{center}
\includegraphics[width=7cm]{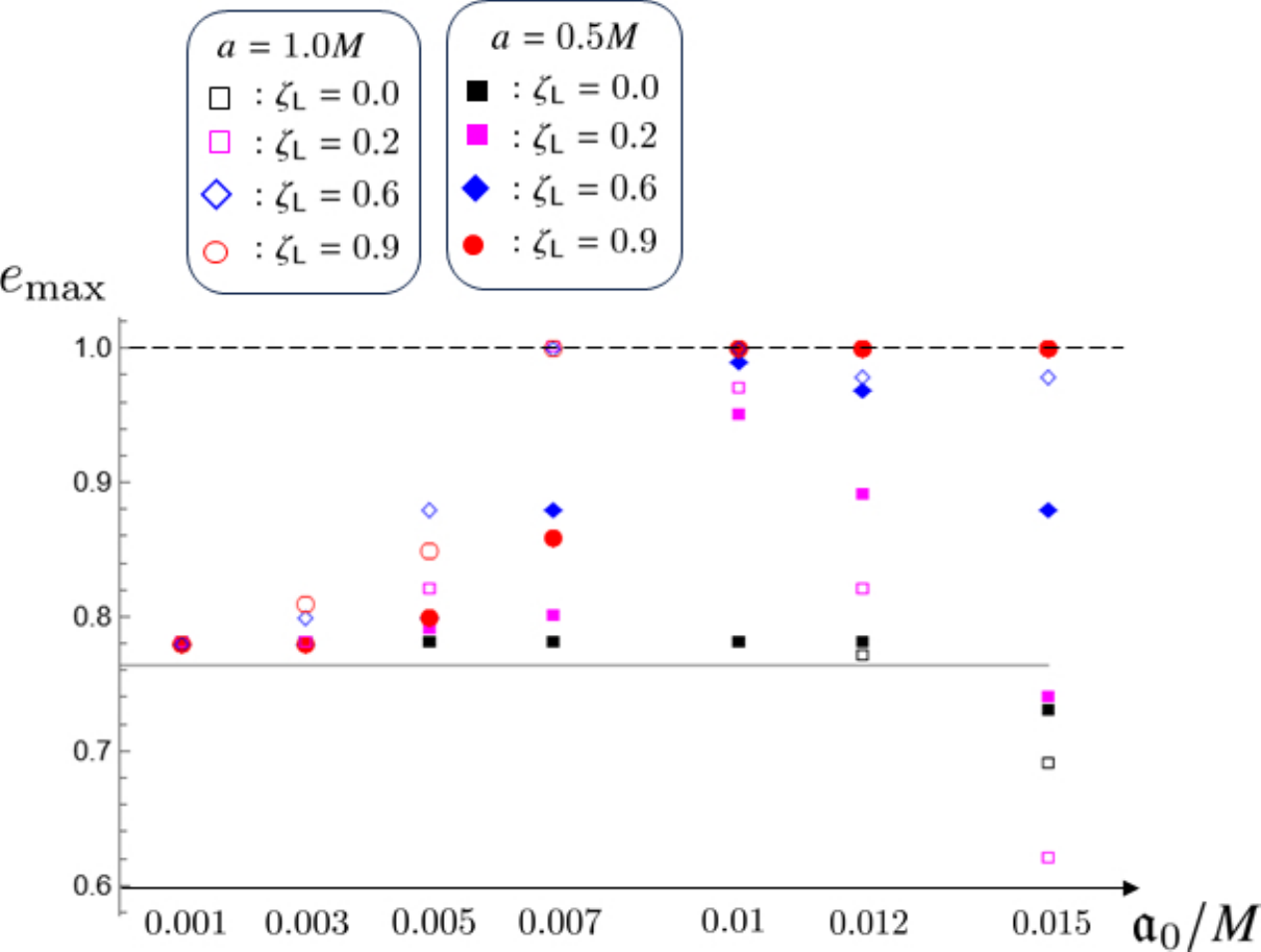}
\\
(a)
\\
\includegraphics[width=7cm]{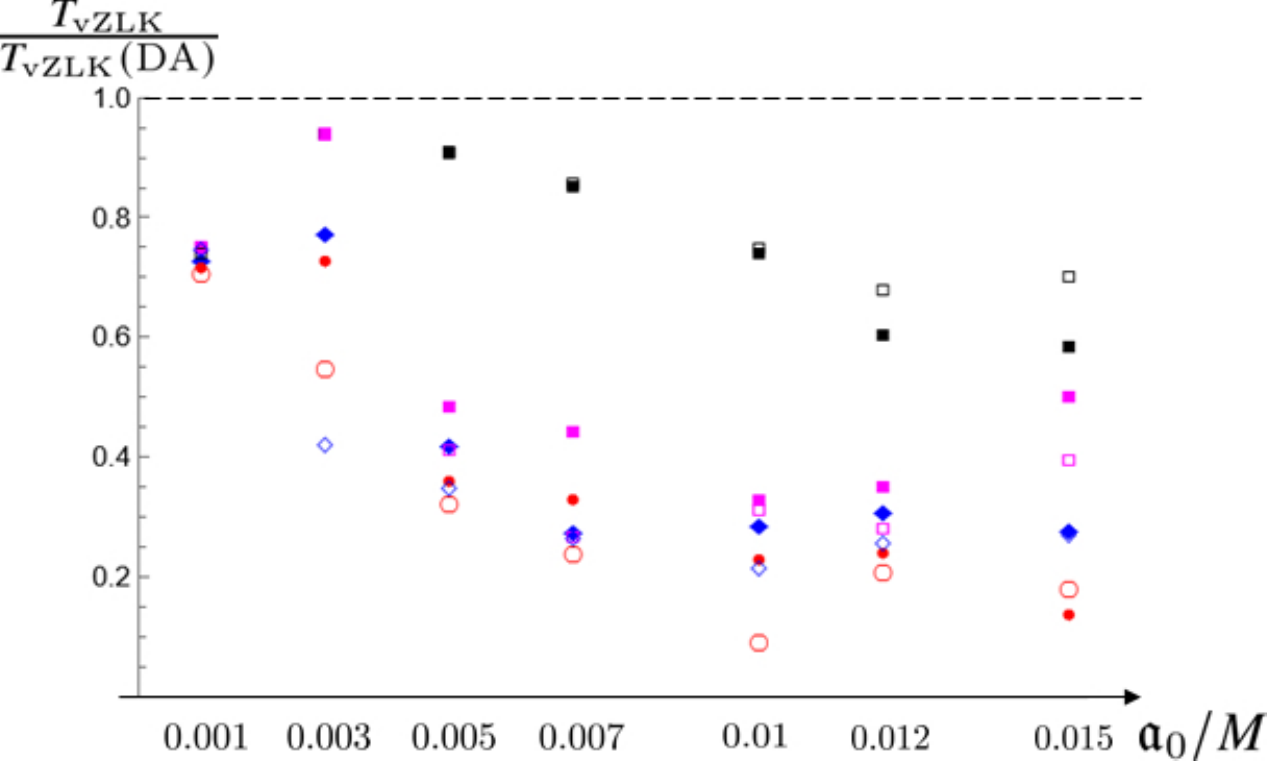}
\\
(b)
\caption{Panel (a) shows the maximum eccentricity, and panel (b) shows the vZLK oscillation period normalized by the value predicted by the DA approximation, for BH spin parameters $a=1.0M$ (empty markers) and $a=0.5M$ (filled markers).
}
\label{fig:10&5960}
\end{center}
\end{figure}

 In Fig. \ref{fig:10&5960} presents the results for $a=0.5M$ and $a=1.0M$, with fixed parameters $\mathfrak{r}_0 = 9M$ and $I_0 = 60^\circ$.
From Fig.~\ref{fig:10&5960}(a), we observe that the maximum eccentricity $e_{\rm max}$ closely matches the prediction of the DA approximation for hard binaries (e.g., $\mathfrak{a}_0 = 0.001M$), with no significant enhancement. However, for softer binaries, $e_{\rm max}$ is notably enhanced, and the degree of enhancement increases as the binary softens. The effect is more pronounced for higher BH spin. For instance, at $\mathfrak{a}_0 = 0.007M$ and $\zeta_{\textsf L} = 0.9$, we find $e_{\rm max} = 0.86$ for $a = 0.5M$, while $e_{\rm max} \approx 1.0$ for $a = 1M$.
 In Fig.~\ref{fig:10&5960}(b), we also show the vZLK oscillation period normalized by 
 that in the DA approximation. We observe a reduction in the vZLK timescale for nonzero libration angles, with the reduction being more significant for the rapidly rotating BH ($a = 1M$) than for the slower one ($a = 0.5M$). 

\subsection{Properties of chaotic vZLK oscillations}
For a soft binary, we find chaotic vZLK oscillations, i.e., the oscillation amplitude and period are irregular. Although these features are  unpredictable, we observe some interesting properties.

\begin{figure}[htbp]
\begin{center}
\includegraphics[width=6cm]{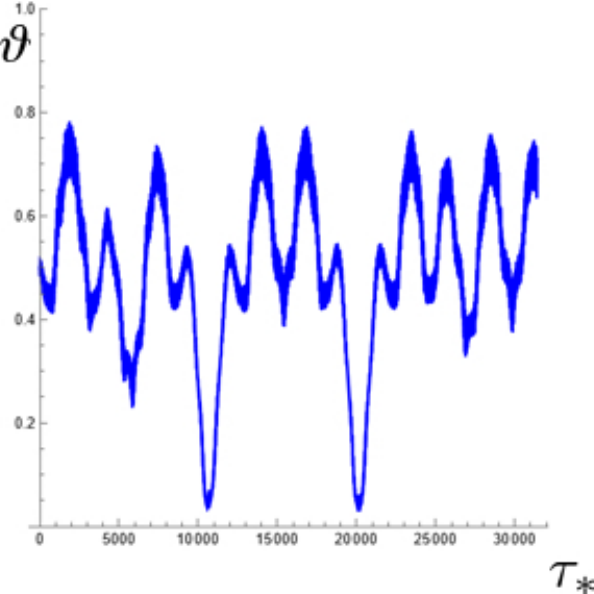}
\caption{The time evolution of the $z$-component of the angular momentum ($\vartheta$) for $a=1.0M, \mathfrak{r}_0=9M, \mathfrak{a}_0=0.007M$. We also set $I_0=60^\circ$ and $\zeta_{\mathsf L}=0.6$
}
\label{fig:vartheta}
\end{center}
\end{figure}

\begin{figure}[htbp]
\begin{center}
\includegraphics[width=7cm]{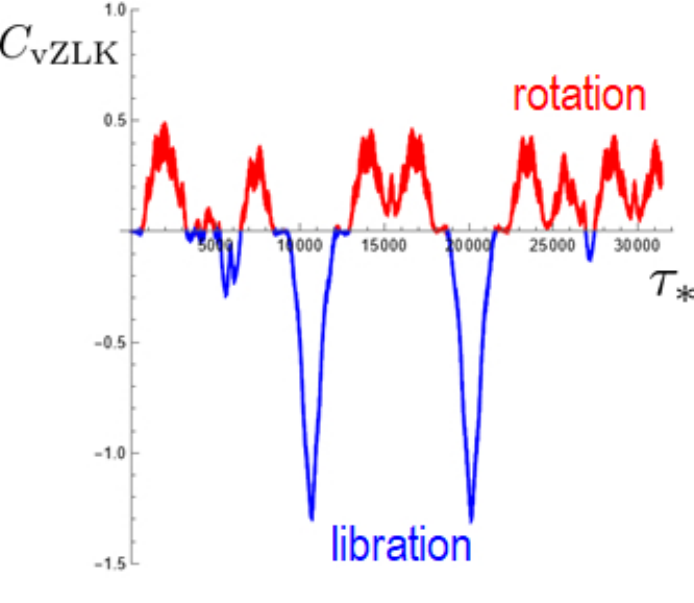}

\caption{The time evolution of $C_{\rm vZLK}$ is shown  for the same model in  Fig. \ref{fig:vartheta}
}
\label{fig:CvZLK}
\end{center}
\end{figure}

In the case of regular vZLK oscillations, the DA approximation is valid for the orbit in the equatorial plane. We find two conserved quantities $\vartheta$ ($z$-component of the angular momentum) and $C_{\rm vZLK}$ (or $v_S$ (tidal potential)). 
Even for the orbit exhibiting latitudinal libration, these two quantities remain approximately  conserved during  regular vZLK oscillations.
On the other hand, for chaotic vZLK oscillations, both quantities vary significantly over time, 
as shown in Figs. \ref{fig:vartheta} and \ref{fig:CvZLK}.
This is because the angular momentum is transferred 
through the interaction between the BH spin and the outer angular momentum. Note that the $z$-component of the total angular momentum is conserved in the Kerr spacetime.

\begin{figure}[htbp]
\begin{center}
\includegraphics[width=6.5cm]{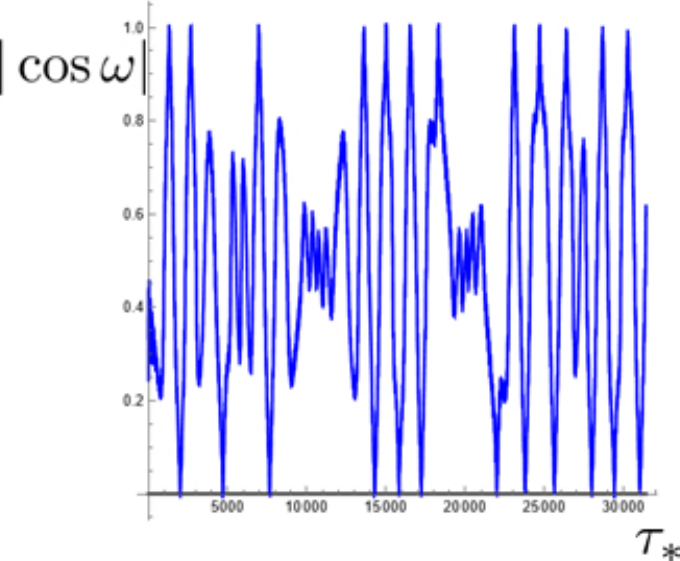}

\caption{The time evolution of $|\cos \omega|$ is shown  for the same model in Fig. \ref{fig:vartheta}.
}
\label{fig:omega}
\end{center}
\end{figure}

Interestingly, although $\vartheta$ and $C_{\rm vZLK}$ are defined within the DA approximation, they can still be used to interpret the chaotic vZLK oscillations.
The sign of $C_{\rm vZLK}$ distinguishes between librating and rotating vZLK oscillations in the DA regime, and this classification also appears to hold in the chaotic case as shown in Fig. \ref{fig:omega}.
Libration of $\omega$ is observed when its evolution is confined to a finite range, $(0<)\omega_{\rm min}\leq \omega\leq \omega_{\rm max} (<2\pi)$,  corresponding to the region where $C_{\rm vZLK}<0$ (the blue curve) in Fig. \ref{fig:CvZLK}.
In contrast, when $\omega$ evolves across 0 (or $2\pi$),
indicating rotation, it corresponds to  $C_{\rm vZLK}>0$ (the red curve) in Fig. \ref{fig:CvZLK}.

\begin{figure}[htbp]
\begin{center}
\includegraphics[width=6cm]{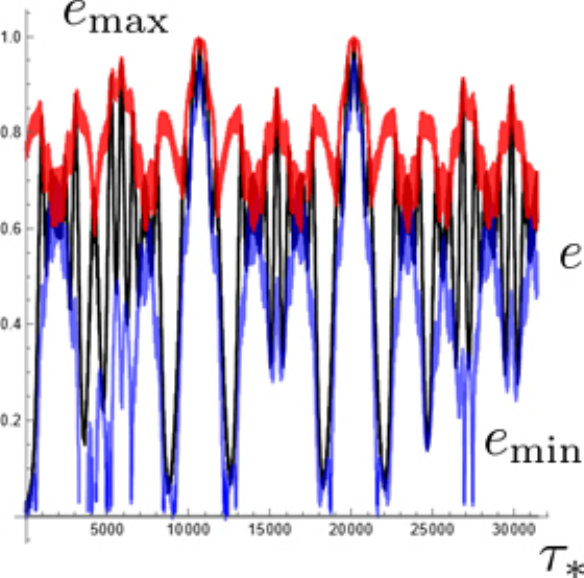}

\caption{
The maximum and minimum eccentricities, 
$e_{\rm max}$ (the red curve) and $e_{\rm min}$
(the blue curve), evaluated using Eq.\eqref{emaxemin_rot} or Eq.\eqref{emaxemin_lib}, are shown for the same model as in Fig.~\ref{fig:vartheta}. The time evolution of the eccentricity is also shown (black curve), lying between the red and blue curves.
}
\label{fig:emaxmin}
\end{center}
\end{figure}

The maximum and minimum values of the eccentricity 
can be computed using the analytic formula \eqref{emaxemin_rot}
or \eqref{emaxemin_lib} within the DA approximation. 
While  the DA approximation formally breaks down in the chaotic regime, 
these formulas remain applicable when we substitute time-dependent values of $\vartheta$ and $C_{\rm vZLK}$.
In Fig. \ref{fig:emaxmin}, we show 
$e_{\rm max}$ (the red curve) and $e_{\rm min}$
(the blue curve) evaluated by these  formula.
The time evolution of the eccentricity during chaotic vZLK oscillations is depicted by the black curve, which lies between the red and blue curves.

From these two examples, we conclude that 
$\vartheta, C_{\rm vZLK}$, and then the formula \eqref{emaxemin_rot} or \eqref{emaxemin_lib} are useful tools for understanding chaotic vZLK oscillations.

\section{Summary and Discussion}
We have investigated the dynamics of a binary system orbiting a rotating SMBH, focusing on the properties of vZLK oscillations. Using Fermi–Walker transport, we have constructed a local inertial frame in Kerr spacetime and modeled the binary with Newtonian dynamics. Assuming the binary's center of mass follows a spherical orbit, we have solved the equations of motion and analyzed the evolution of its orbital parameters.

Our analysis reveals that the latitudinal libration of the binary's orbit—caused by the spin of the SMBH—significantly alters the behavior of vZLK oscillations. As the libration amplitude increases, the oscillation period becomes shorter and the maximum eccentricity increases, especially in chaotic regimes. Remarkably, for sufficiently soft yet bound binaries, the vZLK oscillation timescale transitions from secular to dynamical. This shift arises from the coupling between the SMBH’s spin and the binary’s angular momentum, and becomes more prominent with increasing Kerr spin or decreasing orbital radius.

These effects—the enhancement of eccentricity and the shortening of the oscillation period—may play an important role in the formation and evolution of SMBHs and in gravitational wave astronomy. Highly eccentric binaries are expected to emit strong GW signals, while the reduced vZLK timescale implies shorter binary lifetimes. These factors can influence the evolution of relativistic $N$-body systems and contribute to SMBH formation scenarios.

In the present study, we have focused on spherical orbits of the binary's center of mass. However, extending the analysis to eccentric or unbound orbits will be important. Eccentric orbits may induce long-term modulations of vZLK oscillations~\cite{Lithwick:2011hh,Katz:2011hn,Naoz:2011mb,Li2014ApJ,Liu2015MNRAS}, while unbound orbits can introduce qualitatively different dynamics\cite{Zhang_2024}. Although these extensions involve more complex equations, they are conceptually straightforward and are currently under investigation.

Future work will also explore the gravitational wave signatures from such hierarchical triples. Near the SMBH horizon, black hole perturbation theory may be employed, as the binary's motion is already determined in the present work~\cite{PhysRevD.103.L081501}. However, due to the complexity of the motion, this approach can be challenging. Alternatively, the quadrupole formula may serve as a good approximation when the orbital radius exceeds approximately $10M$~\cite{Shibata:1994qd}, which is a work in progress.

\begin{acknowledgments}
We thank  Vitor Cardoso, Priti Gupta, Masaru Shibata and Haruka Suzuki for
useful discussions. This work was supported in part by the JSPS KAKENHI
Grant Number JP24K07058(K.M.) and JP23K03222(H.O.). 
K.M. would like to acknowledge the Yukawa Institute for Theoretical
Physics at Kyoto University, where the present work  was begun during
the YITP long-term workshop, 
Gravity and Cosmology 2024. 
He would also thank Niels Bohr Institute/Niels Bohr International Academy and The Max Planck Institute for Gravitational Physics (Albert Einstein Institute), where 
this work has been finished.

\end{acknowledgments}

\begin{appendices}
\renewcommand{\theequation}{\Alph{section}.\arabic{equation}}
\begin{widetext}
\section{Binary system in a curved spacetime}
\label{binary_in_curved_ST}
In this Appendix, we summarize how to discuss a binary system in a fixed curved background, which was discussed in details in \cite{Maeda:2023tao,Maeda:2023uyx}.
A binary consists of two point particles with the masses $m_1$ and $m_2$.
The Lagrangian for a binary in a local inertial frame up to 0.5 PN order is given by
\bea
{\cal L}_{\rm binary}={\cal L}_{\rm N}+{\cal L}_{1/2},
\label{Lagrangian_binary}
\ena
where
\bea
{\cal L}_{\rm N}&\equiv& {1\over 2} \left[ m_1 \left({d\vect{x}_1\over d\tau}\right)^2+m_2 \left({d\vect{x}_2\over d\tau}\right)^2\right]
+  {G m_1m_2\over |\vect{x}_1-\vect{x}_2|}
+{\cal L}_{a}+{\cal L}_{\omega}+{\cal L}_{\bar{\cal R}}
\label{Lagrangian_N},
\ena
with
\beann
{\cal L}_{a}
&=&
-\sum_{I=1}^2 m_I a_{\tilde k}x_I ^{\tilde k},
\\
{\cal L}_{\omega}
&=&
-\sum_{I=1}^2 m_I \left[\epsilon_{\tilde j\tilde k\tilde \ell}\omega^{\tilde \ell}
x_I^{\tilde k}{dx_I^{\tilde j}\over d\tau}-
{1\over 2} \left(\vect{\omega}^2 \vect{x}_I^2 -(\vect{\omega} \cdot \vect{x}_I)^2
\right)\right],
\\
{\cal L}_{\bar{\cal R}}
&=&
-{1\over 2}  \sum_{I=1}^2 m_I \bar{\cal R}_{\tilde 0\tilde k\tilde 0\tilde \ell}x_I^{\tilde k} x_I^{\tilde \ell}
\,,
\enann
and 
\bea
{\cal L}_{1/2}&\equiv &-{2\over 3}\sum_{I=1}^2
m_I c^2\bar{\cal R}_{\tilde 0\tilde k \tilde j  \tilde \ell }
x_I^{\tilde k}x_I^{\tilde \ell}\, {v_I^{\tilde j}\over c}
\,.
\label{Lagrangian_0.5PN}
\ena
Here, $
a^{\tilde j} 
$
and 
$
\omega^{\tilde j}
$
are  the acceleration of the observer 
and  its angular velocity in the local inertial frame, respectively.

Introducing the center of mass coordinates and the relative coordinates by
\beann
\vect{R}&=&{m_1\vect{x}_1+m_2\vect{x}_2\over m_1+m_2},
\\
\vect{r}&=& \vect{x}_2-\vect{x}_1
\,,
\enann
we find the Newtonian Lagrangian (Eq.~\eqref{Lagrangian_N}) in terms of  $\vect{R}$  and $\vect{r}$ as
\bea
{\cal L}_{\rm N}={\cal L}_{\rm CM}\left(\vect{R}, {d\vect{R}\over d\tau}\right)+
{\cal L}_{\rm rel}\left(\vect{r}, {d\vect{r}\over d\tau}\right)
\label{Lagrangian_N_Rr}
\,,
\ena
where
\beann
{\cal L}_{\rm CM}\left(\vect{R}, {d\vect{R}\over d\tau}\right)
&=&{1\over 2} (m_1+m_2) \left({d\vect{R}\over d\tau}\right)^2
+{\cal L}_{{\rm CM}\mathchar`-a}\left(\vect{R}, {d\vect{R}\over d\tau}\right)
+{\cal L}_{{\rm CM}\mathchar`-\omega}\left(\vect{R}, {d\vect{R}\over d\tau}\right)
+{\cal L}_{{\rm CM}\mathchar`-\bar{\cal R}}\left(\vect{R}, {d\vect{R}\over d\tau}\right),
\enann
with 
\beann
{\cal L}_{{\rm CM}\mathchar`-a}
&=&-(m_1+m_2)\vect{a}\cdot\vect{R}
\\
{\cal L}_{{\rm CM}\mathchar`-\omega}
&=&-(m_1+m_2)\left[
\epsilon_{\tilde j\tilde k\tilde \ell}\omega^{\tilde \ell}
R^{\tilde k}{dR^{\tilde j}\over d\tau}
-{1\over 2}\left(\vect{\omega}^2 \vect{R}^2
-\left(\vect{\omega}\cdot \vect{R}\right)^2\right)\right],
\\{\cal L}_{{\rm CM}\mathchar`-\bar{\cal R}},
&=&
-{1\over 2}(m_1+m_2)
\bar{\cal R}_{\tilde 0\tilde k \tilde 0 \tilde \ell}R^{\tilde k}R^{\tilde \ell},
\enann
and
 \beann
{\cal L}_{\rm rel}\left(\vect{r}, {d\vect{r}\over d\tau}\right)&=&
{1\over 2}\mu \left({d\vect{r}\over d\tau}\right)^2+  {G m_1m_2\over r}
+{\cal L}_{{\rm rel}\mathchar`-\omega}\left(\vect{r}, {d\vect{r}\over d\tau}\right)
+{\cal L}_{{\rm rel}\mathchar`- \bar{\cal R}}\left(\vect{r}, {d\vect{r}\over d\tau}\right),
\enann
with 
\beann
{\cal L}_{{\rm rel}\mathchar`-\omega}
&=&-\mu \left[
\epsilon_{\tilde j\tilde k\tilde \ell}\omega^{\tilde \ell}
r^{\tilde k}{dr^{\tilde j}\over d\tau}
-{1\over 2}\left(\vect{\omega}^2 \vect{r}^2
-\left(\vect{\omega}\cdot \vect{r}\right)^2\right)\right]\,,
\\
{\cal L}_{{\rm rel}\mathchar`-\bar{\cal R}}
&=&
-{1\over 2}\mu 
\bar{\cal R}_{\tilde 0\tilde k \tilde 0 \tilde \ell}r^{\tilde k}r^{\tilde \ell}
\,.
\enann
Here, $\mu = m_1 m_2/(m_1+m_2)$ is the reduced mass. When we consider only ${\cal L}_{\rm N}$, 
we can separate the variables $\vect{R}$ and $\vect{r}$.
 In particular, when the observer follows the geodesic ($\vect{a}=0$ and $\vect{\omega}=0$),
the orbit of $\vect{R}=0$ is a solution of the equation for $\vect{R}$.
It means that the center of mass (CM) follows the observer's geodesic.
We have only the equation for the relative coordinate $\vect{r}$. However, when we include the 0.5 PN term, it is not the case.
The 0.5PN Lagrangian ${\cal L}_{1/2}$  is rewritten by use of  $\vect{R}$  and    $\vect{r}$ as
\bea
{\cal L}_{1/2}={\cal L}_{1/2\mathchar`-{\rm CM}}
\left(\vect{R}, {d\vect{R}\over d\tau}\right)+
{\cal L}_{1/2\mathchar`-{\rm rel}}\left(\vect{r}, {d\vect{r}\over d\tau}\right)
+{\cal L}_{1/2\mathchar`- {\rm int}}\left(\vect{R}, {d\vect{R}\over d\tau}, \vect{r}, {d\vect{r}\over d\tau}\right),
\label{Lagrangian_1/2_Rr}
\ena
where
\bea
{\cal L}_{1/2\mathchar`-{\rm CM}}\left(\vect{R}, {d\vect{R}\over d\tau}\right)&=&-{2\over 3}(m_1+m_2)R_{\tilde 0\tilde k \tilde j \tilde \ell}
R^{\tilde k}R^{\tilde \ell}{dR^{\tilde j}\over d\tau},
\nn
{\cal L}_{1/2\mathchar`-{\rm rel}}\left(\vect{r}, {d\vect{r}\over d\tau}\right)&=&- {2\over 3} \mu{(m_1-m_2)\over (m_1+m_2)}R_{\tilde 0\tilde k \tilde j \tilde \ell}r^{\tilde k}r^{\tilde \ell}{dr^{\tilde j}\over d\tau},
\nn
{\cal L}_{1/2\mathchar`- {\rm int}}\left(\vect{R}, {d\vect{R}\over d\tau}, \vect{r}, {d\vect{r}\over d\tau}\right)&=&
- {2\over 3} \mu R_{\tilde 0\tilde k \tilde j \tilde \ell}
\left[r^{\tilde k}r^{\tilde \ell}{dR^{\tilde j}\over d\tau}
+\left(R^{\tilde k}r^{\tilde \ell}+r^{\tilde k}R^{\tilde \ell}
\right){dr^{\tilde j}\over d\tau}\right].
\label{interaction_term}
\ena
Due of the interaction term (Eq.~\eqref{interaction_term}), the orbit of $\vect{R}=0$ is no longer a solution even if the acceleration vanishes. The motion of  the CM ($\vect{R}(\tau$)) couples with the relative motion $(\vect{r}(\tau))$. As a result, not only the orbit of a binary but also the motion of the CM will become complicated even if the observer's orbit is a geodesic.

However, if we introduce an appropriate acceleration $\vect{a}$
in 0.5PN order 
 to cancel the interaction terms, $\vect{R}=0$ will become a solution, i.e., 
 the CM can follow the observer's motion as follows:
Integrating by parts the interaction term, we find
 \beann
{\cal L}_{1/2\mathchar`- {\rm int}}\left(\vect{R}, {d\vect{R}\over d\tau}, \vect{r}, {d\vect{r}\over d\tau}\right)
&\approx&
2\mu\left[{1\over 3}{d\bar{\cal R}_{\tilde 0\tilde k \tilde j \tilde \ell}\over d\tau}
r^{\tilde k}r^{\tilde \ell} 
+ \bar{\cal R}_{\tilde 0\tilde k \tilde j \tilde \ell}
r^{\tilde k}{dr^{\tilde \ell}\over d\tau} \right]R^{\tilde j}~~{\rm (integration~by~part)}
\,,
\enann
where the time derivative of the curvature is evaluated along the observer's orbit.

If we define the acceleration by
 \beann
 a_{\tilde j}={2\mu \over m_1+m_2}\left[{1\over 3}{d\bar{\cal R}_{\tilde 0\tilde k \tilde j \tilde \ell}\over d\tau}
r^{\tilde k}r^{\tilde \ell} 
+ \bar{\cal R}_{\tilde 0\tilde k \tilde j \tilde \ell}
r^{\tilde k}{dr^{\tilde \ell}\over d\tau}  \right]
\label{0.5PN_acceleration}
\,,
 \enann
two terms ${\cal L}_{1/2\mathchar`- {\rm int}}$ and ${\cal L}_{{\rm CM}\mathchar`-a}$ cancel each other.
As a result, the Lagrangians for $\vect{R}$ and $\vect{r}$ are decoupled,
and  $\vect{R}=0$ becomes an exact solution of the equation for $\vect{R}$, which is
derived from the Lagrangian
(${\cal L}_{\rm CM}+{\cal L}_{1/2\mathchar`-{\rm CM}}$).
The CM follows the observer's orbit and therefore, we obtain the decoupled equation for the relative coordinate $\vect{r}$.
 
In order to obtain the proper observer's orbit, 
which is not a geodesic but may be close to the geodesic, 
we have to solve the equation of motion including small acceleration such that
 \bea
 {Du_{\rm CM}^\mu\over d\tau}=a^\mu=e^{\mu\tilde j } a_{\tilde j}=
{2\mu \over m_1+m_2}e^{\mu\tilde j }\left[{1\over 3}{d\bar{\cal R}_{\tilde 0\tilde k \tilde j \tilde \ell}\over d\tau}
r^{\tilde k}r^{\tilde \ell} 
+ \bar{\cal R}_{\tilde 0\tilde k \tilde j \tilde \ell}
r^{\tilde k}{dr^{\tilde \ell}\over d\tau} \right]
\,.
\label{eq_CM}
 \ena
 
As a result, we first solve the equation for the relative coordinate $\vect{r}$, which is 
obtained only by the Lagrangian ${\cal L}_{\rm rel}(\vect{r})
 +{\cal L}_{1/2\mathchar`-{\rm rel}}(\vect{r})$.
 Note that when $m_1=m_2$, we have only Newtonian 
 Lagrangian ${\cal L}_{\rm rel}$ because ${\cal L}_{1/2\mathchar`-{\rm rel}}=0$ vanishes.
After obtaining the solution of $\vect{r}(\tau)$, 
we find the  motion for the CM
 (or the observer)  in the background spacetime
 by solving Eq.~\eqref{eq_CM}. Using  the relative motion $\vect{r}(\tau)$ with the solution of the CM motion 
 ($x_{\rm CM}^\mu(\tau)$), 
 we will obtain a binary motion in a given curved background spacetime ($x_1^\mu(\tau)\,,x_2^\mu(\tau)$).

\end{widetext}
\section{Riemann curvature in Carter's tetrad system}
There is one convenient tetrad system in Kerr spacetime,
which is the so-called 
Carter's tetrad given by
\beann
e_{~\bar{t}}^\mu&=&{r^2+a^2\over \sqrt{\Sigma\Delta}}\left(1, 0, 0, {a\over r^2+a^2}\right)
\\
e_{~\bar{\mathfrak{r}}}^\mu&=&\sqrt{\Delta\over \Sigma}\left(0,1,0,0\right)
\\
e_{~\bar \zeta}^\mu&=& \sqrt{1-\zeta^2\over \Sigma}\left(0, 0, 1, 0\right)
\\
e_{~\bar \phi}^\mu&=&{1\over \sqrt{\Sigma(1-\zeta^2)}}\left(a(1-\zeta^2), 0, 0, 1\right)
\enann
and its dual
\beann
\omega^{~\bar{t}}_\mu&=&\sqrt{\Delta\over \Sigma }\left(1, 0, 0, -a(1-\zeta^2)\right)
\\
\omega^{~\bar{\mathfrak{r}}}_\mu&=&\sqrt{ \Sigma\over\Delta}\left(0,1,0,0\right)
\\
\omega^{~\bar \zeta}_\mu&=&\sqrt{\Sigma\over 1-\zeta^2}\left(0, 0,  1, 0\right)
\\
\omega^{~\bar \phi}_\mu&=& (r^2+a^2)\sqrt{1-\zeta^2\over \Sigma }\left(-{a\over (r^2+a^2)}
, 0, 0, 1\right)
\,.
\enann

We  then obtain the non-trivial tetrad components of the Riemann curvature  as follows:
 \beann
\bar{\cal R}_{\bar{t}\bar \phi\bar{t}\bar \phi}&=&-\bar{\cal R}_{\bar{\mathfrak{r}}\bar \zeta\bar{\mathfrak{r}}\bar \zeta}\,=\, {\cal Q}_1
\\
\bar{\cal R}_{\bar{t}\bar \phi\bar{\mathfrak{r}}\bar \zeta}&=& -{\cal Q}_2
\\
\bar{\cal R}_{\bar{t}\bar{\mathfrak{r}}\bar{t}\bar{\mathfrak{r}}}&=&-R_{\bar \zeta\bar \phi\bar \zeta\bar \phi}\,=\, -2{\cal Q}_1
\\
\bar{\cal R}_{\bar{t}\bar{\mathfrak{r}}\bar \phi\bar \zeta}&=&-2 {\cal Q}_2
\\
\bar{\cal R}_{\bar{t}\bar \zeta\bar{t}\bar \zeta}&=&-\bar{\cal R}_{\bar{\mathfrak{r}}\bar \phi\bar{\mathfrak{r}}\bar \phi}\,=\, {\cal Q}_1
\\
\bar{\cal R}_{\bar{t}\bar \zeta\bar \phi\bar{\mathfrak{r}}}&=&-{\cal Q}_2
\,,
\enann
where
\beann
{\cal Q}_1&=&
{M\mathfrak{r}(\mathfrak{r}^2-3a^2\zeta^2)\over \Sigma^3}
\\~
{\cal Q}_2&=&
{Ma\zeta(3\mathfrak{r}^2-a^2\zeta^2)\over \Sigma^3}
\,.
\enann

On the equatorial plane ($\zeta=0$), 
we find much simpler expression as
\bea
&&
\bar{\cal R}_{\bar{t}\bar \zeta\bar{t}\bar \zeta}
\,=\,\bar{\cal R}_{\bar{t}\bar \phi\bar{t}\bar \phi}
\,=\,-\bar{\cal R}_{\bar{\mathfrak{r}}\bar \zeta\bar{\mathfrak{r}}\bar \zeta}
\,=\,-\bar{\cal R}_{\bar{\mathfrak{r}}\bar \phi\bar{\mathfrak{r}}\bar \phi}
\,=\, {M\over \mathfrak{r}^3}
\nonumber
\\
&&
\bar{\cal R}_{\bar \zeta\bar \phi\bar \zeta\bar \phi}
\,=\,
-\bar{\cal R}_{\bar{t}\bar{\mathfrak{r}}\bar{t}\bar{\mathfrak{r}}}
\,=\,{2M\over \mathfrak{r}^3}
\,,
\label{Riemann_Carter}
\ena
which are independent of the BH spin parameter $a$.

\begin{widetext}

\section{Lagrange planetary equations for a binary system near SMBH}
\label{planetary_equations}

To comprehend our numerical findings better, we should consider the Lagrange planetary equations. These equations provide the evolution of orbital parameters, such as the semi-major axis, eccentricity, and inclination. To derive these planetary equations, we work with the proper Hamiltonian, where the mass parameter is set to $\mu=1$. The proper Hamiltonian is defined as follows:
\beann
\bar{{\cal H}}=\bar{{\cal H}}_0+\bar{{\cal H}}_1,
\enann
where
\beann
\bar{{\cal H}}_0
&=&{1\over 2} \bar{\vect{\mathsf{p}}}^2- {G (m_1+m_2)\over \mathsf{r}},
\\
\bar{{\cal H}}_1
&=&-
\bar{{\cal L}}_{{\rm rel}\mathchar`-\bar{\cal R}}
=
{1\over 2}
\bar{\cal R}_{\tilde 0\tilde k \tilde 0 \tilde \ell}\mathsf{r}^{\tilde k}\mathsf{r}^{\tilde \ell}
 \enann

The position $\bf{\mathsf{r}}=(\mathsf{x},\mathsf{y},\mathsf{z})$ 
of a binary 
should be described in the non-rotating proper reference frame.
\\

The unperturbed Hamiltonian, denoted as $\bar{{\cal H}}_0$, is equivalent to that of a binary system in Newtonian dynamics. It leads to an elliptical orbit, described by the equation
\bea
\mathsf{r}= {\mathfrak{a}(1-e^2)\over 1+e\cos f}
\label{elliptic_r}
\,,
\ena
Here, $\mathsf{r}$ represents the radial distance from the center of mass, while $a$, $e$, and $f$ are the semi-major axis, eccentricity, and true anomaly, respectively. This orbital plane is inclined at an angle $I$ relative to the equatorial plane in the proper reference frame. Consequently, the relative position vector $\bf{\mathsf{r}}=(\mathsf{x},\mathsf{y},\mathsf{z})$ of the binary system can be determined by the orbital parameters $(\omega\,,\Omega\,,  \mathfrak{a}\,, e\,, I\,, f)$ as 
\bea
\begin{pmatrix}
\mathsf{x} \\
\mathsf{y} \\
\mathsf{z} \\
\end{pmatrix}
&=&
\mathsf{r}\begin{pmatrix}
\cos \Omega\cos(\omega+f)-\sin\Omega\sin(\omega+f)\cos I \\
\sin \Omega\cos(\omega+f)+\cos\Omega\sin(\omega+f)\cos I\\
\sin(\omega+f)\sin I \\
\end{pmatrix}
\label{orbital_parameters2}
\ena
 with Eq. (\ref{elliptic_r}).
 
 \end{widetext}

The introduction of Delaunay variables further refines this description as follows:

\beann
\left\{
\begin{array}{l}
\mathfrak{l}=n(t-t_0)
\\
\mathfrak{g}=\omega
\\
\mathfrak{h}=\Omega
\\
\end{array}
\right.
\hskip 3.3cm
\\
{\rm and}\hskip 6cm
\\
\left\{
\begin{array}{l}
\mathfrak{L}=\sqrt{G(m_1+m_2)\mathfrak{a}}
\\
\mathfrak{G}=\sqrt{G(m_1+m_2)\mathfrak{a}(1-e^2)}
\\
\mathfrak{H}=\sqrt{G(m_1+m_2)\mathfrak{a}(1-e^2)}\cos I
\,,
\\
\end{array}
\right.
\enann
where
\beann
n\equiv {2\pi \over  P}=\sqrt{{G(m_1+m_2)\over \mathfrak{a}^3}},
\enann
is the mean motion,
we find new unperturbed Hamiltonian as
\beann
\widetilde{\bar{\cal H}}_0=-{G^2(m_1+m_2)^2\over 2\mathfrak{L}^2}.
\enann
\\[-1em]

Including the perturbations $\bar{\cal H}_1$,
we obtain the Hamiltonian the Delaunay variables as
\beann
\widetilde{\bar{\cal H}}=\widetilde{\bar{\cal H}}_0+\bar{\cal H}_1
\,.
\enann

The proper Hamiltonian is described by the orbital parameters by inserting the relation given in Eq.(\ref{orbital_parameters2}) with Eq. (\ref{elliptic_r}).
We then obtain the planetary equations for the present hierarchical triple system, which is mathematically equivalent to our basic equations in the text.

\subsubsection{Double-averaging (DA) approach}
\label{DA_approach}

Rather than directly solving the above Lagrange planetary equations, 
one possible approach involves averaging the perturbed Hamiltonian over two periods: the inner and outer orbital periods, when we are interested in understanding the long-term behavior of the system, particularly phenomena like the vZLK mechanism. 
This allows us to simplify the equations for analysis.

The doubly-averaged Hamiltonian is defined by
\beann
\langle\langle\bar{\cal H}_1 \rangle\rangle\equiv {1\over 2\pi}\int_0^{2\pi}
d\mathfrak{l}_{\rm out}\left( {1\over 2\pi}\int_0^{2\pi}
d\mathfrak{l}\, \bar{\cal H}_1\right)
\enann
This is described as
\bea
\langle\langle\bar{\cal H}_1 \rangle\rangle
={1\over 2}\langle \bar{\cal R}_{\tilde 0\tilde k \tilde 0 \tilde \ell}  \rangle
 \langle\mathsf{r}^{\tilde k}\mathsf{r}^{\tilde \ell}  \rangle
 \label{DA_perturbed_H1}
\ena
where
\beann
\langle \bar{\cal R}_{\tilde 0\tilde k \tilde 0 \tilde \ell}  \rangle
 &=&{1\over 2\pi}\int_0^{2\pi}
d\mathfrak{l}_{\rm out}\bar{\cal R}_{\tilde 0\tilde k \tilde 0 \tilde \ell} 
 \\
 \langle\mathsf{r}^{\tilde k}\mathsf{r}^{\tilde \ell}  \rangle
 &=&{1\over 2\pi}\int_0^{2\pi}
d\mathfrak{l}\mathsf{r}^{\tilde k}\mathsf{r}^{\tilde \ell} 
\enann

If  the inner and outer orbits are approximated by the elliptic orbit, we find 
\beann
d\mathfrak{l}_{\rm out}&=&{1\over \sqrt{1-e_{\rm out}^2}}\left({\mathsf{r}_{\rm out}\over \mathfrak{a}_{\rm out}}\right)^2 df_{\rm out}
\\
d\mathfrak{l}&=&{1\over \sqrt{1-e^2}}\left({\mathsf{r}\over \mathfrak{a}}\right)^2 df
\,.
\enann
\begin{widetext}
For the inner orbit, we obtain
\beann
\langle\mathsf{x}^2 \rangle &=&{\mathfrak{a}^2\over 4}\left[(2+3e^2)
\left(\cos^2\Omega+\sin^2\Omega\cos^2I\right)+5e^2\left((\cos^2\Omega-\sin^2\Omega\cos^2 I)\cos 2\omega-2\cos\Omega\sin\Omega\cos I\sin2\omega\right)\right]
\\
\langle\mathsf{y}^2 \rangle &=&{\mathfrak{a}^2\over 4}\left[(2+3e^2)
\left(\sin^2\Omega+\cos^2\Omega\cos^2I\right)+5e^2\left((\sin^2\Omega-\cos^2\Omega\cos^2 I)\cos 2\omega+2\cos\Omega\sin\Omega\cos I\sin2\omega\right)\right]
\\
\langle\mathsf{z}^2 \rangle &=&{\mathfrak{a}^2\over 4}\sin^2 I
 \left[2+3e^2-5e^2\cos 2\omega\right]
\\
\langle\mathsf{x}\mathsf{y} \rangle &=&
{\mathfrak{a}^2\over 4}\left[(2+3e^2)\cos\Omega\sin\Omega \sin^2 I+5e^2\left(
\cos\Omega\sin\Omega(1+\cos^2 I)\cos 2\omega+
(\cos^2\Omega-\sin^2\Omega)\cos I \sin 2\omega\right)\right]
\\
\langle\mathsf{y}\mathsf{z}  \rangle &=&
{\mathfrak{a}^2\over 4}\left[(2+3e^2)\cos\Omega\sin I\cos I  +5e^2 \sin I \left(
\sin\Omega\sin 2\omega
-\cos \Omega\cos I \cos 2\omega\right)\right]
\\
\langle\mathsf{z}\mathsf{x}  \rangle &=&
{\mathfrak{a}^2\over 4}\left[-(2+3e^2)\sin\Omega\sin I\cos I  +5e^2 \sin I \left(
\cos\Omega\sin 2\omega
+\sin \Omega\cos I \cos 2\omega\right)\right]
\enann
Since the outer orbit is gven, we find
$\langle \bar{\cal R}_{\tilde 0\tilde k \tilde 0 \tilde \ell}  \rangle$ are constants desribed by the  parameters of the  outer orbit.

Using the double-averaged Hamiltonian Eq.(\ref{DA_perturbed_H1}), we obtain the double-averaged Lagrange planetary equations as
\bea
\dot e&=&
{\sqrt{1-e^2}\over n\mathfrak{a}^2e}{\partial \langle\langle\bar{\cal H}_1 \rangle\rangle\over \partial \omega}
\label{PE_Sch_e}
\\
\dot I&=&
{1\over n\mathfrak{a}^2\sin I \sqrt{1-e^2}}{\partial \langle\langle\bar{\cal H}_1 \rangle\rangle\over \partial \Omega}
-{\cos I\over n\mathfrak{a}^2\sin I \sqrt{1-e^2}}{\partial \langle\langle\bar{\cal H}_1 \rangle\rangle\over \partial \omega}
\label{PE_Sch_I}
\\
\dot \omega&=&
-{\sqrt{1-e^2}\over n\mathfrak{a}^2e}{\partial \langle\langle\bar{\cal H}_1 \rangle\rangle\over \partial e}
+{\cos I\over n\mathfrak{a}^2\sin I \sqrt{1-e^2}}{\partial \langle\langle\bar{\cal H}_1 \rangle\rangle\over \partial I}
\label{PE_Sch_omega}
\\
\dot \Omega&=&
-{1\over n\mathfrak{a}^2\sin I \sqrt{1-e^2}}{\partial \langle\langle\bar{\cal H}_1 \rangle\rangle\over \partial I}
\label{PE_Sch_Omega}
\ena
where
\beann
\langle\langle\bar{\cal H}_1 \rangle\rangle
={1\over 2}\left[\langle \bar{\cal R}_{\tilde 0\tilde{\mathsf{x}} \tilde 0 \tilde{\mathsf{x}}}  \rangle \langle\mathsf{x}^2 \rangle
+\langle \bar{\cal R}_{\tilde 0\tilde{\mathsf{y}} \tilde 0 \tilde{\mathsf{y}}}  \rangle \langle\mathsf{y}^2 \rangle
+\langle \bar{\cal R}_{\tilde 0\tilde{\mathsf{z}} \tilde 0 \tilde{\mathsf{z}}}  \rangle \langle\mathsf{z}^2 \rangle
+\langle \bar{\cal R}_{\tilde 0\tilde{\mathsf{x}} \tilde 0 \tilde{\mathsf{y}}}  \rangle \langle\mathsf{x}\mathsf{y} \rangle
+\langle \bar{\cal R}_{\tilde 0\tilde{\mathsf{y}} \tilde 0 \tilde{\mathsf{z}}}  \rangle \langle\mathsf{y}\mathsf{z} \rangle
+\langle \bar{\cal R}_{\tilde 0\tilde{\mathsf{z}} \tilde 0 \tilde{\mathsf{x}}}  \rangle \langle\mathsf{z}\mathsf{x} \rangle
\right]
\enann

The semi-major axis $\mathfrak{a}$ is constant in the present approximation.
From Eqs. (\ref{PE_Sch_e}) and (\ref{PE_Sch_I}), 
we find
\beann
{d\over d\tau}\left(\sqrt{1-e^2}\cos I\right)=-{1\over n\mathfrak{a}^2}{\partial \langle\langle\bar{\cal H}_1 \rangle\rangle\over \partial \Omega}
\,.
\enann

Hence, the $z$ component of the  angular momentum, which is proportional to $\sqrt{1-e^2}\cos I$,  is not conserved in general.
However, it is conserved when a binary orbit is in the equatorial plane
because $\langle\langle\bar{\cal H}_1 \rangle\rangle$ becomes independent of $\Omega$.

\end{widetext}

\subsubsection{Analytic solutions of a binary motion in the equatorial plane under DA approximation } 
\label{DA_solution}
Here as for a reference solution, 
we analyze a binary motion in the equatorial plane using DA approximation.

Introducing a ``tidal potential" by
$V_S\equiv - \langle\langle\bar{\cal H}_1\rangle\rangle
$,
we rewrite the above DA planetary equations as
\bea
\dot{e}&=&-{\sqrt{1-e^2}\over n\mathfrak{a}^2 e}{\partial {V_S}\over \partial \omega},
\label{LPeq_e1}
\\
\dot I&=&{\cos I\over n\mathfrak{a}^2\sin I\sqrt{1-e^2}}
{\partial {V_S}\over \partial \omega},
\label{LPeq_I1}
\\
\dot \omega&=&{\sqrt{1-e^2}\over n\mathfrak{a}^2 e}{\partial {V_S}\over \partial e}-{\cos I\over n\mathfrak{a}^2\sin I\sqrt{1-e^2}}{\partial {V_S}\over \partial I},~~~~~~~
\label{LPeq_om1}
\\
\dot \Omega&=&{1\over n\mathfrak{a}^2\sin I\sqrt{1-e^2}}{\partial {V_S}\over \partial I}
\label{LPeq_Om1}
\,.
\ena
We derive closed-form differential equations for the variables $e$, $I$, and $\omega$ using Eqs. (\ref{LPeq_e1}), (\ref{LPeq_I1}), and (\ref{LPeq_om1}). These equations provide insights into various properties of vZLK oscillations, including the oscillation amplitude of eccentricity and the oscillation time scale. This analysis is consistent with previous studies on Newtonian and 1PN hierarchical triple systems, as discussed in~\cite{Suzuki:2020zbg}. 

The tidal potential $V_S$ is written by use of 
$\eta \equiv  \sqrt{1-e^2}$ and $\mu_I \equiv  \cos I$
as
\beann
V_S&\equiv &
-\langle\langle\bar{\cal H}_1\rangle\rangle
\nn
&=&
{\alpha M  \mathfrak{a}^2  \over 16\mathfrak{r}_0^3} v_S(\eta, \mu_I),
\enann
where
\beann
\alpha&\equiv& {\mathfrak{r}_0^2
+3a^2-4a\sigma\sqrt{M\mathfrak{r}_0}
\over \mathfrak{r}_0^2
-3M\mathfrak{r}_0+2 a\sigma\sqrt{M\mathfrak{r}_0}},
\\
v_S(\eta, \mu_I)&\equiv & 2(-1+3\mu_I^2\eta^2)+12 C_{\rm vZLK}
\enann
with
\beann
C_{\rm vZLK}&\equiv& (1-\eta^2)
\Big{[}1-{5\over 2}(1-\mu_I^2)\sin^2\omega \Big{]}.
\enann
The sign $\sigma=\pm 1$ corresponds to prograde and retrograde orbits, respectively.

Note that $\alpha$ gives relativistic corrections, and we recover the same equations for Newtonian hierarchical triple system with quadrupole approximation when $\alpha=1$.

\begin{figure}[htbp]
\begin{center}
\includegraphics[width=6.0cm]{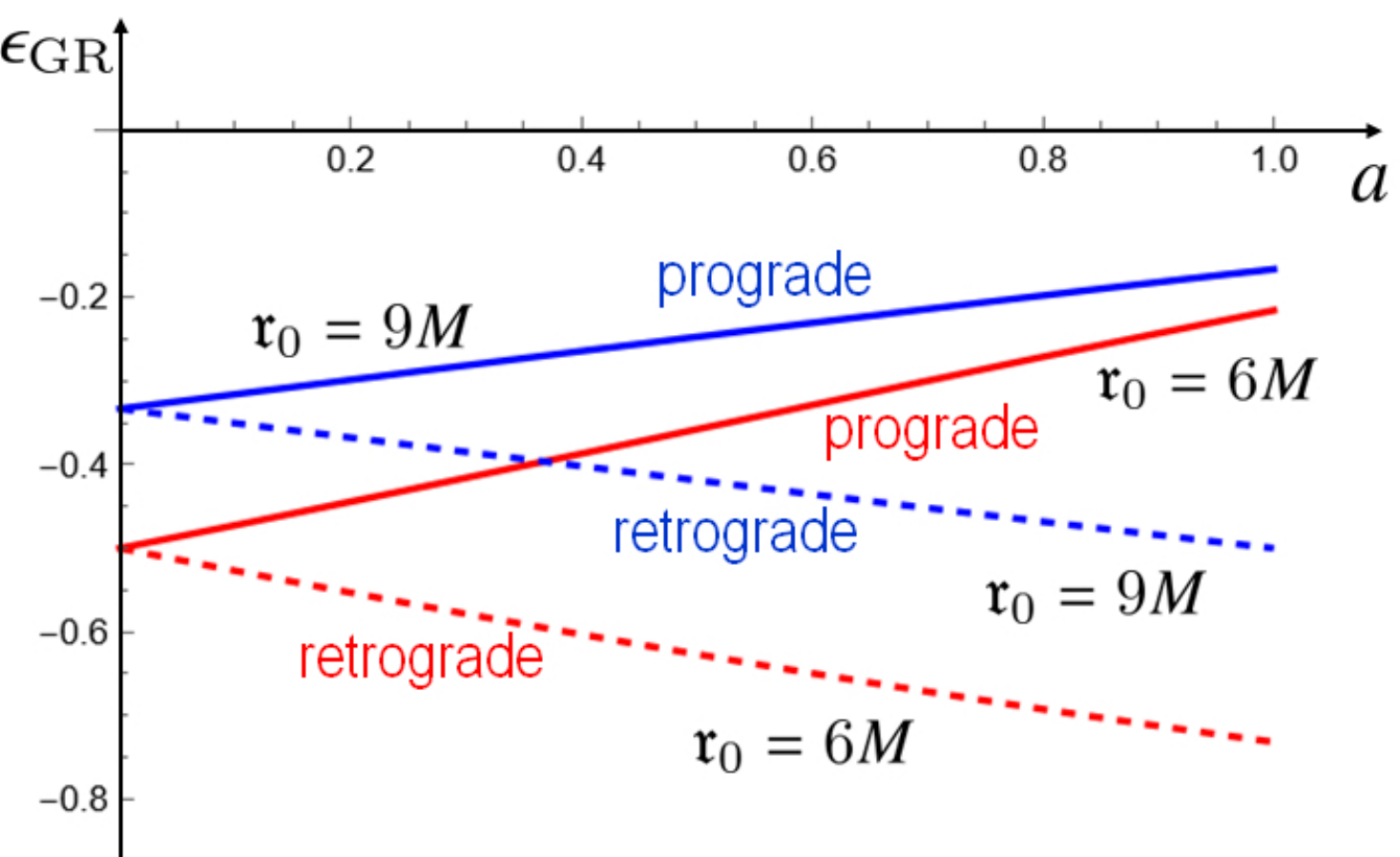}
\caption{The relativistic correction $\epsilon_{\rm GR}$ to the vZLK oscillation period  as a function of the Kerr parameter $a$. The blue and red lines represent the cases of $\mathfrak{r}_0=9M$ and $6M$, respectively. Solid and dashed lines correspond to prograde and retrograde orbits, respectively.}
\label{fig:relativistic_effect}
\end{center}
\end{figure}

We introduce the typical vZLK time scale by 
\beann
\tau_{\rm vZLK}\equiv {16 n \mathfrak{r}_0^3 \over \alpha M}=\tau_{\rm vZLK}^{(N)}\left(1+\epsilon_{\rm GR}\right)
\,,
\enann
where $\tau_{\rm vZLK}^{(N)}\equiv  {16 n \mathfrak{r}_0^3 /M}$ is the Newtonian vZLK time scale, and
\beann
\epsilon_{\rm GR}\equiv -3{(\sqrt{M\mathfrak{r}_0}-\sigma a)^2\over \mathfrak{r}_0^2+3a^2-4 \sigma a\sqrt{M\mathfrak{r}_0}}
\enann
denotes the relativistic correction. Since it is negative definite, the relativistic effect reduces the vZLK period
by several tens of percent as shown in Fig. \ref{fig:relativistic_effect}.

Using the normalized time 
\beann
\tilde \tau\equiv {\tau\over \tau_{\rm vZLK}}\,, 
\enann
the above planetary equation is rewritten as
\beann
{d\eta\over d\tilde \tau}&=&{\partial {v_S}\over \partial \omega},
\\
{1\over \mu_I}{d\mu_I\over d\tilde \tau}&=&
-{1\over \eta}{\partial {v_S}\over \partial \omega},
\\
{d\omega\over d\tilde \tau}&=&
-{\partial {v_S}\over \partial \eta}+{\mu_I\over \eta}
{\partial {v_S}\over \partial \mu_I}.
\enann
From these equations, we can easily show that 
\beann
{d(\mu_I\eta) \over d\tilde \tau}=0
~\,,~~~
{dv_S\over d\tilde \tau}=0
\,,
\enann
which means there exist two conserved quantities $\vartheta\equiv \mu_I\eta$ and $C_{\rm vZLK}$ just as the Newtonian and 1PN hierarchical triple system under dipole approximation.
Using these two conserved quantities, we obtain a single equation for $\xi\equiv \eta^2$ as
\beann
{d\xi\over d\tilde \tau}=-24\sqrt{2}\sqrt{f(\xi)g(\xi)},
\enann
with
\beann
f(\xi)&\equiv&\left(1-C_{\rm vZLK}\right)-\xi\,,
\\
g(\xi)&\equiv&-5\vartheta^2
+\left(5\vartheta^2+3+2C_{\rm vZLK}\right)\xi-3\xi^2
\enann

\begin{widetext}

We then find
\bea
{d\xi\over d\tilde \tau}=-24\sqrt{6}\sqrt{(\xi-\xi_0)(\xi-\xi_+)(\xi-\xi_-)},
\label{eq_for_xi}
\ena
where
\beann
\xi_0&=&1-C_{\rm vZLK},
\\
\xi_\pm&=&{1\over 2}\Big[\left(1+{5\over 3}
\vartheta^2+{2\over 3}C_{\rm vZLK}\right)
\pm \sqrt{\left(1+{5\over 3}
\vartheta^2+{2\over 3}C_{\rm vZLK}\right)^2-{20\over 3}\vartheta^2
}\Big],
\enann
are the solutions of $f(\xi)=0$ and $g(\xi)=0$, respectively.
\\

\end{widetext}

Analyzing the above equation, we find that there exists vZLK oscillations in this system just the same as in Newtonian hierarchical triple system, and we can classify the vZLK oscillations by the sign of $C_{\rm vZLK}$ into two cases:\\[.5em] 
\indent $1$. $C_{\rm vZLK}>0$ (rotation) \\[.5em] 
\indent $2$. $C_{\rm vZLK}<0$ (libration).

\subsubsection{$C_{\rm vZLK}>0$ {\rm (rotation)}}
In this case, $0<\xi_-<\xi_0<1<\xi_+$.
This is possible if 
\beann
0<C_{\rm vZLK}<1
\,.
\enann

We can integrate Eq. (\ref{eq_for_xi}) by use of the elliptic function as
\beann
{\xi_0-\xi\over \xi_+-\xi}={\xi_0-\xi_-\over \xi_+-\xi_-}{\rm sn}^2\left(\beta_{\rm (rot)}
(\tilde \tau-\tilde \tau_0), k_{\rm (rot)}\right)
\,,
\enann
where ${\rm sn}(x,k)$ is the Jacobi elliptic function sn with the elliptic modulus $k$, and
\beann
\beta_{\rm (rot)}&\equiv& 12\sqrt{6(\xi_+-\xi_-)}
\\
k_{\rm (rot)}&\equiv &\sqrt{\xi_0-\xi_-\over \xi_+-\xi_-}
\enann
\begin{widetext}
This solution gives the evolution of the eccentricity $e$ as
\beann
e^2={(1-\xi_0)(\xi_+-\xi_-)-(1-\xi_+)(\xi_0-\xi_-){\rm sn}^2\left(\beta_{\rm (rot)}
(\tilde \tau-\tilde \tau_0), k_{\rm (rot)}\right)
\over
(\xi_+-\xi_-)-(\xi_0-\xi_-){\rm sn}^2\left(\beta_{\rm (rot)}
(\tilde \tau-\tilde \tau_0), k_{\rm (rot)}\right)}
\enann
\end{widetext}

Since $0\leq {\rm sn}^2(x,k)\leq 1$,
 we find the maximum and minimum values of the eccentricity as
\bea
e_{\rm max}=\sqrt{1-\xi_-}
\,,
e_{\rm min}=\sqrt{1-\xi_0}.
\label{emaxemin_rot}
\ena

The vZLK oscillation  time scale is given by 
\bea
T_{\rm vZLK}=\tau_{\rm vZLK}\, \mathfrak{T}_{\rm vZLK}^{\rm (rot)},
\label{TKL_rotation}
\ena
where
\beann
\mathfrak{T}_{\rm vZLK}^{\rm (rot)}&\equiv& 
{2K\left(k_{\rm (rot)}\right)\over \beta_{\rm (rot)}}
\,.
\enann
$K(k)$ is the complete elliptic integral of the first kind with the elliptic modulus $k$.

\subsubsection{$C_{\rm vZLK}<0$ {\rm (libration)}}
The libration oscillations occur when 
\beann
&&
 -{3\over 2}<C_{\rm vZLK}<0\,,~
\enann
 and
\beann
\vartheta<{\sqrt{3}-\sqrt{-2C_{\rm vZLK}}\over \sqrt{5}}
\,.
\enann
Since $0<\xi_-<\xi_+<1<\xi_0$  in this case, 
 we find 
 \beann
{\xi_+-\xi\over \xi_0-\xi}={\xi_+-\xi_-\over \xi_0-\xi_-}{\rm sn}^2\left(\beta_{\rm (lib)}
(\tilde \tau-\tilde \tau_+), k_{\rm (lib)}\right)
\,,
\enann
where 
\beann
\beta_{\rm (lib)}&\equiv& 12\sqrt{6(\xi_0-\xi_-)}
\\
k_{\rm (lib)}&\equiv &\sqrt{\xi_+-\xi_-\over \xi_0-\xi_-}
\,.
\enann

\begin{widetext}
We then find the evolution of the eccentricity $e$ as
\beann
e^2={(1-\xi_+)(\xi_0-\xi_-)-(1-\xi_0)(\xi_+-\xi_-){\rm sn}^2\left(\beta_{\rm (lib)}
(\tilde \tau-\tilde \tau_0), k_{\rm (lib)}\right)
\over
(\xi_0-\xi_-)-(\xi_+-\xi_-){\rm sn}^2\left(\beta_{\rm (lib)}
(\tilde \tau-\tilde \tau_0), k_{\rm (lib)}\right)}
\,,
\enann
\end{widetext}
which gives 
\bea
e_{\rm max}=\sqrt{1-\xi_-}\,,
e_{\rm min}=\sqrt{1-\xi_+}.
\label{emaxemin_lib}
\ena

The vZLK time scale is given by 
\bea
T_{\rm vZLK}=\tau_{\rm vZLK}\, \mathfrak{T}_{\rm vZLK}^{\rm (lib)}\,,
\label{TKL_libration}
\ena
where
\beann
\mathfrak{T}_{\rm vZLK}^{\rm (lib)}&\equiv& 
{2 K\left(k_{\rm (lib)}\right)\over \beta_{\rm (lib)}}
\enann

The maximum and minimum values of the eccentricity in the vZLK oscillations
are determined by two conserved parameters, $\vartheta$ and $C_{\rm vZLK}$.
Note that the maximum eccentricity in vZLK oscillations is important, especially when we discuss emission of GWs.

The time scale of the vZLK oscillations is also important for 
observation of the gravitational waves.
Since 
$\mathfrak{T}_{\rm vZLK}^{\rm (rot)}$ and $\mathfrak{T}_{\rm vZLK}^{\rm (lib)}$
are order of unity, the time scale is almost determined by $\tau_{\rm vZLK}$,
which is rewritten by
\bea
n\tau_{\rm vZLK}=16 \mathfrak{f}\,\left(1+\epsilon_{\rm GR}\right)
\,,
\ena
where 
\beann
\mathfrak{f}\equiv {m_1+m_2\over M}\left({\mathfrak{r}_0\over \mathfrak{a}}\right)^3
\enann
denotes the firmness parameter of a binary.

\begin{figure}[htbp]
\begin{center}
\includegraphics[width=6cm]{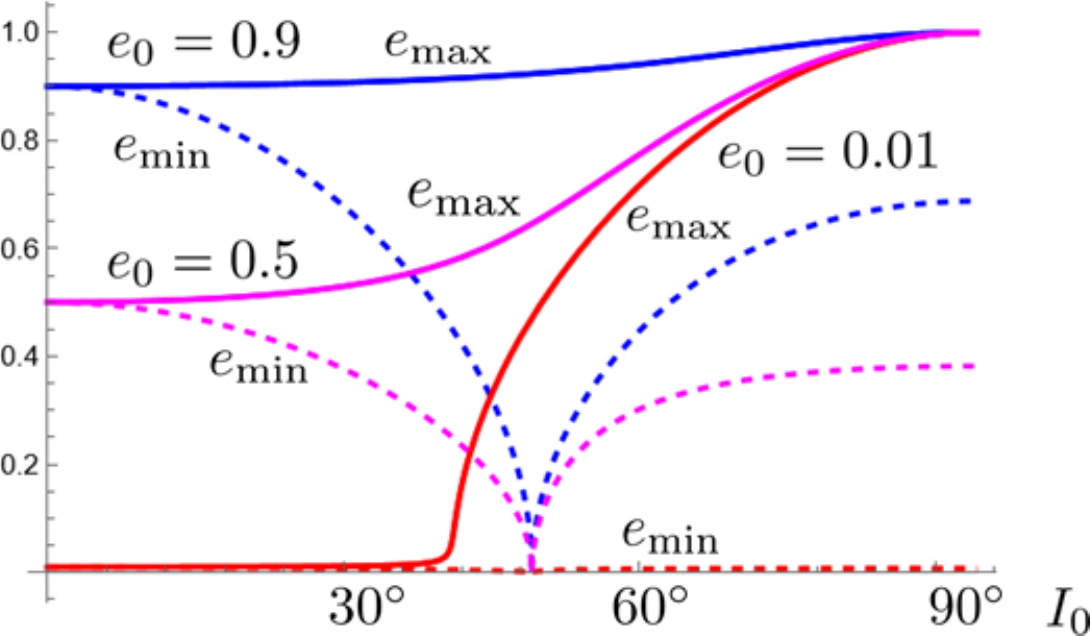}
\\
(a)
\\[1em]
\includegraphics[width=6cm]{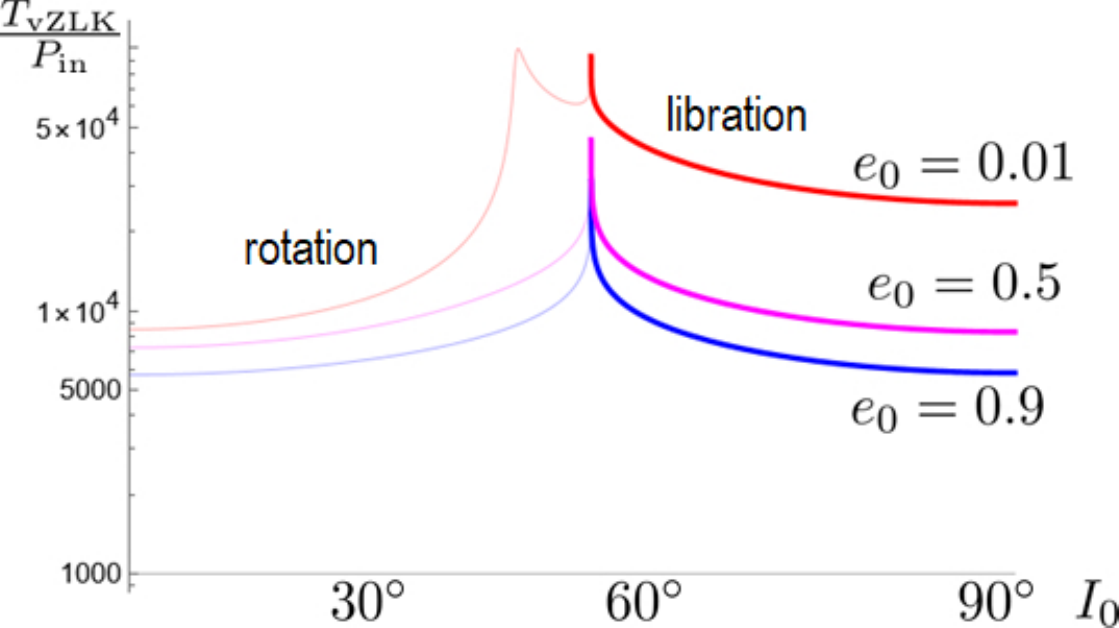}
\\
(b)
\caption{(a) Maximum (solid) and minimum (dashed) eccentricities, and 
(b) vZLK oscillation periods normalized by $P_{\rm in}$ in the DA approximation, for 
$e_0=0.01$ (red), $0.5$ (magenta), and $0.9$ (blue).
We fix $a=0.5M$, $\mathfrak{r}_0=9M$, $\mathfrak{a}_0=0.003M$.
}
\label{fig:emaxminTvZLK}
\end{center}
\end{figure}

Some example of $e_{\rm max}\,,
e_{\rm min}$ and 
$T_{\rm vZLK}$ is given in Fig. \ref{fig:emaxminTvZLK}.

\end{appendices}

\newpage

\bibliography{refer}

\end{document}